\documentclass[aps, prd, amsmath, amssymb, floats, floatfix, twocolumn, nofootinbib, superscriptaddress]{revtex4-1}

\usepackage{graphicx}
\usepackage{color}
\usepackage{latexsym}
\usepackage{epsfig}
\usepackage[colorlinks]{hyperref}
\usepackage{array}
\usepackage{enumitem}
\usepackage{textcomp}
\usepackage{footmisc}
\usepackage{natbib}
\usepackage{cleveref}
\setlist{noitemsep,leftmargin=*}

\newcommand{\be}{\begin{eqnarray}}
\newcommand{\ee}{\end{eqnarray}}

\newcommand{\bitem}{\begin{itemize}}
\newcommand{\eitem}{\end{itemize}}
\newcommand{\bwide}{\begin{widetext}}
\newcommand{\ewide}{\end{widetext}}

\begin{document}

\title{Constraining the Konoplya-Rezzolla-Zhidenko deformation parameters I:\\limits from supermassive black hole X-ray data}

\author{Askar~B.~Abdikamalov}
\affiliation{Center for Field Theory and Particle Physics and Department of Physics, Fudan University, 200438 Shanghai, China}
\affiliation{Ulugh Beg Astronomical Institute, Tashkent 100052, Uzbekistan}

\author{Dimitry~Ayzenberg}
\affiliation{Theoretical Astrophysics, Eberhard-Karls Universit\"at T\"ubingen, D-72076 T\"ubingen, Germany}

\author{Cosimo~Bambi}
\email[Corresponding author: ]{bambi@fudan.edu.cn}
\affiliation{Center for Field Theory and Particle Physics and Department of Physics, Fudan University, 200438 Shanghai, China}

\author{Sourabh~Nampalliwar}
\affiliation{Theoretical Astrophysics, Eberhard-Karls Universit\"at T\"ubingen, D-72076 T\"ubingen, Germany}

\author{Ashutosh~Tripathi}
\affiliation{Center for Field Theory and Particle Physics and Department of Physics, Fudan University, 200438 Shanghai, China}

\begin{abstract}
X-ray reflection spectroscopy is a powerful technique for probing the nature of gravity around black holes in the so-called strong-field regime. One popular approach is to look at theory-agnostic deviations away from the Kerr solution, which is the only astrophysically relevant black hole solution within classical general relativity, in order to verify whether astrophysical black holes are described by the Kerr metric. We have recently extended our X-ray reflection spectroscopy framework to a class of very general axisymmetric non-Kerr black holes proposed by Konoplya, Rezzolla \& Zhidenko (Phys. Rev. D93, 064015, 2016). Here, we analyze \textsl{XMM-Newton} and \textsl{NuSTAR} observations of the supermassive black hole in the Seyfert~1 galaxy MCG--06--30--15 with six different deviation parameters of this extended model. We recover the Kerr solution in all cases, but some deformation parameters are poorly constrained. We discuss the implications of this verification and future possibilities. 
\end{abstract}

\maketitle

\section{Introduction \label{s-intro}}
\begin{figure}[!htb]
		\includegraphics[width=0.48\textwidth]{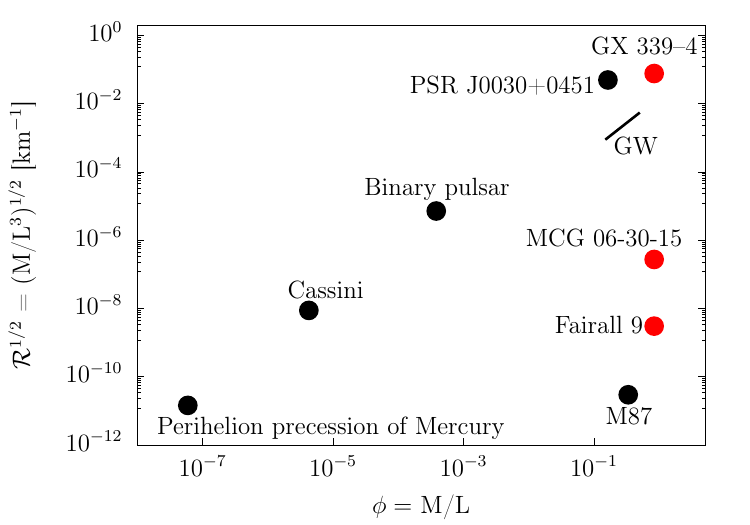}
		\caption{A potential-curvature plot showing several astrophysical systems which have been used to test GR. Sources analyzed with X-rays have been marked in red. A GW event appears as a dynamic system in this plot, and is denoted with a line instead of a point. See Tab~\ref{tab:scale} for details on the characteristic mass and length scales of the systems shown, and the text for discussion.}
		\label{f-scale}
	\end{figure}

One of the most important predictions of Einstein's theory of gravity, also known as general relativity (GR), is the existence of black holes (BHs). While originally thought of as mathematical idealizations, BHs are now expected to be present in myriad numbers throughout the Universe. With developments in technology, the ability to detect astrophysical systems has progressed remarkably over the last decade. Although there existed a quite general and powerful framework to study the behavior of gravity in the \textit{weak-field} regime from long ago, these new developments have enabled novel and much more precise (than before) probes of the behavior of strong-field gravity. BHs provide the best environments to perform such probes. One way to see this is to use the potential-curvature plot~\cite{Psaltis:2008bb,Baker:2014zba,Yunes:2016jcc,Cardenas-Avendano:2019zxd}, where we can classify astrophysical systems according to their characteristic curvature scale and characteristic potential scale. Following Refs.~\cite{Yunes:2016jcc,Cardenas-Avendano:2019zxd}, we define the characteristic curvature $\mathcal{R} = M/L^3$ and the characteristic potential $\phi = M/L$, where $M$ is the characteristic mass scale and $L$ the characteristic length scale of the astrophysical system under consideration. Fig.~\ref{f-scale} shows a range of astrophysical systems, which have been used to test GR, on such a plot. Among all the systems in the right half of the plot, corresponding to the strong-field regime, a majority have one or more BHs in the system.

As a consequence of the ``no-hair'' theorems (see, for instance, \cite{Chrusciel:2012jk} and references therein for their assumptions), four-dimensional GR predicts that isolated BHs in our Universe are described by only two parameters, which are refereed to as its mass and spin angular momentum, and defined by the Kerr solution.\footnote{A third parameter, the electric charge, though allowed within GR, is expected to be negligible in macroscopic astrophysical BHs~\cite{Bambi:2008hp}.}\footnote{The inverse is not true, i.e., the existence of BHs which satisfy the Kerr solution does not automatically validate GR, since there are theories that differ from GR but whose BH solutions coincide with those of GR~\cite{Psaltis:2007cw}.} This is known as the \textit{Kerr hypothesis}. The hypothesis is expected to hold even for BHs surrounded by accretion disks, since the gravitational effects of the disk are normally negligible compared to those of the BH~\cite{Bambi:2014koa}. This gives rise to an interesting possibility for testing GR in the strong-field regime with BHs -- consider a metric which parametrically deviates away from Kerr, i.e., the deviation, or \textit{deformation}, away from Kerr is controlled with a set of (possibly infinite) parameters. This new metric may or may not be the solution of a known theory of gravity (but see Ref.~\cite{Suvorov:2020bvk} for an interesting approach to mapping parametrically deformed metrics to some scalar-tensor theory of gravity). By analyzing astrophysical data against this new metric, one can try to constrain the deviation parameters and perform verification tests of GR in this theory-agnostic approach~\cite{Bambi2015}.

There are several techniques in vogue today which probe BHs and their environments. Theory-agnostic tests of gravity have been performed with gravitational waves (GWs)~\cite{LIGOScientific:2019fpa,Abbott:2020jks,Cardenas-Avendano:2019zxd}, X-ray spectroscopy~\cite{Cao:2017kdq,Tripathi:2020dni,Tripathi:2020yts,Bambi:2021hxv}, BH imaging~\cite{PhysRevLett.125.141104,Psaltis:2020ctj,Volkel:2020xlc}, and infrared observations of the Galactic Center~\cite{Abuter:2018drb}. Our focus in this work is on X-ray reflection spectroscopy (XRS). XRS is based on the idea of extracting information about the BH from relativistic reflection spectrum of accretion disks. The technique is well established for measuring the spin of Kerr BHs, and has recently been extended to perform both theory-agnostic and theory-specific tests of GR. In the presence of accurate models and high-quality data, XRS can be a very powerful technique for constraining deviations from the Kerr metric. One of the most attractive aspects of XRS that sets it apart from other techniques, is its applicability in both stellar-mass and supermassive BHs. This means that the whole of the right half of the potential-curvature phase space shown in Fig.~\ref{f-scale} is accessible to XRS-based tests. This is highlighted by marking one representative low-mass X-ray binary (GX~339--4~\cite{Tripathi:2020dni}), a typical AGN (MCG--06--30--15~\cite{Tripathi2019}) and a \textit{heavy} AGN (Fairall~9~\cite{Liu:2020fpv}). GWs from ground-based detectors, on the other hand, only cover the upper-right quadrant of this phase space, and BH imaging techniques only the lower-right quadrant.

One of the most popular theory-agnostic metrics in the market today is the metric proposed in Ref.~\cite{Konoplya2016} by Konoplya, Rezzolla \& Zhidenko (KRZ metric hereafter). The KRZ metric is a stationary axisymmetric metric written in Boyer-Lindquist-like coordinates. Notably, it does not always possess a Killing tensor and, as such, the equations of motion are not always separable. This makes it a better choice for verification tests of GR than those metrics that always have a Carter constant, since it captures a larger variety of deviations from Kerr. In addition, the metric deformation functions are expressed in terms of continued-fraction expansions, which has superior convergence compared to the more common $M/r$-based power series expansion. This feature provides significant advantage when dealing with rapidly rotating BHs where the characteristic length scales (the innermost stable circular orbit, the photon orbit, etc.) are $\sim M$ and higher-orders terms in the $M/r$ expansion become non-negligible. In a recent work, we implemented this metric in the XRS framework and put constraints on possible deviations from GR in terms of six distinct deviation parameters of the KRZ metric~\cite{Nampalliwar:2019iti}.

In the present work, we analyze the X-ray spectra of the Seyfert~1 galaxy MCG--06--30--15 as observed simultaneously by \textsl{XMM-Newton} and \textsl{NuSTAR} telescopes in 2013. We use the reflection model {\tt relxill\_nk}, a public model developed by us, to model the reflection component and constrain parameters of the KRZ metric. Our aim is to verify the Kerr hypothesis, namely, to verify whether, and how well, we can constrain the deviations to the Kerr solution using the KRZ metric. The presence of a very prominent and broad iron line in the spectrum and the unprecedented high quality of simultaneous \textsl{XMM-Newton} and \textsl{NuSTAR} observations make the 2013 data of MCG--06--30--15 particularly suitable for our test.

\begin{table}[tbp]
\renewcommand\arraystretch{1.5}
\centering
        \begin{tabular}{l c c}
        \hline\hline
System and main reference & $M~[M_{\odot}]$ & $L$ \\
        \hline
Cassini~\cite{Bertotti:2003rm} & $1$ & $1.1\times10^6$~km \\
Mercury's perihelion~\cite{Will2014} & $1$ & $5.8\times10^7$~km \\
Binary pulsar (Shapiro)~\cite{Yunes:2016jcc} & $1.34$ & $1.04\times10^4$~km \\
PSR~J0030+0451~\cite{Miller:2019cac,Silva:2020acr} & $1.44$ & 13~km \\
GX~339--4~\cite{Tripathi:2020dni} & $10$ & $R_{\rm ISCO}$  \\
GW150914~\cite{TheLIGOScientific:2016src} & $65.3$ & $385-1300$~km \\
MCG--06--30--15~\cite{Tripathi2019} & $2.8\times10^6$ & $R_{\rm ISCO}$  \\
Fairall~9~\cite{Liu:2020fpv} & $2.55\times10^8$  & $R_{\rm ISCO}$  \\
M87~\cite{PhysRevLett.125.141104} & $6.5\times10^9$ & $R_{\rm ph}$ \\
        \hline
        \end{tabular}
         \caption{\label{tab:scale} The characteristic mass and length scales of astrophysical systems shown in Fig.~\ref{f-scale}. For BHs analyzed with X-rays, the characteristic length scale is taken to be $R_{\rm ISCO}$, i.e, the radius of the innermost stable circular orbit (ISCO), while those analyzed with imaging have their characteristic length scale at $R_{\rm ph}$, i.e., the photon orbit. The binary pulsar data point is related to the Shapiro delay at the impact parameter~\cite{Yunes:2016jcc}.}
\end{table}

This article is organized as follows. Sec.~\ref{s-model} gives a review of the reflection model and the KRZ metric and its deformation parameters. Sec.~\ref{s-obs} presents the source properties and the details of the observation. Details of data analysis and results are given in Sec.~\ref{s-analysis}, and the results are discussed in Sec.~\ref{s-discuss}. Through the article, we use geometrized units, namely $c=G=1$, and use the metric signature $(-+++)$. Additionally, since XRS is independent of the mass of the BH and its distance from Earth, we set the BH mass $M=1$.  

\section{The {\tt relxill\_nk} model \label{s-model}}
{\tt relxill\_nk} is an extension of {\tt relxill}, the standard X-ray reflection model for Kerr BHs~\cite{Garcia2013,Dauser2014}, to metrics beyond the Kerr solution~\cite{relxillnk,Abdikamalov:2019yrr,Abdikamalov:2020oci}. {\tt relxill} itself combines the radiative transfer code {\tt xillver} that balances the microphysics inside the accretion disk in a rigorous way and provides a local spectrum~\cite{Garcia2010} and the relativistic blurring code {\tt relconv} that evolves the local spectrum along null geodesics, on a Kerr background, to calculate the spectrum as seen by a distant observer~\cite{Dauser:2010ne,Dauser:2013xv}. {\tt relxill\_nk} modifies {\tt relconv} to evolve the local spectrum on non-Kerr backgrounds.

\begin{figure}[!htb]
		\includegraphics[width=0.48\textwidth]{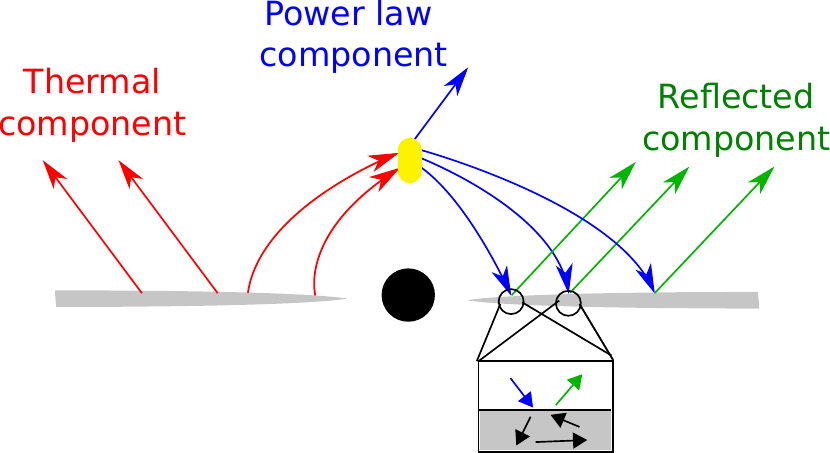}
		\caption{Sketch of the disk-corona model at the core of {\tt relxill\_nk}. Shown are the BH (black circle), the accretion disk (grey) and the corona (yellow). Various components of the total radiation are shown, though {\tt relxill\_nk} models the power-law and the reflection components. The inset shows conversion of incident radiation into reflected radiation.}
		\label{f-disk}
	\end{figure}
	
The fundamental morphology of the neighborhood of a BH in {\tt relxill\_nk} is the idealized disk-corona model~\cite{Bambi:2020jpe}. Although the model has been extended to include many more features, for the sake of simplicity we will present the most basic features here. Fig.~\ref{f-disk} presents a sketch of the system. At the center of the system is the BH, and a geometrically-thin and optically-thick accretion disk lies in its equatorial plane (the latter being defined relative to the BH spin axis)~\cite{Novikov1973}. The disk emits thermal radiation according to Planck's law. This radiation is upscattered via inverse Compton scattering in the corona, a weakly-understood region of extremely hot electron plasma, and appears as a hard X-ray power-law continuum. Some of these hard X-rays irradiate, get reprocessed, and are re-emitted from the disk, giving rise to a \textit{reflection} component.

After emission, the radiation travels along null geodesics towards the observer. The total flux received at the observer is given as
\be\label{eq-Fobs}
F_o (\nu_o) 
= \int I_o(\nu_o, X, Y) d\tilde{\Omega},
\ee
where $I_{ o}$ is the intensity received by the observer and depends on the photon frequency at the observer, $X$ and $Y$ are Cartesian coordinates on the plane of the observer, and $d\tilde{\Omega}$ is the integration element on this plane. Since the intensity is known at the point of emission (given by Planck's law in the case of the thermal component and by {\tt xillver}, for instance, in the case of the reflection component), we relate $I_{ o}$ to the intensity at emission $I_{ e}$ with Liouville's theorem as follows
\be
	I_o (\nu_o) = g^3 I_e (\nu_e)
\ee
where $g = \nu_o/\nu_e$ is the redshift the photons experience on their way from emission to observation. 

At this stage, calculation of flux involves raytracing photons every time the flux has to be calculated. This can be extremely time-consuming, especially for non-Kerr metric backgrounds where any simplification of the geodesic evolution equations may not be possible, and cumbersome for data analysis. The {\tt relxill} and {\tt relxill\_nk} suites of models use a transfer function which acts as an integration kernel and considerably speeds up computation of the flux. It is defined as~\cite{Cunningham1975}
\be\label{eq-trf}
f(g^*,r_e,\iota) = \frac{1}{\pi r_e} g 
\sqrt{g^* (1 - g^*)} \left| \frac{\partial \left(X,Y\right)}{\partial \left(g^*,r_e\right)} \right| \, ,
\ee
where $r_e$ is the radial coordinate on the disk, $\iota$ is the inclination of the observer relative to the BH spin axis, $g^*$ is the normalized redshift factor, defined as
\be
g^* = \frac{g - g_{\rm min}}{g_{\rm max} - g_{\rm min}} \, ,
\ee
where $g_{\rm min}$ and $g_{\rm max}$ are, respectively, the minimum and maximum redshift at a constant $r_e$ and $\iota$, and $\frac{\partial \left(X,Y\right)}{\partial \left(g^*,r_e\right)}$ is the Jacobian relating quantities at the observer and the disk.

The metric we use here to test the Kerr hypothesis is given in Boyer-Lindquist-like coordinates $(t, r, \theta, \phi)$ as~\cite{Konoplya2016, Ni2016, Nampalliwar:2019iti}
\begin{align}\label{eq:metric}
ds^2 =& - \frac{N^2 - W^2 \sin^2\theta}{K^2} \, dt^2 - 2 W r \sin^2\theta \, dt \, d\phi\nonumber\\
&+ K^2 r^2 \sin^2\theta \, d\phi^2 + \frac{\Sigma \, B^2}{N^2} \, dr^2 + \Sigma \, r^2 \, d\theta^2 \, ,
\end{align}
where the metric functions are defined as
\bwide
\be\label{eq:deffunc}
N^2 &=& \left(1 - \frac{r_0}{r}\right) \left (1 - \frac{\epsilon_0 r_0}{r} 
+ \left(k_{00} - \epsilon_0\right)\frac{r_0^2}{r^2} + \frac{\delta_1 r^3_0}{r^3}\right ) 
+ \left( \frac{a_{20} r^3_0}{r^3}
+ \frac{a_{21} r^4_0}{r^4} + \frac{k_{21} r^3_0}{r^3 \left ( 1+ \frac{k_{22}(1-\frac{r_0}{r}) }{1+k_{23}(1-\frac{r_0}{r})}\right)}\right) \cos^2\theta  \, ,\nonumber\\
K^2 &=& 1 + \frac{a_* W}{r} + \frac{1}{\Sigma} \left( \frac{k_{00} r^2_0}{r^2} 
+ \left ( \frac{k_{20} r^2_0}{r^2} + \frac{k_{21} r^3_0}{r^3 \left ( 1+ \frac{k_{22}(1-\frac{r_0}{r})}{1+k_{23}(1-\frac{r_0}{r})}\right ) } \right ) \cos^2\theta \right)\, \\
W &=& \frac{1}{\Sigma} \left(\frac{w_{00} r^2_0}{r^2} + \frac{\delta_2 r^3_0}{r^3}
+ \frac{\delta_3 r^3_0}{r^3} \cos^2\theta \right) \, , \qquad B = 1 + \frac{\delta_4 r^2_0}{r^2} + \frac{\delta_5 r^2_0}{r^2} \cos^2\theta \, , \qquad
\Sigma = 1 + \frac{a_*^2}{r^2} \cos^2\theta \,.\nonumber
\ee
\ewide
Here $a_*=J/M^2$ is the dimensionless BH spin, and
\begin{gather}
r_0 = 1 + \sqrt{1 - a_*^2} \, ,\\ a_{21} = - \frac{a_*^4}{r^4_0} + \delta_6 \, , \qquad k_{21} = \frac{a_*^4}{r_0^4} - \frac{2 a_*^2}{r^3_0} - \delta_6 \, .
\end{gather}
The metric contains six parameters, denoted by $\{ \delta_i \}$ ($i = 1, 2, \dots, 6$), quantifying deviations away from the Kerr solution. The remaining parameters are defined such that Eq.~\ref{eq:metric} reduces to the Kerr metric when all $\{ \delta_i \}$ are identically zero. Their exact expressions can be found in Ref.~\cite{Nampalliwar:2019iti}. (Note that the expressions given in Ref.~\cite{Konoplya2016} do not reduce to the Kerr metric, and the correct expressions are given in Ref.~\cite{Nampalliwar:2019iti}.) Therein are also given bounds on $\{ \delta_i \}$ that are required to ensure regularity of the spacetime outside the horizon (e.g., a negative definite metric determinant, a positive definite $g_{\phi\phi}$, and a nonzero $N^2$). In particular, 
\begin{gather}\label{eq-regularity}
	\delta_1 > \frac{4r_0 - 3r_0^2 - a_*^2}{r_0^2}, \\\nonumber
	\delta_2, \delta_3 \left\{\begin{array}{l}
		> \\
		<\\
		\end{array} -\frac{4}{a_*^3}(1-\sqrt{1-a_*^2}) \quad \begin{array}{l}
		 \rm{if} \; a_* > 0\\ 
		 \rm{if} \; a_* < 0,
		\end{array} \right. \\\nonumber
	\delta_4, \delta_5 > -1.
\end{gather}
The bounds on $\delta_6$ turn out to be stronger than what is reported in Ref.~\cite{Nampalliwar:2019iti}. While the new bound cannot be expressed analytically, it is easily evaluated numerically. The following analysis takes this new bound into account and restricts the parameter exploration to only the allowed region.

We note that $\delta_1$ is associated to a deformation of the metric coefficient $g_{tt}$, $\delta_2$ and $\delta_3$ to deformations related to the BH rotation, $\delta_4$ and $\delta_5$ to deformations of $g_{rr}$, and $\delta_6$ alters the shape of the BH event horizon. Since the structure of an infinitesimally thin disk (in particular, ISCO radius and orbital velocity of the gas) are determined by $g_{tt}$, $g_{t\phi}$, and $g_{\phi\phi}$, only $\delta_1$, $\delta_2$, $\delta_3$, and $\delta_6$ can modify the motion of the gas in the disk. However, $\delta_1$ and $\delta_2$ have a large impact on the disk, while the effect of $\delta_3$ and $\delta_6$ is quite weak. $\delta_4$ and $\delta_5$ do not have any effect on the disk and only change the motion of the X-ray photons from the emission point in the disk to the detection point far from the source. The impact of the deformation parameters $\{ \delta_i \}$ on the reflection spectrum of a disk was shown in Ref.~\cite{Nampalliwar:2019iti}.

The {\tt relxill\_nk} model has two parameters that control the non-Kerr nature of the BH. One parameter is used to decide the type of deviation (e.g.~the $i$ in $\delta_i$), and the other decides the size of the deviation. Since {\tt relxill\_nk} allows for one type of deviation at a time, the analysis is performed for each $\delta_i$ separately. In the following sections, we will use the model to analyze some X-ray data.
 
\section{Source and observation overview \label{s-obs}}

MCG--06--30--15 is a bright Active Galactic Nucleus (AGN) in which a broad
iron line was clearly detected by \textsl{ASCA} for the first time \cite{Tanaka1995}. 
The iron K$\alpha$ line was extended to lower energies which indicates its origin in the innermost regions
of the BH \cite{Iwasawa1996,Iwasawa1999}. MCG--06--30--15 has been observed by many
X-ray missions like \textsl{BeppoSAX}~\cite{Guainazzi1998}, \textsl{RXTE}~\cite{Lee2000, Vaughan2001}, 
\textsl{XMM-Newton}~\cite{Wilms2001, Fabian2002, Fabian2003,
Vaughan2004, Brenneman2006}, \textsl{Suzaku}~\cite{Miniutti2007, Noda2011}, and \textsl{NuSTAR}~\cite{Marinucci2014}. 
The observation of MCG--06--30--15  by \textsl{NuSTAR}, along with the simultaneous \textsl{XMM-Newton} observation, displays a prominent Compton hump 
around 20-30~keV and the iron K$\alpha$ line peaked at 6-7-keV. The presence of these features make this source suitable for testing 
general relativity using X-ray reflection spectroscopy. Ref.~\cite{Tripathi2019} analyzed the same dataset for testing the Kerr 
hypothesis using the Johannsen metric~\cite{Johannsen2015} as the background metric. 
The spectrum of this source at lower energies is very complex due to absorption by warm ionized winds \cite{Otani1996}. 
High resolution \textsl{Chandra} and \textsl{XMM-Newton} studies confirmed the presence of absorbers around the source~\cite{Brand2001, 
Lee2001, Young2005, Turner2003, Turner2004}. Besides these complexities, the source is also found to be extremely variable~\cite{Tripathi2019}. 

\subsection{Observations and Data Reduction}

{\textsl{ XMM-Newton}}~\cite{Jansen2001} with its EPIC CCD detectors Pn~\cite{struder2001} and MOS1/2~\cite{Turner2001} 
observed MCG--06--30--15 for three consecutive revolutions
(obs. ID 0693781201, 0693781301 and 0693781401) starting 2013 January 29 for about 315 ks. The Pn raw data for these
revolutions are downloaded from the HEASARC website and is processed into cleaned event files using Science Analysis 
Software (SAS) v16.0.0. MOS data is not included in this analysis because it is severely affected by pileup. 
TABTIGEN is used to generate good time intervals (GTIs). A source region of radius 40 arcsec is taken around the center of the source.
A background region of 50 arcsec is taken as far as possible from the source to avoid any contamination
from source photons. The corresponding ancillary and response files are generated using the SAS routines
ARFGEN and RMFGEN, respectively. Finally, the source spectra is rebinned such that it oversamples the
instrumental resolution by a factor of 3 and has a minimum of 50 counts per bin. 

{\textsl{ NuSTAR}}~\cite{Harrison2013} with its two detectors FPMA and FPMB observed this source simultaneously 
with {\textsl{ XMM-Newton}} for about 360 ks (obs. ID 60001047002, 60001047003, and 60001047005). The raw data from both detectors are processed into cleaned event files using the
NUPIPELINE routine of the {\textsl{NuSTAR}} data analysis software (NuSTARDAS), which is distributed as part of the high
energy analysis software (HEASOFT). We use the latest Calibration files from the Calibration database (CALDB) v20180312. 
A source region of 70 arcsec is extracted from the cleaned event files around the center of the source. 
A background region of radius 100 arcsec is taken on the same detector and as far as possible from the source. 
Source spectra, background spectra, and response files are generated using the NUPRODUCTS
routine. The source spectra is rebinned to 70 counts per bin to improve the signal-to-noise ratio
and to apply the $\chi^2$ statistic. 
\begin{figure*}[t]
\vspace{-1.0cm}
\begin{center}
\includegraphics[width=8.0cm,trim={2.2cm 0 3.2cm 15cm},clip]{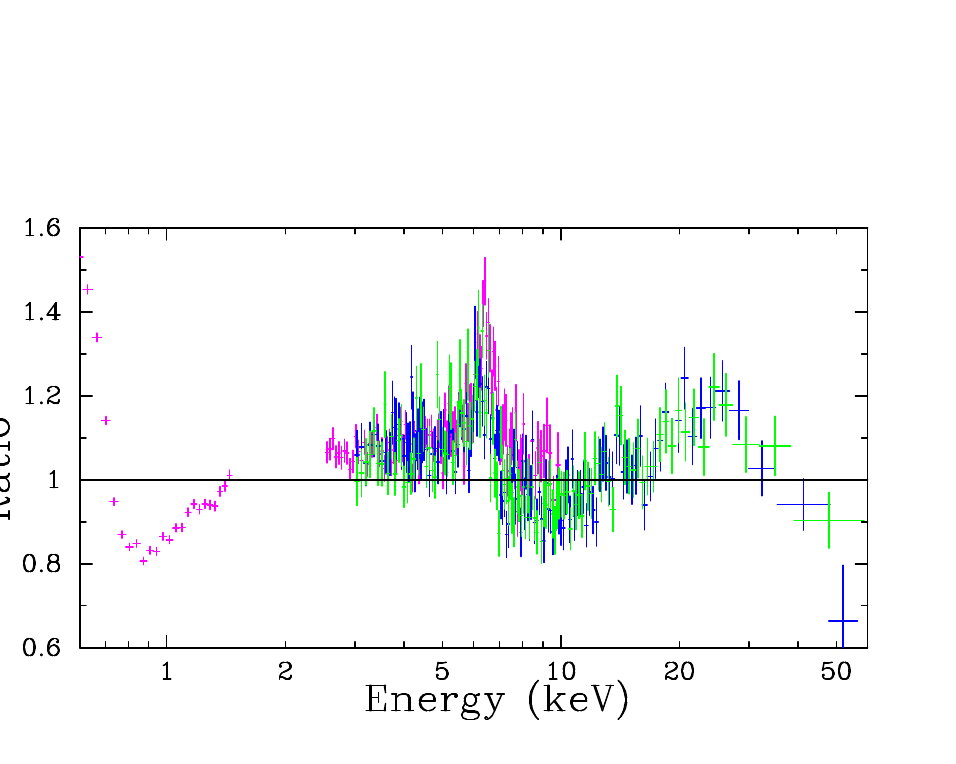}
\end{center}
\vspace{-0.7cm}
\caption{Data to model ratio for the absorbed power-law in the low flux state. Magenta, blue, and green
curves correspond to Pn, FPMA and FPMB data, respectively. \label{f-ratio-pow}}
\end{figure*}

Due to the extreme variability of the source using strictly simultaneous flux resolved data \cite{Tripathi2019} 
is necessary. We combined the GTIs from both {\textsl{XMM-Newton}} and {\textsl{NuSTAR}} cleaned event 
files using the ftool MGTIME. The data from EPIC-Pn, FPMA, and 
FPMB are divided into four flux states. These flux states are divided such that the counts in each state for each instrument 
is similar. 

 \section{Spectral analysis \label{s-analysis}}
 
 For further work, we used the X-ray spectral analysis package XSPEC v12.11.1 \cite{Arnaud1996}, WILMS abundance \cite{Wilms2000}, and VERN cross-section \cite{Verner1996} distributed as part of HEASOFT v6.28. 
 
 For each of the Pn, FPMA, and FPMB instruments, we have four spectra corresponding to four flux states 
 (low, medium, high, very high). So, there are twelve spectra in total which are fit simultaneously. 
 For each flux state, the cross-calibration constant for \textsl{XMM-Newton} is frozen to 1 leaving the 
 cross-calibration constant for FPMA ($C_{\rm FPMA}$) and FPMB ($C_{\rm FPMB}$) free to vary. Throughout 
 our analysis, the values of $C_{\rm FPMA}$ and $C_{\rm FPMB}$ are within 5\% of each other which is in agreement
 with the standard calibration of instruments. For \textsl{NuSTAR}, we used the energy range of 
 3.0-80.0 keV where the quality of the data is considered to be suitable for spectral studies. 
 For \textsl{XMM-Newton}, data in the 0.5-10.0 keV energy range is used. Due to poor data quality below 0.5 keV and 
 background domination above 10 keV, these energy ranges are excluded during analysis. 
 The energy range 1.5-2.5 of Pn data is not used because of the calibration issues 
 discussed in \cite{Marinucci2014} and \cite{Tripathi2019}.  
 
\begin{figure*}[t]
\vspace{-1.0cm}
\begin{center}
\includegraphics[width=8.0cm,trim={2.2cm 0 3.2cm 15cm},clip]{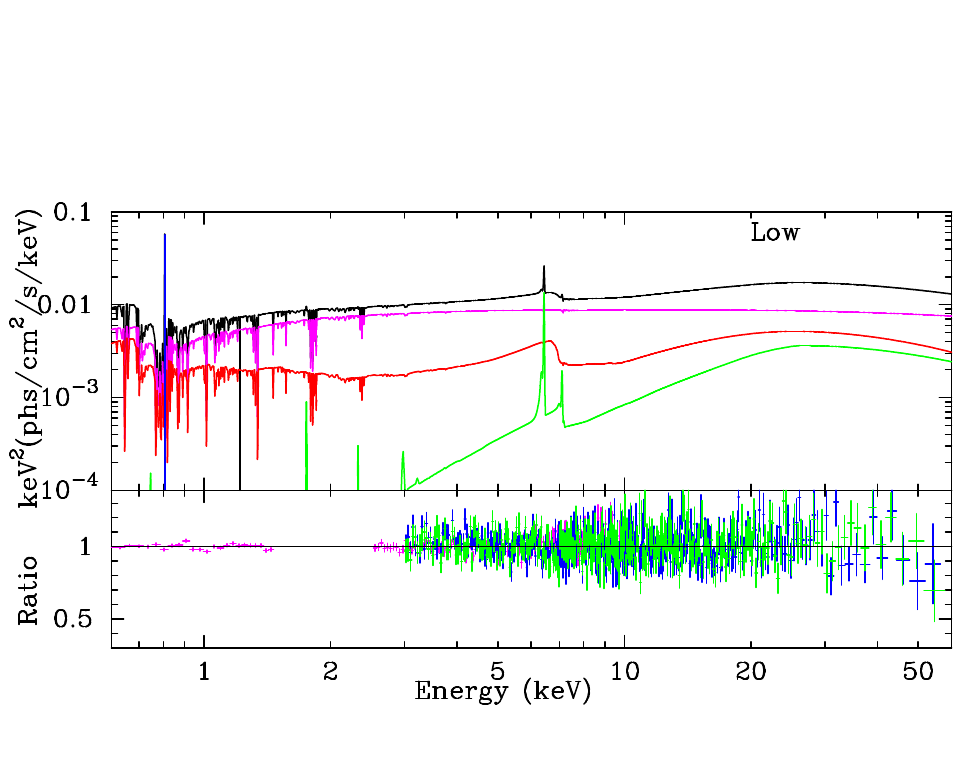}
\hspace{0.7cm}
\includegraphics[width=8.0cm,trim={2.2cm 0 3.2cm 15cm},clip]{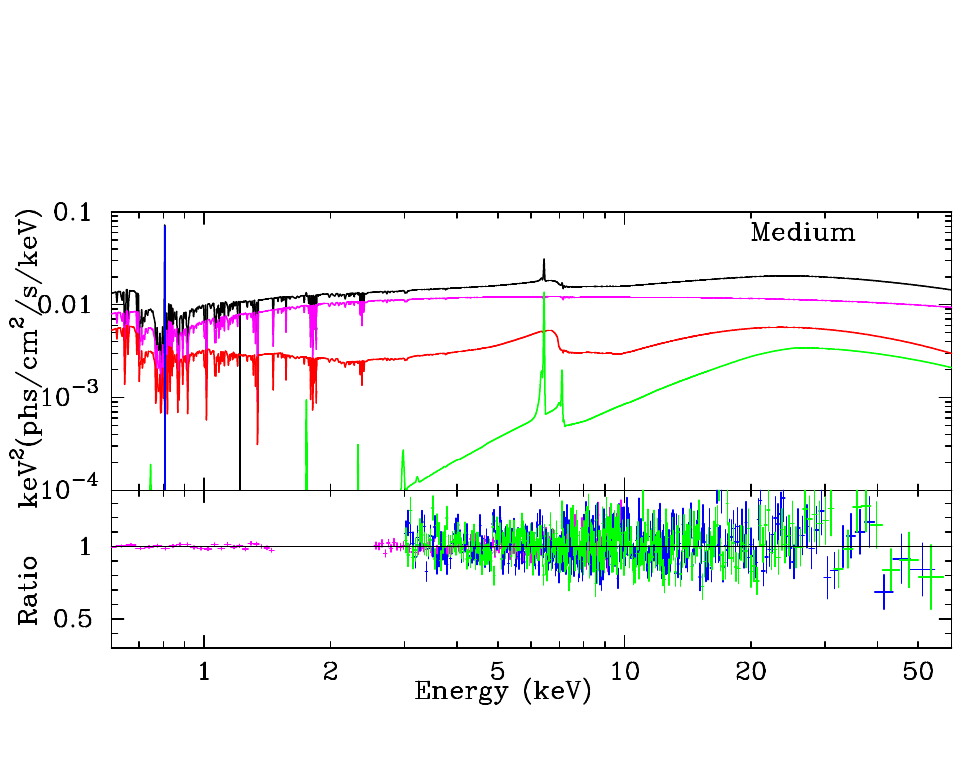} \\ 
\includegraphics[width=8.0cm,trim={2.2cm 0 3.2cm 18cm},clip]{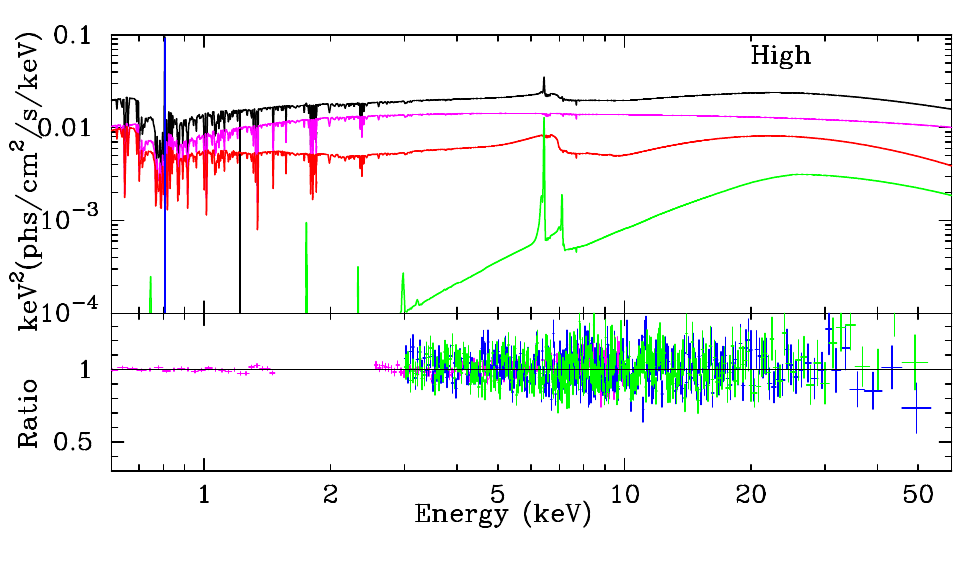}
\hspace{0.7cm}
\includegraphics[width=8.0cm,trim={2.2cm 0 3.2cm 18cm},clip]{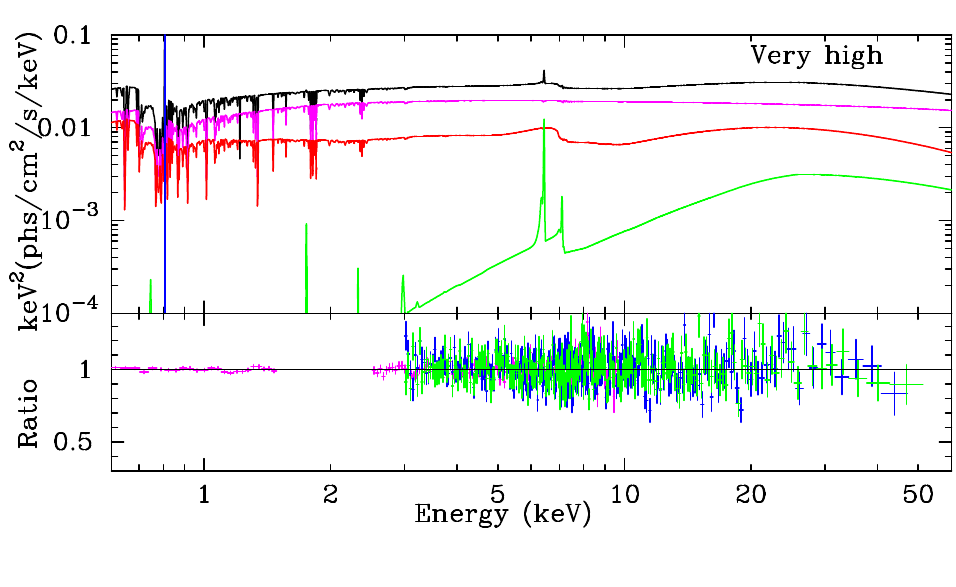}
\end{center}
\vspace{-0.7cm}
\caption{ The best-fit model (upper quadrant) and data to best-fit model ratio (lower quadrant) for
different flux states. In the upper quadrant, the black, magenta, red and green curves corresponds to 
total theoretical model, {\tt cutoffpl}, {\tt relxill\_nk} and {\tt xillver} respectively. In the lower
quadrant, magenta, blue and green crosses represent Pn, FPMA and FPMB data respectively.\label{f-ratio}}
\end{figure*}

 To display the features present in the observation, we fit the data with the absorbed power-law.
 Fig.~\ref{f-ratio-pow} shows the ratio of the lowest flux state data to the model {\tt tbabs*cutoffpl}. 
 Broad iron K$\alpha$ line and Compton hump are clearly visible around 6.5 keV and 30 keV,
 respectively. At lower energies (below 3 keV), residuals are present which could be due to absorption by warm 
 ionized clouds \cite{Lee2001, Sako2003}. It is quite common in AGN to have these features along with the excess photon 
 counts at lower energies \cite{Gierlinski2004, Crummy2006, Miniutti2009, Walton2014}. 
 
\begin{table*}[t]
\centering
\vspace{0.5cm}
\begin{tabular}{l|cccc|cccc}
\hline\hline
Model & \multicolumn{4}{c}{$\delta_1$} & \multicolumn{4}{c}{$\delta_2$} \\
\hline
Group & 1 & 2 & 3 & 4 & 1 & 2 & 3 & 4 \\
\hline
{\tt tbabs} &&&& &&&& \\
$N_{\rm H} / 10^{22}$ cm$^{-2}$ & \multicolumn{4}{c}{$0.039^\star$} & \multicolumn{4}{c}{$0.039^\star$} \\
\hline
{\tt warmabs$_1$} &&&& \\
$N_{\rm H \, 1} / 10^{22}$ cm$^{-2}$ & $0.47^{+0.11}_{-0.11}$ & $1.189^{+0.017}_{-0.056}$ & $1.017^{+0.019}_{-0.036}$ & $0.75^{+0.03}_{-0.04}$ 
& $0.63^{+0.07}_{-0.06}$ & $1.182^{+0.012}_{-0.052}$ & $1.010^{+0.012}_{-0.030}$ & $0.74^{+0.07}_{-0.03}$ \\
$\log\xi_1$ & $1.84^{+0.04}_{-0.02}$ & $1.954^{+0.012}_{-0.018}$ & $1.920^{+0.015}_{-0.019}$ & $1.828^{+0.012}_{-0.017}$ 
& $1.91^{+0.71}_{-0.05}$ & $1.96^{+0.01}_{-0.02}$ & $1.922^{+0.010}_{-0.022}$ & $1.830^{+0.009}_{-0.019}$ \\
\hline
{\tt warmabs$_2$} &&&& \\
$N_{\rm H \, 2} / 10^{22}$ cm$^{-2}$ & $0.64^{+0.11}_{-0.21}$ & $0.02^{+0.02}_{-0.02}$ & $0.52^{+0.14}_{-0.12}$ & $0.25^{+0.04}_{-0.04}$ 
& $0.48^{+0.14}_{-0.06}$ & $0.02^{+0.02}_{-0.02}$ & $0.53^{+0.18}_{-0.09}$ & $0.26^{+0.03}_{-0.06}$ \\
$\log\xi_2$ & $1.92^{+0.02}_{-0.08}$ & $3.1_{-0.6}$ & $3.23^{+0.05}_{-0.06}$ & $2.48^{+0.09}_{-0.05}$ 
& $1.86^{+0.03}_{-0.04}$ & $3.1_{-0.6}$ & $3.23^{+0.04}_{-0.04}$ & $2.49^{+0.09}_{-0.15}$ \\
\hline
{\tt dustyabs} &&&& \\
$\log \big( N_{\rm Fe} / 10^{21}$ cm$^{-2} \big)$ & \multicolumn{4}{c}{$17.403^{+0.007}_{-0.022}$} & \multicolumn{4}{c}{$17.412^{+0.019}_{-0.029}$} \\
\hline
{\tt cutoffpl} &&&& \\
$\Gamma$ & $1.954^{+0.008}_{-0.004}$ & $1.973^{+0.005}_{-0.004}$ & $2.010^{+0.004}_{-0.011}$ & $2.029^{+0.005}_{-0.011}$ & $1.956^{+0.006}_{-0.003}$ & $1.975^{+0.005}_{-0.003}$ & $2.016^{+0.004}_{-0.010}$ & $2.029^{+0.004}_{-0.011}$ \\
$E_{\rm cut}$ [keV] & $199^{+28}_{-28}$ & $159^{+24}_{-19}$ & $167^{+26}_{-20}$ & $287^{+112}_{-57}$
& $196^{+12}_{-27}$ & $150^{+15}_{-12}$ & $165^{+16}_{-16}$ & $281^{+93}_{-35}$ \\
norm~$(10^{-3})$ & $8.42^{+0.15}_{-0.12}$ & $12.24^{+0.20}_{-0.27}$ & $15.4^{+0.3}_{-0.3}$ & $21.3^{+0.4}_{-1.8}$ & $8.51^{+0.08}_{-0.30}$ & $12.6^{+0.7}_{-0.4}$ & $15.65^{+0.14}_{-0.11}$ & $21.68^{+0.16}_{-0.52}$ \\ 
\hline
{\tt relxill\_nk} &&&& \\
$q_{\rm in}$ & $6.8^{+0.9}_{-1.1}$ & $7.5^{+0.8}_{-0.9}$ & $7.7^{+0.4}_{-0.4}$ & $8.6^{+0.3}_{-0.4}$
& $6.8^{+0.4}_{-1.2}$ & $7.02^{+0.36}_{-0.18}$ & $7.81^{+0.36}_{-0.18}$ & $8.36^{+0.63}_{-0.15}$ \\
$q_{\rm out}$ & \multicolumn{4}{c}{$3^\star$} & \multicolumn{4}{c}{$3^\star$} \\
$R_{\rm br}$ [$M$] & $2.99^{+0.18}_{-0.29}$ & $3.03^{+0.15}_{-0.18}$ & $3.38^{+0.31}_{-0.11}$ & $3.41^{+0.19}_{-0.09}$
& $2.78^{+0.07}_{-0.22}$ & $2.83^{+0.68}_{-0.19}$ & $3.15^{+0.11}_{-0.04}$ & $3.24^{+0.47}_{-0.04}$ \\
$i$ [deg] & \multicolumn{4}{c}{$31.1^{+1.2}_{-1.4}$} & \multicolumn{4}{c}{$31.5^{+1.2}_{-1.5}$} \\
$a_*$ & \multicolumn{4}{c}{$0.956^{+0.006}_{-0.006}$} & \multicolumn{4}{c}{$0.959^{+0.011}_{-0.032}$} \\
$\delta$ & \multicolumn{4}{c}{$-0.14^{+0.09}_{-0.15}$} & \multicolumn{4}{c}{$0.14^{+0.34}_{-0.19}$} \\
$z$ & \multicolumn{4}{c}{$0.007749^\star$} & \multicolumn{4}{c}{$0.007749^\star$} \\
$\log\xi$ & $2.88^{+0.04}_{-0.08}$ & $3.01^{+0.03}_{-0.04}$ & $3.053^{+0.021}_{-0.013}$ & $3.133^{+0.021}_{-0.023}$ 
& $2.86^{+0.03}_{-0.05}$ & $2.95^{+0.04}_{-0.04}$ & $3.064^{+0.017}_{-0.022}$ & $3.125^{+0.009}_{-0.021}$ \\
$A_{\rm Fe}$ & \multicolumn{4}{c}{$3.15^{+0.21}_{-0.17}$} & \multicolumn{4}{c}{$3.28^{+0.26}_{-0.14}$} \\
norm~$(10^{-3})$ & $0.050^{+0.003}_{-0.004}$ & $0.063^{+0.004}_{-0.004}$ & $0.103^{+0.004}_{-0.010}$ & $0.13^{+0.08}_{-0.05}$ & $0.049^{+0.004}_{-0.004}$ & $0.059^{+0.002}_{-0.005}$ & $0.102^{+0.002}_{-0.005}$ & $0.128^{+0.002}_{-0.007}$ \\ 
\hline
{\tt xillver} &&&& \\
$\log\xi'$ & \multicolumn{4}{c}{$0^\star$} & \multicolumn{4}{c}{$0^\star$} \\
norm~$(10^{-3})$ & \multicolumn{4}{c}{$0.057^{+0.004}_{-0.004}$} & \multicolumn{4}{c}{$0.058^{+0.004}_{-0.003}$} \\
\hline
{\tt zgauss} &&&& &&&&\\
$E_{\rm line}$ [keV] & \multicolumn{4}{c}{$0.814^{+0.001}_{-0.003}$} & \multicolumn{4}{c}{$0.814^{+0.001}_{-0.003}$} \\
\hline
{\tt zgauss} &&&& &&&&\\
$E_{\rm line}$ [keV] & \multicolumn{4}{c}{$1.225^{+0.011}_{-0.009}$} & \multicolumn{4}{c}{$1.226^{+0.013}_{-0.009}$} \\
\hline
$\chi^2$/dof & \multicolumn{4}{c}{$3027.34/2685 = 1.12750$} & \multicolumn{4}{c}{$3028.04/2685 = 1.12776$} \\
\hline\hline
\end{tabular}
\vspace{0.2cm}
\caption{Summary of the best-fit values for the models with $\delta_{1}$ and $\delta_{2}$. The ionization parameter $\xi$ is in units erg~cm~s$^{-1}$. The reported uncertainties correspond to the 90\% confidence level for one relevant parameter. $^\star$ indicates that the parameter is frozen. The data sets 1, 2, 3, and 4 correspond, respectively, to the low, medium, high, and very-high flux states. See the text for more details. \label{t-fita}}
\end{table*}

\begin{table*}[t]
\centering
\vspace{0.5cm}
\begin{tabular}{l|cccc|cccc}
\hline\hline
Model & \multicolumn{4}{c}{$\delta_3$} & \multicolumn{4}{c}{$\delta_4$} \\
\hline
Group & 1 & 2 & 3 & 4 & 1 & 2 & 3 & 4 \\
\hline
{\tt tbabs} &&&& &&&& \\
$N_{\rm H} / 10^{22}$ cm$^{-2}$ & \multicolumn{4}{c}{$0.039^\star$} & \multicolumn{4}{c}{$0.039^\star$} \\
\hline
{\tt warmabs$_1$} &&&& \\
$N_{\rm H \, 1} / 10^{22}$ cm$^{-2}$ & $0.49^{+2.68}_{-0.02}$ & $1.177^{+0.026}_{-0.021}$ & $1.01^{+0.03}_{-0.05}$ & $0.74^{+0.06}_{-0.05}$ 
& $0.52^{+2.21}_{-0.06}$ & $1.183^{+0.024}_{-0.042}$ & $1.01^{+0.03}_{-0.05}$ & $0.74^{+0.04}_{-0.04}$ \\
$\log\xi_1$ & $1.86^{+1.20}_{-0.04}$ & $1.956^{+0.021}_{-0.019}$ & $1.922^{+0.024}_{-0.022}$ & $1.83^{+0.03}_{-0.03}$ 
& $1.86^{+0.05}_{-0.04}$ & $1.955^{+0.015}_{-0.020}$ & $1.921^{+0.018}_{-0.022}$ & $1.83^{+0.03}_{-0.03}$ \\
\hline
{\tt warmabs$_2$} &&&& \\
$N_{\rm H \, 2} / 10^{22}$ cm$^{-2}$ & $0.63^{+3.02}_{-0.02}$ & $0.02^{+0.02}_{-0.02}$ & $0.53^{+0.19}_{-0.17}$ & $0.25^{+0.06}_{-0.06}$ 
& $0.59^{+2.42}_{-0.06}$ & $0.02^{+0.02}_{-0.02}$ & $0.53^{+0.19}_{-0.17}$ & $0.25^{+0.06}_{-0.05}$ \\
$\log\xi_2$ & $1.91^{+1.14}_{-0.04}$ & $3.1_{-0.8}$ & $3.23^{+0.06}_{-0.01}$ & $2.49^{+0.09}_{-0.09}$ 
& $1.91^{+0.04}_{-0.08}$ & $3.1_{-0.8}$ & $3.23^{+0.06}_{-0.10}$ & $2.48^{+0.12}_{-0.15}$ \\
\hline
{\tt dustyabs} &&&& \\
$\log \big( N_{\rm Fe} / 10^{21}$ cm$^{-2} \big)$ & \multicolumn{4}{c}{$17.410^{+0.022}_{-0.033}$} & \multicolumn{4}{c}{$17.410^{+0.019}_{-0.016}$} \\
\hline
{\tt cutoffpl} &&&& \\
$\Gamma$ & $1.954^{+0.008}_{-0.008}$ & $1.972^{+0.009}_{-0.010}$ & $2.015^{+0.009}_{-0.011}$ & $2.028^{+0.007}_{-0.006}$ & $1.955^{+0.010}_{-0.009}$ & $1.974^{+0.010}_{-0.011}$ & $2.015^{+0.012}_{-0.010}$ & $2.028^{+0.012}_{-0.011}$ \\
$E_{\rm cut}$ [keV] & $196^{+50}_{-48}$ & $153^{+45}_{-25}$ & $160^{+38}_{-26}$ & $269^{+135}_{-72}$
& $198^{+50}_{-36}$ & $156^{+41}_{-27}$ & $165^{+39}_{-30}$ & $280^{+165}_{-76}$ \\
norm~$(10^{-3})$ & $8.5^{+0.5}_{-0.3}$ & $12.2^{+0.4}_{-0.8}$ & $15.5^{+1.1}_{-0.7}$ & $21.5^{+0.8}_{-1.9}$ & $8.48^{+0.22}_{-0.31}$ & $12.3^{+0.5}_{-0.4}$ & $15.4^{+0.5}_{-1.3}$ & $21.3^{+0.7}_{-1.3}$ \\ 
\hline
{\tt relxill\_nk} &&&& \\
$q_{\rm in}$ & $6.1^{+1.9}_{-2.9}$ & $7.2^{+1.6}_{-2.9}$ & $7.3^{+1.2}_{-1.2}$ & $8.2^{+0.6}_{-0.7}$
& $6.1^{+1.7}_{-3.1}$ & $7.2^{+1.8}_{-4.1}$ & $7.7^{+0.7}_{-0.8}$ & $8.36^{+0.63}_{-0.15}$ \\
$q_{\rm out}$ & \multicolumn{4}{c}{$3^\star$} & \multicolumn{4}{c}{$3^\star$} \\
$R_{\rm br}$ [$M$] & $2.99^{+0.47}_{-0.71}$ & $3.0^{+0.3}_{-0.4}$ & $3.43^{+0.44}_{-0.23}$ & $3.40^{+0.19}_{-0.18}$
& $3.0^{+0.7}_{-0.5}$ & $2.9^{+0.7}_{-0.5}$ & $3.30^{+0.20}_{-0.19}$ & $3.40^{+0.36}_{-0.19}$ \\
$i$ [deg] & \multicolumn{4}{c}{$31.4^{+1.5}_{-1.5}$} & \multicolumn{4}{c}{$31.3^{+1.1}_{-1.6}$} \\
$a_*$ & \multicolumn{4}{c}{$0.959^{+0.015}_{-0.031}$} & \multicolumn{4}{c}{$0.958^{+0.015}_{-0.020}$} \\
$\delta$ & \multicolumn{4}{c}{$5_{-7}^{\rm +(B)}$} & \multicolumn{4}{c}{$0.6^{+2.1}_{\rm -(B)}$} \\
$z$ & \multicolumn{4}{c}{$0.007749^\star$} & \multicolumn{4}{c}{$0.007749^\star$} \\
$\log\xi$ & $2.87^{+0.07}_{-0.05}$ & $3.007^{+0.017}_{-0.095}$ & $3.054^{+0.022}_{-0.022}$ & $3.13^{+0.04}_{-0.03}$ 
& $2.87^{+0.03}_{-0.06}$ & $3.00^{+0.04}_{-0.11}$ & $3.057^{+0.022}_{-0.021}$ & $3.13^{+0.04}_{-0.03}$ \\
$A_{\rm Fe}$ & \multicolumn{4}{c}{$3.14^{+0.28}_{-0.18}$} & \multicolumn{4}{c}{$3.19^{+0.25}_{-0.40}$} \\
norm~$(10^{-3})$ & $0.050^{+0.004}_{-0.004}$ & $0.062^{+0.004}_{-0.005}$ & $0.101^{+0.007}_{-0.012}$ & $0.126^{+0.019}_{-0.010}$ & $0.050^{+0.004}_{-0.004}$ & $0.062^{+0.004}_{-0.004}$ & $0.104^{+0.011}_{-0.006}$ & $0.129^{+0.009}_{-0.008}$ \\ 
\hline
{\tt xillver} &&&& \\
$\log\xi'$ & \multicolumn{4}{c}{$0^\star$} & \multicolumn{4}{c}{$0^\star$} \\
norm~$(10^{-3})$ & \multicolumn{4}{c}{$0.058^{+0.006}_{-0.005}$} & \multicolumn{4}{c}{$0.058^{+0.008}_{-0.007}$} \\
\hline
{\tt zgauss} &&&& &&&&\\
$E_{\rm line}$ [keV] & \multicolumn{4}{c}{$0.8142^{+0.0007}_{-0.0016}$} & \multicolumn{4}{c}{$0.814^{+0.001}_{-0.005}$} \\
\hline
{\tt zgauss} &&&& &&&&\\
$E_{\rm line}$ [keV] & \multicolumn{4}{c}{$1.225^{+0.011}_{-0.010}$} & \multicolumn{4}{c}{$1.226^{+0.011}_{-0.008}$} \\
\hline
$\chi^2$/dof & \multicolumn{4}{c}{$3027.40/2685 = 1.12752$} & \multicolumn{4}{c}{$3027.69/2685 = 1.12763$} \\
\hline\hline
\end{tabular}
\vspace{0.2cm}
\caption{Same as in table~\ref{t-fita} for $\delta_{3}$ and $\delta_{4}$.
(B) means that the 90\% uncertainty reaches the boundary of the regular spacetimes. 
\label{t-fitb}}
\end{table*}
 
\begin{table*}[t]
\centering
\vspace{0.5cm}
\begin{tabular}{l|cccc|cccc}
\hline\hline
Model & \multicolumn{4}{c}{$\delta_5$} & \multicolumn{4}{c}{$\delta_6$} \\
\hline
Group & 1 & 2 & 3 & 4 & 1 & 2 & 3 & 4 \\
\hline
{\tt tbabs} &&&& &&&& \\
$N_{\rm H} / 10^{22}$ cm$^{-2}$ & \multicolumn{4}{c}{$0.039^\star$} & \multicolumn{4}{c}{$0.039^\star$} \\
\hline
{\tt warmabs$_1$} &&&& \\
$N_{\rm H \, 1} / 10^{22}$ cm$^{-2}$ & $0.46^{+0.19}_{-0.09}$ & $1.18^{+0.04}_{-0.04}$ & $1.01^{+0.04}_{-0.05}$ & $0.74^{+0.10}_{-0.06}$ 
& $0.47^{+0.04}_{-0.10}$ & $1.191^{+0.025}_{-0.047}$ & $1.02^{+0.04}_{-0.05}$ & $0.74^{+0.02}_{-0.06}$ \\
$\log\xi_1$ & $1.85^{+0.13}_{-0.03}$ & $1.954^{+0.016}_{-0.018}$ & $1.919^{+0.021}_{-0.027}$ & $1.83^{+0.05}_{-0.03}$ 
& $1.85^{+0.07}_{-0.03}$ & $1.954^{+0.016}_{-0.019}$ & $1.920^{+0.020}_{-0.021}$ & $1.83^{+0.04}_{-0.04}$ \\
\hline
{\tt warmabs$_2$} &&&& \\
$N_{\rm H \, 2} / 10^{22}$ cm$^{-2}$ & $0.66^{+0.18}_{-0.43}$ & $0.02^{+0.02}_{-0.02}$ & $0.52^{+0.19}_{-0.15}$ & $0.25^{+0.06}_{-0.05}$ 
& $0.66^{+0.10}_{-0.06}$ & $0.02^{+0.02}_{-0.02}$ & $0.51^{+0.19}_{-0.18}$ & $0.25^{+0.06}_{-0.05}$ \\
$\log\xi_2$ & $1.91^{+0.03}_{-0.08}$ & $3.1_{-0.8}$ & $3.23^{+0.05}_{-0.08}$ & $2.48^{+0.16}_{-0.13}$ 
& $1.91^{+0.04}_{-0.09}$ & $3.1_{-0.8}$ & $3.23^{+0.07}_{-0.10}$ & $2.48^{+0.16}_{-0.13}$ \\
\hline
{\tt dustyabs} &&&& \\
$\log \big( N_{\rm Fe} / 10^{21}$ cm$^{-2} \big)$ & \multicolumn{4}{c}{$17.40^{+0.03}_{-0.03}$} & \multicolumn{4}{c}{$17.404^{+0.029}_{-0.015}$} \\
\hline
{\tt cutoffpl} &&&& \\
$\Gamma$ & $1.954^{+0.012}_{-0.013}$ & $1.971^{+0.018}_{-0.011}$ & $2.014^{+0.012}_{-0.011}$ & $2.026^{+0.011}_{-0.011}$ & $1.955^{+0.008}_{-0.010}$ & $1.974^{+0.008}_{-0.011}$ & $2.015^{+0.008}_{-0.010}$ & $2.029^{+0.009}_{-0.012}$ \\
$E_{\rm cut}$ [keV] & $200^{+51}_{-33}$ & $155^{+40}_{-25}$ & $164^{+41}_{-29}$ & $280^{+114}_{-77}$
& $198^{+50}_{-38}$ & $157^{+45}_{-28}$ & $166^{+41}_{-30}$ & $284^{+169}_{-84}$ \\
norm~$(10^{-3})$ & $8.39^{+0.16}_{-0.34}$ & $12.12^{+0.80}_{-0.20}$ & $15.2^{+1.3}_{-0.6}$ & $20.9^{+1.8}_{-0.6}$ & $8.46^{+0.23}_{-0.27}$ & $12.4^{+0.4}_{-0.8}$ & $15.3^{+1.3}_{-0.9}$ & $21.3^{+0.9}_{-0.8}$ \\ 
\hline
{\tt relxill\_nk} &&&& \\
$q_{\rm in}$ & $6.5^{+1.5}_{-1.9}$ & $7.6^{+0.9}_{-3.2}$ & $7.5^{+0.4}_{-0.4}$ & $8.4^{+0.5}_{-0.7}$
& $6.2^{+1.8}_{-1.8}$ & $7.0^{+1.7}_{-3.1}$ & $7.6^{+0.6}_{-0.8}$ & $8.2^{+0.6}_{-0.8}$ \\
$q_{\rm out}$ & \multicolumn{4}{c}{$3^\star$} & \multicolumn{4}{c}{$3^\star$} \\
$R_{\rm br}$ [$M$] & $2.8^{+0.9}_{-0.3}$ & $2.89^{+0.68}_{-0.22}$ & $3.30^{+0.80}_{-0.11}$ & $3.32^{+0.19}_{-0.14}$
& $2.9^{+0.8}_{-0.6}$ & $2.9^{+0.3}_{-0.5}$ & $3.27^{+0.25}_{-0.24}$ & $3.3^{+0.4}_{-0.4}$ \\
$i$ [deg] & \multicolumn{4}{c}{$31.4^{+1.5}_{-1.6}$} & \multicolumn{4}{c}{$31.2^{+1.5}_{-1.5}$} \\
$a_*$ & \multicolumn{4}{c}{$0.960^{+0.010}_{-0.014}$} & \multicolumn{4}{c}{$0.962^{+0.011}_{-0.010}$} \\
$\delta$ & \multicolumn{4}{c}{$-0.2^{+2.2}_{\rm -(B)}$} & \multicolumn{4}{c}{$-0.2^{\rm +(B)}_{\rm -(B)}$} \\
$z$ & \multicolumn{4}{c}{$0.007749^\star$} & \multicolumn{4}{c}{$0.007749^\star$} \\
$\log\xi$ & $2.88^{+0.07}_{-0.05}$ & $3.01^{+0.03}_{-0.05}$ & $3.059^{+0.019}_{-0.016}$ & $3.14^{+0.05}_{-0.04}$ 
& $2.874^{+0.066}_{-0.020}$ & $3.00^{+0.04}_{-0.12}$ & $3.059^{+0.021}_{-0.031}$ & $3.13^{+0.04}_{-0.03}$ \\
$A_{\rm Fe}$ & \multicolumn{4}{c}{$3.1^{+0.5}_{-0.3}$} & \multicolumn{4}{c}{$3.2^{+0.3}_{-0.3}$} \\
norm~$(10^{-3})$ & $0.049^{+0.006}_{-0.002}$ & $0.063^{+0.010}_{-0.006}$ & $0.104^{+0.019}_{-0.010}$ & $0.131^{+0.002}_{-0.005}$ & $0.050^{+0.004}_{-0.004}$ & $0.062^{+0.006}_{-0.006}$ & $0.104^{+0.006}_{-0.005}$ & $0.129^{+0.009}_{-0.010}$ \\ 
\hline
{\tt xillver} &&&& \\
$\log\xi'$ & \multicolumn{4}{c}{$0^\star$} & \multicolumn{4}{c}{$0^\star$} \\
norm~$(10^{-3})$ & \multicolumn{4}{c}{$0.057^{+0.007}_{-0.007}$} & \multicolumn{4}{c}{$0.058^{+0.008}_{-0.007}$} \\
\hline
{\tt zgauss} &&&& &&&&\\
$E_{\rm line}$ [keV] & \multicolumn{4}{c}{$0.8130^{+0.0015}_{-0.0009}$} & \multicolumn{4}{c}{$0.814^{+0.001}_{-0.005}$} \\
\hline
{\tt zgauss} &&&& &&&&\\
$E_{\rm line}$ [keV] & \multicolumn{4}{c}{$1.225^{+0.012}_{-0.009}$} & \multicolumn{4}{c}{$1.226^{+0.012}_{-0.010}$} \\
\hline
$\chi^2$/dof & \multicolumn{4}{c}{$3027.76/2685 = 1.12766$} & \multicolumn{4}{c}{$3027.81/2685 = 1.12767$} \\
\hline\hline
\end{tabular}
\vspace{0.2cm}
\caption{Same as in table~\ref{t-fita} for $\delta_{5}$ and $\delta_{6}$.
(B) means that the 90\% uncertainty reaches the boundary of the regular spacetimes.  \label{t-fitc}}
\end{table*}

 We add the relativistic reflection model {\tt relxill\_nk} to the absorbed power-law to fit the
 reflection component. To address the residuals at lower energies, we add two warm absorbers and 
 one dusty neutral absorber. Narrow line emissions around 7 keV are also present and are modeled with a distant reflector 
 that is non-relativistic in nature. A narrow emission line and absorption line can also be seen after adding 
 these components. In XSPEC, the model describing the source is written as :
 
 \vspace{0.2cm}

\noindent {\tt tbabs$\times$dustyabs$\times$warmabs$_1$$\times$warmabs$_2$$\times$(cutoffpl \\
+relxill\_nk+xillver+zgauss+zgauss)} .

\vspace{0.2cm}

 {\tt tbabs} accounts for galactic absorption along the
 line of sight of the observer and has column density ($N_{\rm H}$) as its only free parameter \cite{Wilms2000}. We freeze its 
 value to $3.9\times 10^{20}$cm$^{-3}$ obtained by other independent measurements \cite{Dickey1990}. {\tt dustyabs} accounts for the 
 neutral dust absorber and has iron density ({$\log N_{Fe}$}) as a free parameter. This multiplicative 
 table has been made especially for this source using the high-resolution \textsl{Chandra} data. Please see \cite{Lee2001} for
 more details about absorption by dust in MCG--06--30--15. {\tt warmabs$_1$} and {\tt warmabs$_2$} describes the two warm absorbers
 modeled with the multiplicative table constructed using {\tt xstar}. Each warm absorber is modeled 
 as an ionized zone characterized by column density ($N_{\rm H}$) and ionization parameter ($\log\xi$). {\tt cutoffpl} corresponds 
 to the power-law continuum with the photon index $\Gamma$, cut-off energy of the continuum ($E_{\rm cut}$), and the normalization
 as free parameters. {\tt relxill\_nk} describes  the reflection coming from the inner regions of the accretion disk where the 
 relativistic effects are significant~\cite{relxillnk, Abdikamalov:2019yrr}. In this work, we used 
 {\tt relxill\_nk} using the KRZ metric as the background metric. {\tt xillver} describes the reflection from the region far 
 away from the source where the relativistic effects are negligible \cite{Garcia2010}. {\tt zgauss} models the red-shifted Gaussian 
 line. Here, one of the Gaussians represents the emission line at 0.81 keV
 which is believed to be oxygen line emission due to relativistic outflow \cite{Leighly1997}. 
 The other Gaussian corresponds to the absorption line at 1.24 keV which is most likely the blue-shifted oxygen absorption.

\begin{figure*}[t]
\begin{center}
\includegraphics[width=8.5cm,trim={1cm 2.5cm 0cm 1cm},clip]{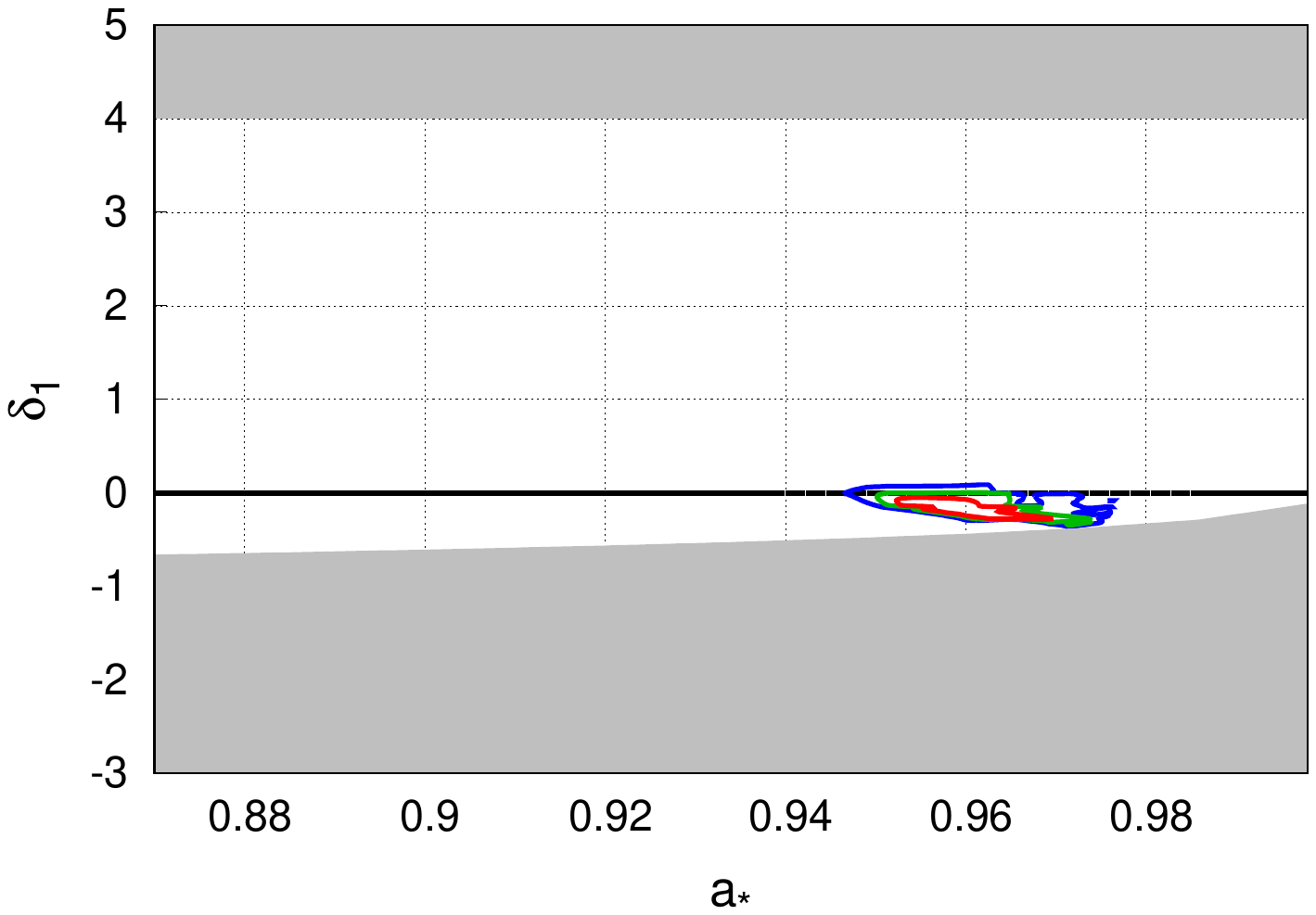}
\includegraphics[width=8.5cm,trim={1cm 2.5cm 0cm 1cm},clip]{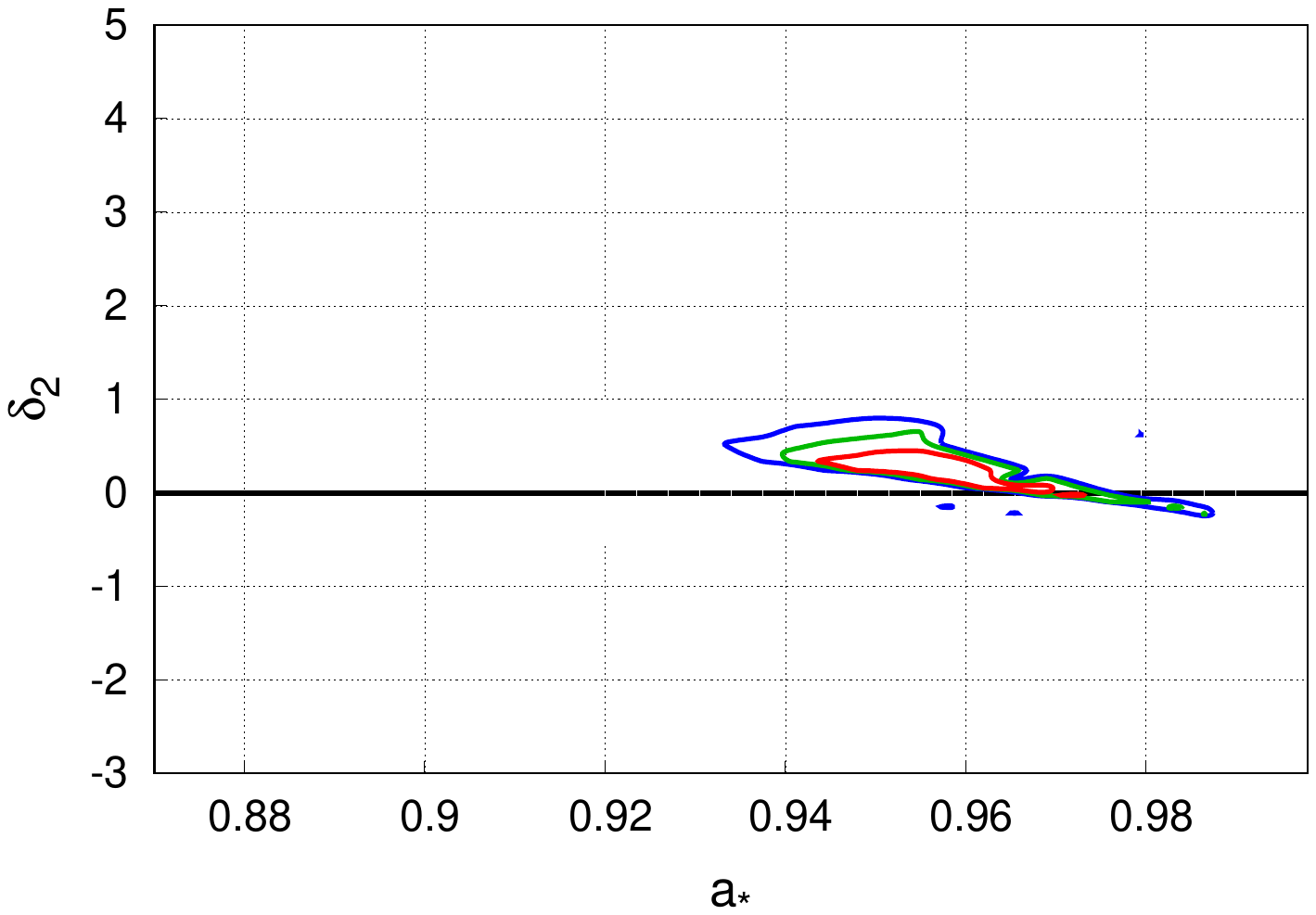} \\ 
\includegraphics[width=8.5cm,trim={1cm 2.5cm 0cm 1cm},clip]{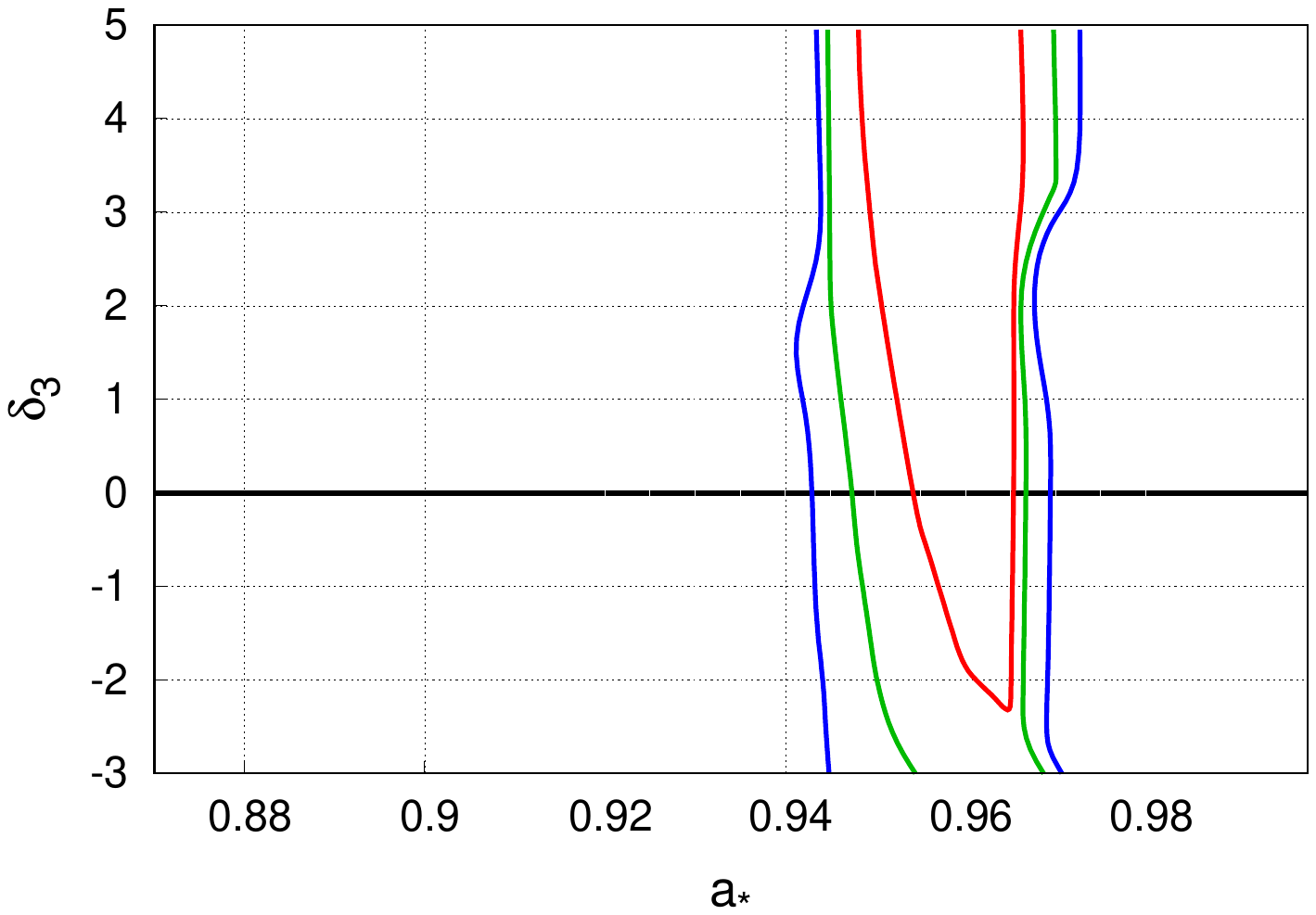}
\includegraphics[width=8.5cm,trim={1cm 2.5cm 0cm 1cm},clip]{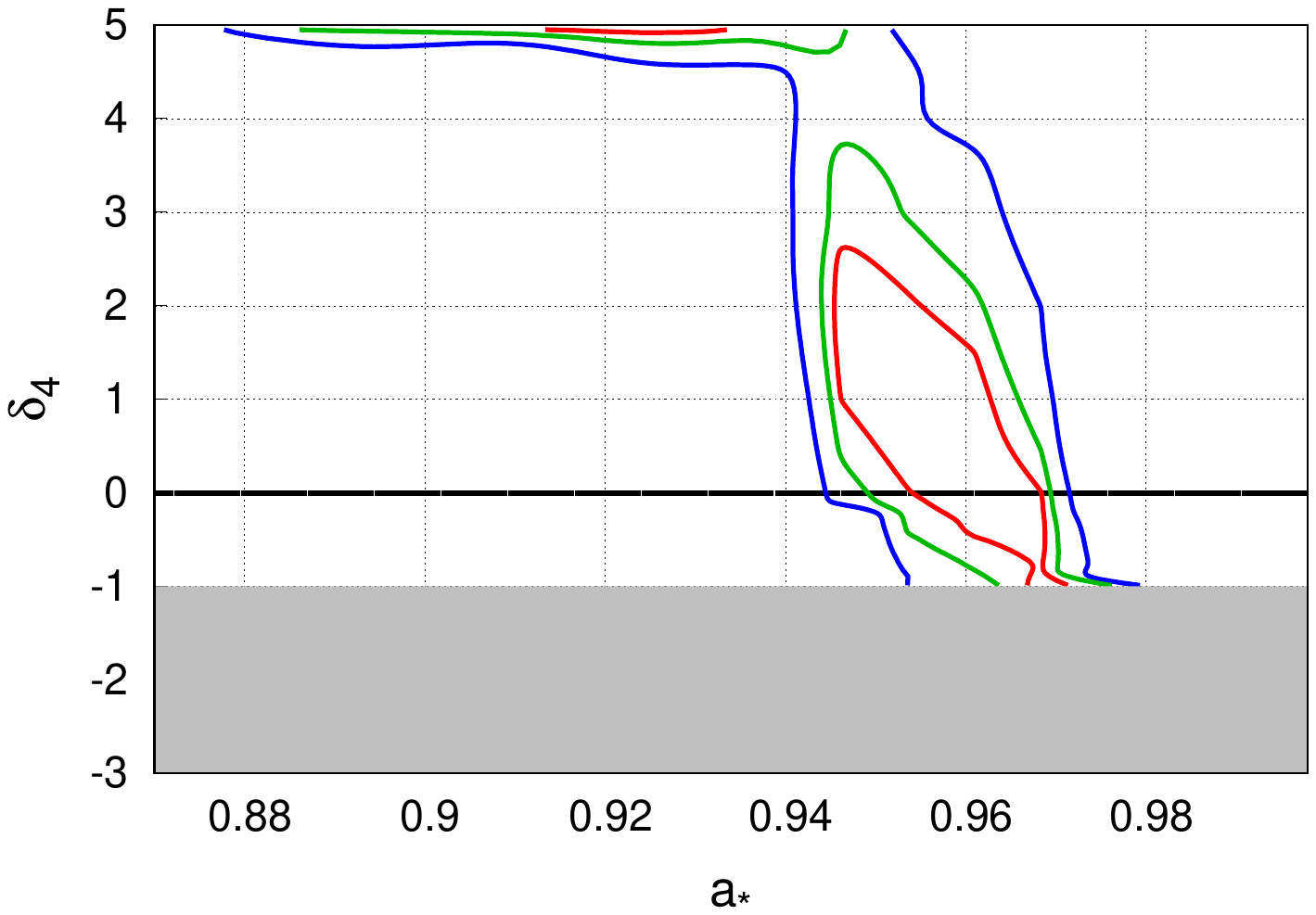}\\
\includegraphics[width=8.5cm,trim={1cm 2.5cm 0cm 1cm},clip]{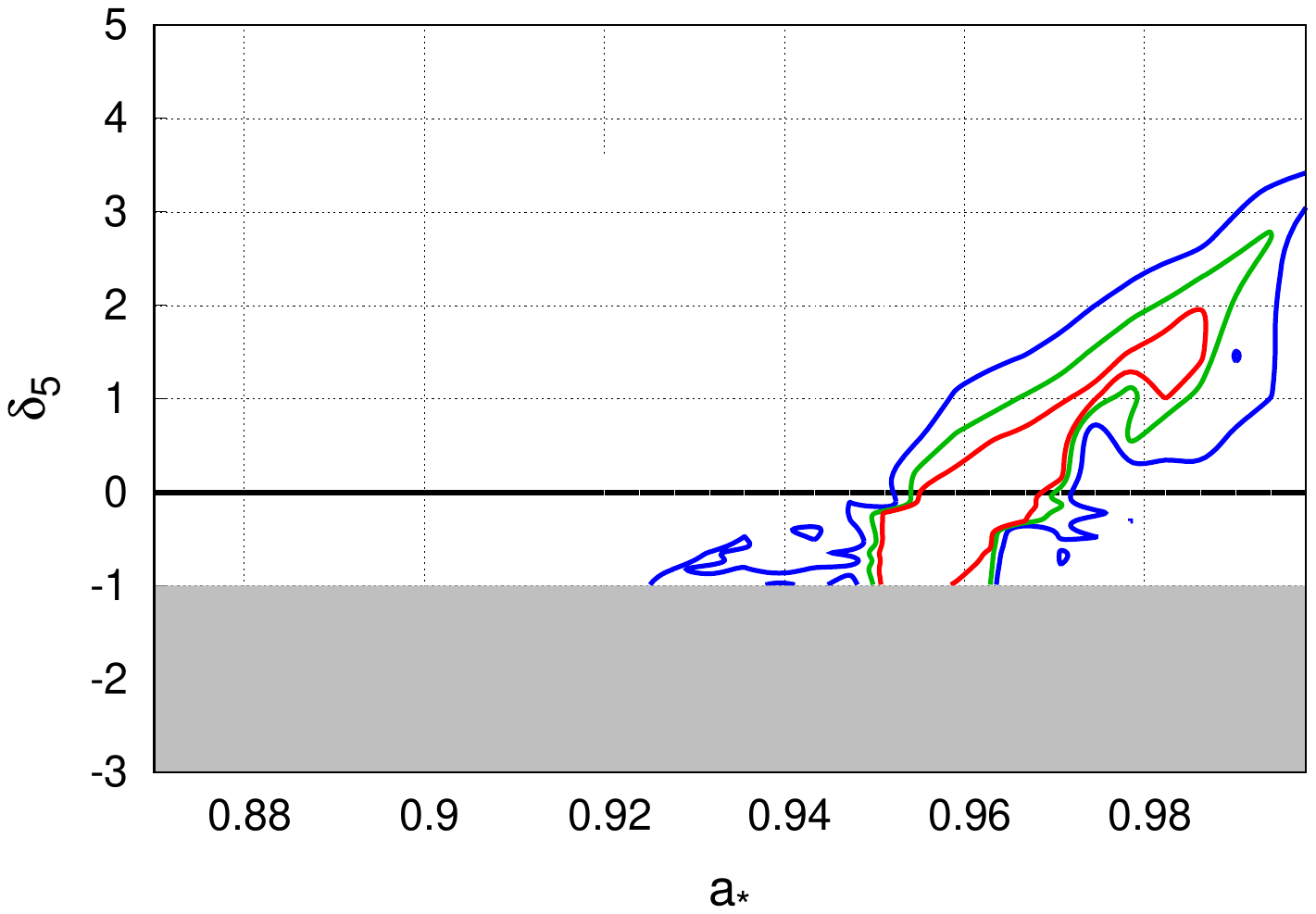}
\includegraphics[width=8.5cm,trim={1cm 2.5cm 0cm 1cm},clip]{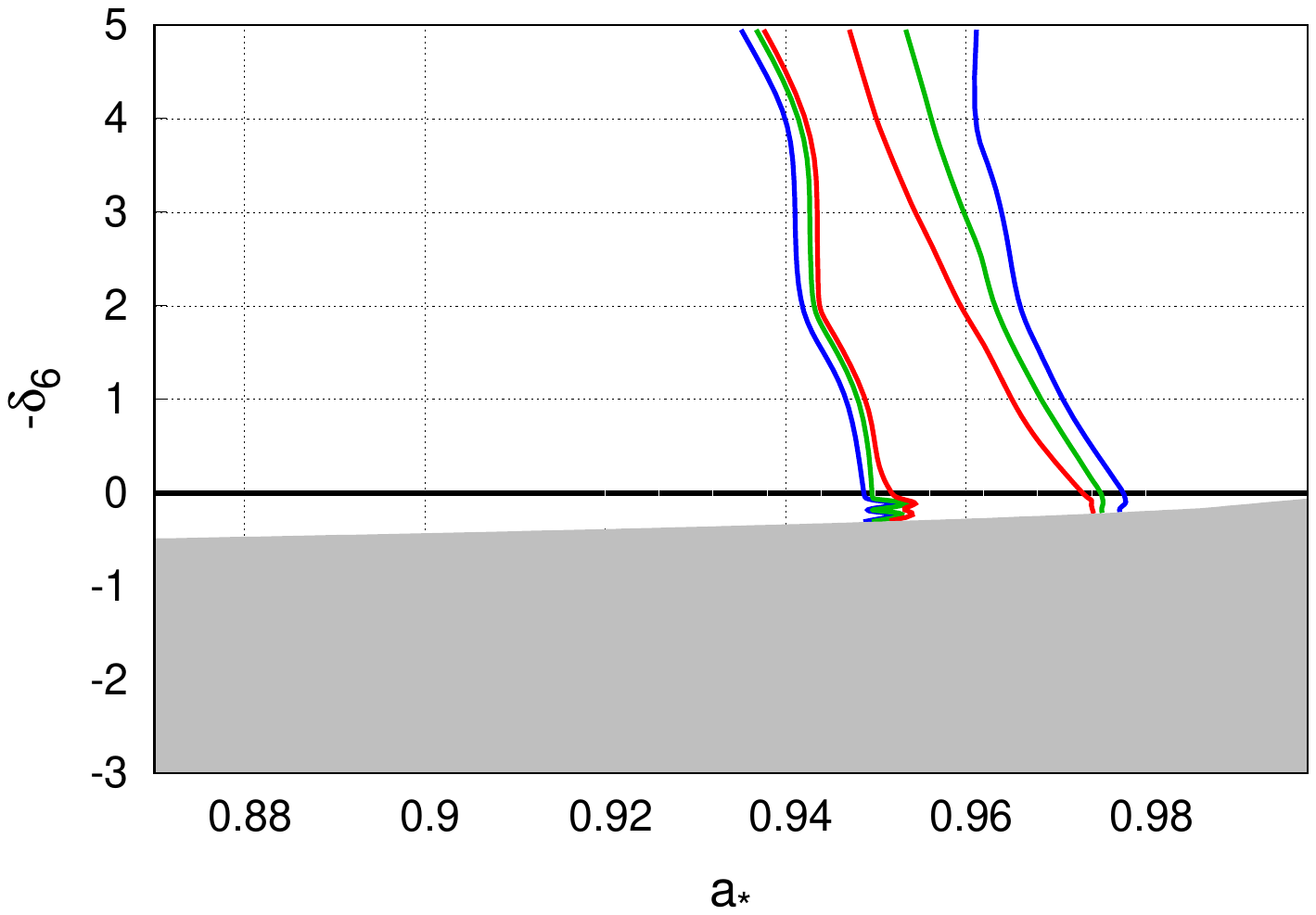}
\end{center}
\vspace{-0.3cm}
\caption{Constraints on the BH spin and on the deformation parameters. The red, green, and blue curses correspond, respectively, to the 68\%, 90\%, and 99\% confidence level contours for two relevant parameters. The black horizontal line at $\delta_i = 0$ corresponds to the Kerr solution. The gray region is not included in our analysis because the spacetime is not regular there, see Eq.~(\ref{eq-regularity}).
\label{f-contour}}
\end{figure*}

\begin{figure*}[t]
\begin{center}
\includegraphics[width=8.5cm,trim={1cm 2.5cm 0cm 1cm},clip]{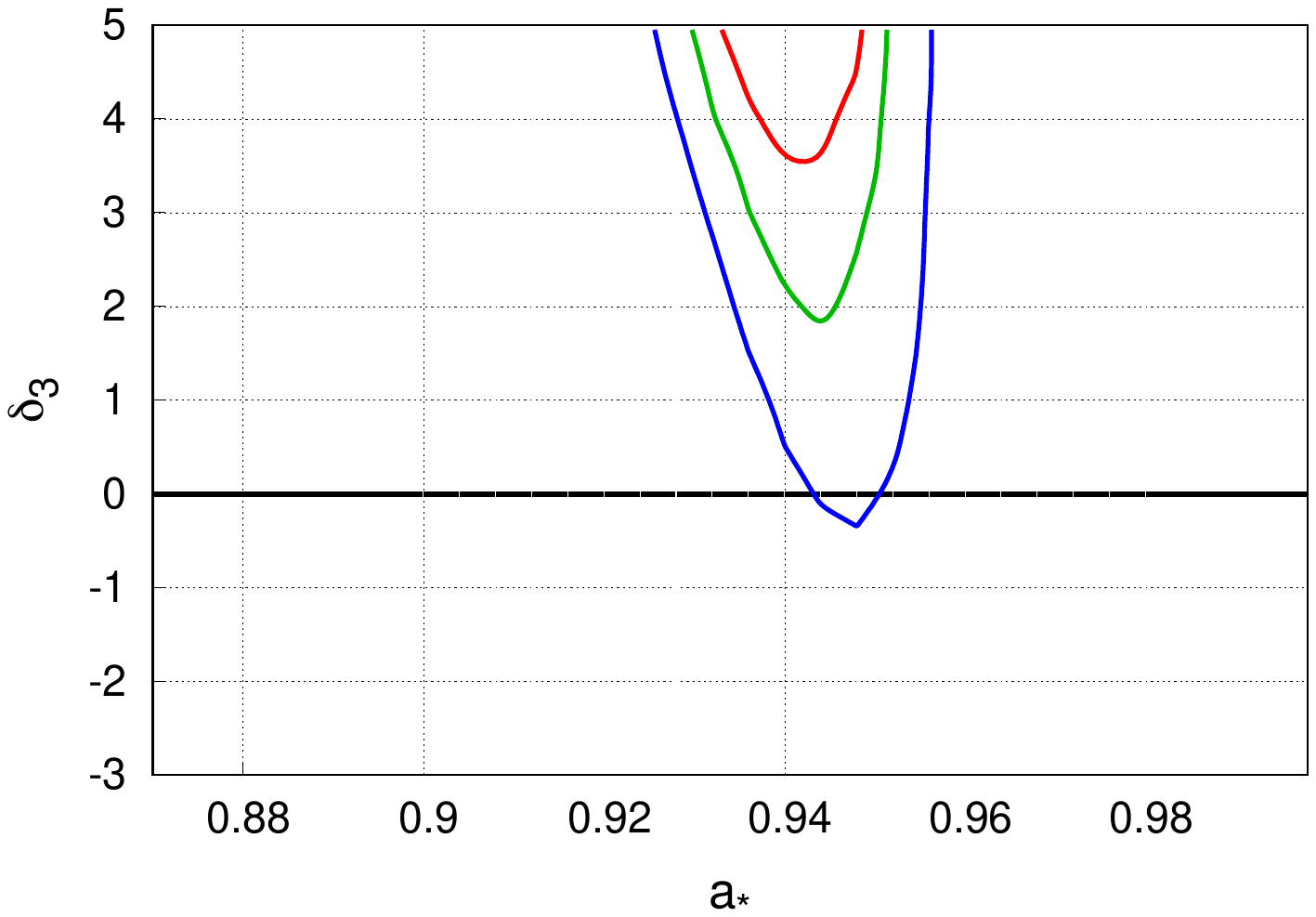}
\includegraphics[width=8.5cm,trim={1cm 2.5cm 0cm 1cm},clip]{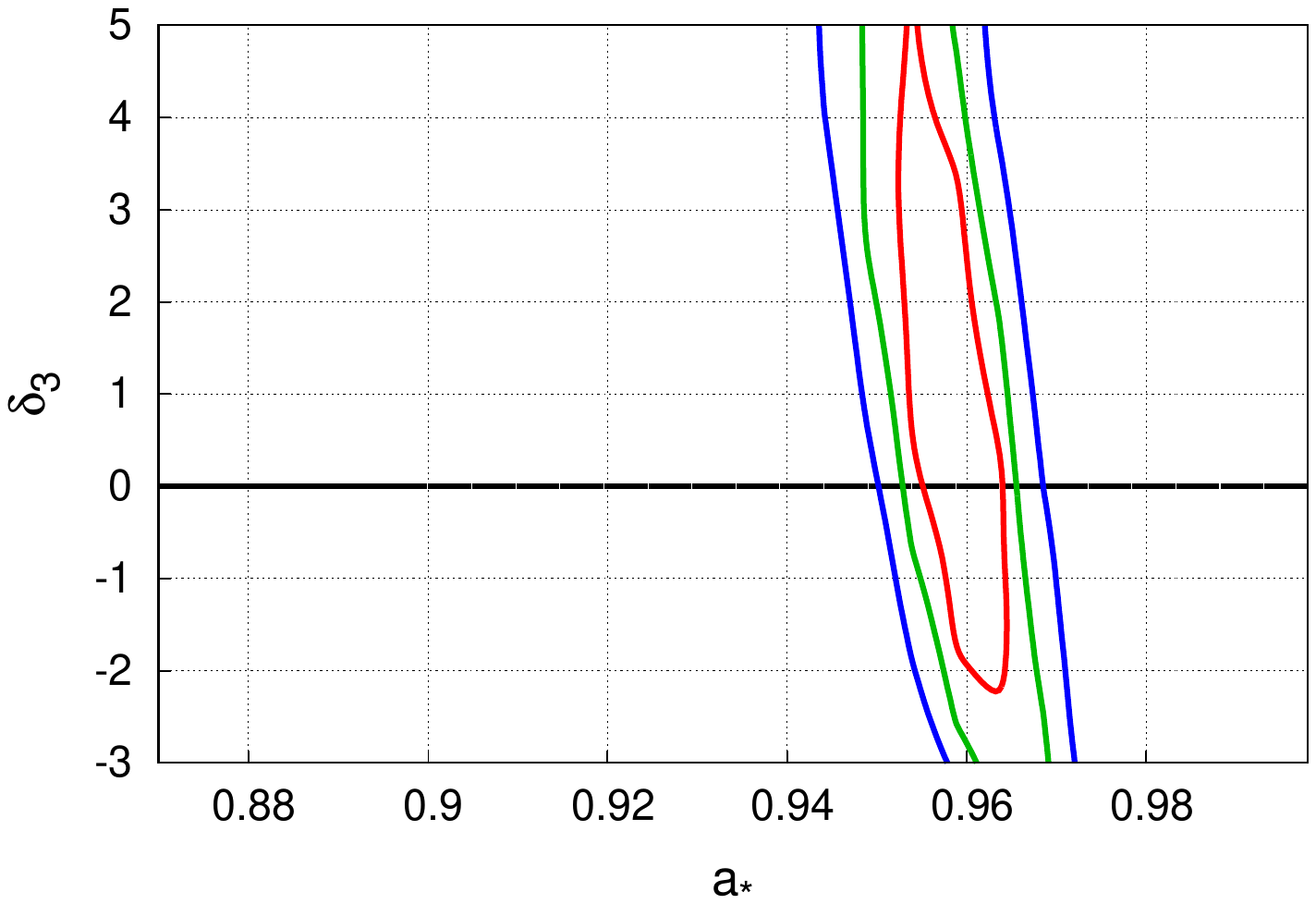} \\
\includegraphics[width=8.5cm,trim={1cm 2.5cm 0cm 1cm},clip]{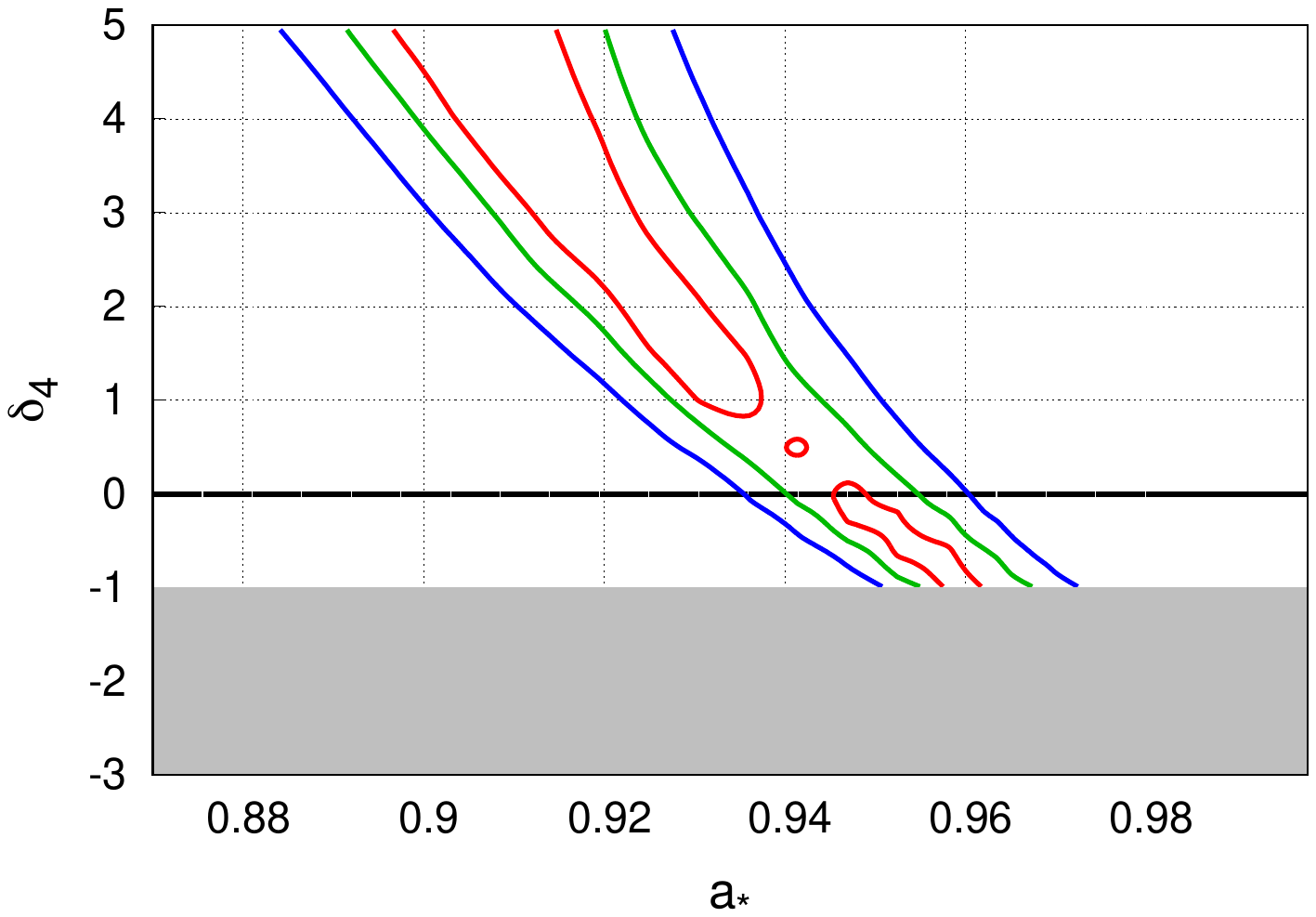}
\includegraphics[width=8.5cm,trim={1cm 2.5cm 0cm 1cm},clip]{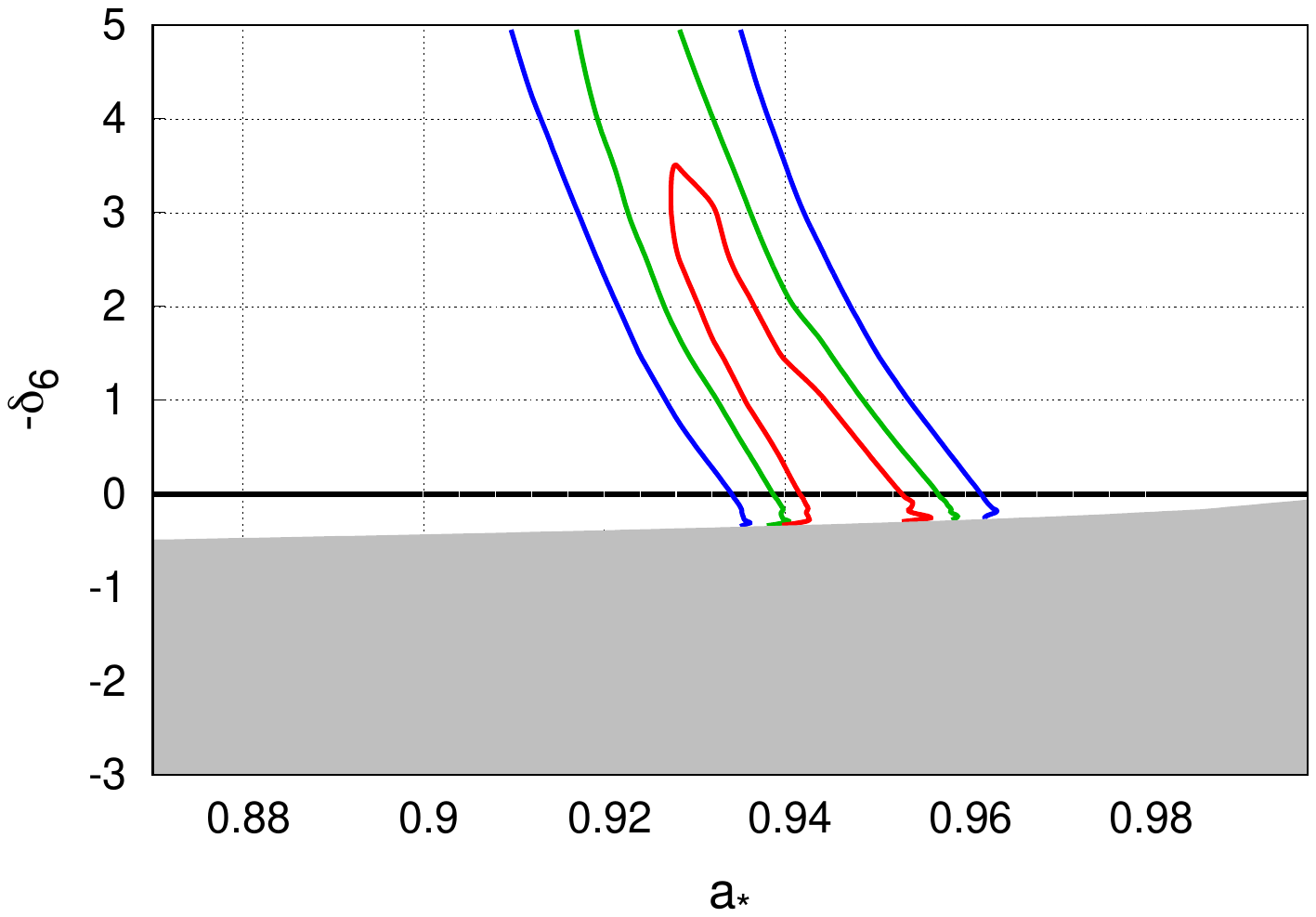}
\end{center}
\vspace{-0.3cm}
\caption{Constraints on the BH spin and on the deformation parameters $\delta_3$, $\delta_4$, and $\delta_6$ from our simulations of a simultaneous 300~ks observation with X-IFU/\textsl{Athena} and LAD/\textsl{eXTP}. The red, green, and blue curses correspond, respectively, to the 68\%, 90\%, and 99\% confidence level contours for two relevant parameters. The black horizontal line corresponds to the Kerr solution. The gray region is not included in our analysis because the spacetime is not regular there, see Eq.~(\ref{eq-regularity}). In the case of $\delta_3$, we show the results of the simulation of the full model (top-left panel) and of the simulation without absorbers (top-right panel); see the text for more details.
\label{f-simu}}
\end{figure*}

Column density, the only parameter in {\tt tbabs} model, is kept frozen and constant for all flux states. The column density and ionization
 parameter of {\tt warmabs$_1$} and {\tt warmabs$_1$} model are free to vary among flux states as the warm absorbers are expected 
 to vary over small timescales. The iron density of {\tt dustyabs} is free to vary 
 but tied among the four flux states. The power-law emission, represented by {\tt cutoffpl}, also varies among flux states 
 because coronal emission is expected to vary in order to produce the different flux states.
 
The reflection component varies over small timescales when calculated near the
 BH due to relativistic effects. 
 The relativistic reflection model {\tt relxill\_nk} assumes the emissivity profile in the form of a broken power-law modeled with three parameters: inner emissivity $q_{\rm in}$, outer emissivity $q_{\rm out}$, and break radius $R_{\rm br}$. This is the standard description for a corona of unknown geometry.  
 These three parameters vary among the four flux states because
 different reflected flux is likely to be the result of different emissivity profiles. Spin, inclination, and iron abundance are
 tied among the different states as these parameters are not expected to vary over such small timescales. The ionization parameter varies among the flux states as it is the property associated with flux. Deformation parameter is a property of spacetime and is not expected to change over the flux variations. 
 So, it is linked among the flux states. The reflection fraction is frozen to $-1$ in order to return only the 
 reflected component as the power-law emission is modeled with {\tt cutoffpl}.

$\Gamma$ and $E_{\rm cut}$ of {\tt xillver} are tied to the coronal emission of 
 the corresponding flux state. As we are considering only the reflected component far away from BH, we freeze the 
 reflection fraction to $-1$. $\log\xi$ is frozen to 0 as it is assumed that there will be no ionization far away from BH. 
 The iron abundance is assumed to be solar. The emission line at 0.81 keV and absorption line at 
 1.24 keV are modeled with {\tt zgauss}. 
 
 Fig.~\ref{f-ratio} shows the best-fit model (upper quadrant) and data to best-fit model ratio (lower quadrant) for all
 four flux states (low, medium, high, very high) for deformation parameter $\delta_1$. We do not show the corresponding 
 ratio plots for other deformation parameters as they are very similar to Fig.~\ref{f-ratio}.
 The best fit parameters values obtained for the best fit model are given in  \crefrange{t-fita}{t-fitc} 
 for all six deformation parameters. Fig.~\ref{f-contour} shows the
 confidence contours in the spin and deformation parameter plane for all six cases. The red, green, and blue curves show 68\%, 90\%, and 99\% confidence,
 respectively. The black horizontal line corresponds to the Kerr solution.

\section{Discussion \label{s-discuss}}

As shown in Fig.~\ref{f-contour}, the results are consistent with a Kerr BH solution. While MCG--06--30--15 has been used to verify the Kerr solution before, this work is significant for a few reasons. This is the first time that a test of the Kerr hypothesis using MCG--06--30--15 has been performed in the context of the KRZ metric. The KRZ metric is quite generic and has \textit{fewer} symmetries than the Kerr solution. It is, thus, capable of capturing a larger variety of potential violations of the Kerr solution and of GR. This also makes it a better proxy for the BHs of some of the most popular modified theories of gravity, which do not possess all the symmetries of the Kerr solution. A verification of the Kerr solution in this context is, therefore, an important step forward towards testing modified theories of gravity. Considering the properties of MCG--06--30--15 and the excellent quality of the 2013 data of \textsl{XMM-Newton} and \textsl{NuSTAR}, the analysis reported in this paper is presumably the best we can do today for testing the Kerr hypothesis with supermassive BHs using XRS. The dataset we studied here is quite complex, requiring as many as three absorption components and split across four different flux states. That the Kerr solution is recovered, in most cases at the 1-$\sigma$ confidence level itself, is remarkable and adds to the robustness of the result.

From Fig.~\ref{f-contour}, we see that the deformation parameters $\delta_3$, $\delta_4$, and $\delta_6$ are poorly constrained: eventually, their constraints are set by the boundaries of the regular spacetime region rather than by our fits. In order to figure out whether better data than those available can constrain these parameters or whether the reflection spectrum is not very sensitive to these deformations from the Kerr metric, we simulated a 300~ks simultaneous observation of MCG--06--30--15 with the X-IFU instrument on board of \textsl{Athena} \cite{Nandra:2013jka} and the LAD instrument on board of \textsl{eXTP} \cite{Zhang:2016ach}. We note that X-IFU/\textsl{Athena} has an exquisite energy resolution in the iron line region (at the level of 2.5~eV, while the Pn energy resolution is around 150~eV) and LAD/\textsl{eXTP} covers a wide energy band to include the Compton hump: a simultaneous observation of these two instruments is supposed to be particularly suitable for the study of reflection features and will represent the counterpart of what we can do today with \textsl{XMM-Newton} and \textsl{NuSTAR}. The input values used in the simulations are the best-fit values found in the previous section (but setting $\delta_i = 0$) for the low flux state, which is the state in which the reflection spectrum is more prominent and should thus more easily constrain the deformation parameters. The results of our simulations are summarized in Fig.~\ref{f-simu}, where we see the constraints on the BH spin and of the three deformation parameters\footnote{We note that the fit for $\delta_3$ (top-left panel of Fig.~\ref{f-simu}) does not seem to recover the Kerr solution well even if the input model assumes the Kerr metric ($\delta_3 = 0$). We investigated the reason and it seems related to the combination of the complicated absorption model of the source and the response of the instruments. Repeating the simulation without absorbers, we find the situation in the top-right panel of Fig.~\ref{f-simu}, which is the result that we would expect from a simulation.}. As we can see, even an optimistic observation with the next generation of X-ray mission cannot constrain these deformation parameters well. We thus conclude that XRS, or at least XRS when applied to a source with the properties of MCG--06--30--15, is unsuitable to test the deformations produced by the parameters $\delta_3$, $\delta_4$, and $\delta_6$. Other techniques, or other sources with different properties, are necessary.

We note that the errors reported in \crefrange{t-fita}{t-fitc} and Fig.~\ref{f-contour} are only the statistical errors. Systematic errors, in particular those related to the theoretical model, are not included~\cite{Bambi:2020jpe}. However, most modeling uncertainties are quite under control and are expected to be subdominant for the quality of the data available today, where the statistical error is the main source of uncertainty. Our model assumes that the disk is infinitesimally thin, with the inner edge at the ISCO, and that there is no emission of radiation inside the ISCO\footnote{We note that there are attempts to construct more sophisticated models, where the accretion disk is obtained from GRMHD simulations and the corona and the illumination of the disk are calculated self-consistently; see, e.g., Refs.~\cite{Kinch:2016ipi,Kinch:2018ceh}. However, these models are not yet suitable to analyze data and can only simulate some spectra.}. The impact of the thickness of the disk was studied in Ref.~\cite{Tripathi:2021wap} for this dataset, with the conclusion that the infinitesimally thin disk approximation does not produce any significant bias in the estimate of the properties of the source. The material in the plunging region is expected to be fully ionized and therefore its reflection spectrum has no features: neglecting the radiation from the plunging region in the analysis of MCG--06--30--15 should not affect our measurements~\cite{Cardenas-Avendano:2020xtw}. The ionization parameter is constant over the whole disk in our analysis, while it would be natural to expect a non-vanishing ionization gradient. However, even the assumption of a constant ionization parameter should not affect our capability of constraining deformations from the Kerr solution~\cite{Abdikamalov:2021rty}. As of now, the impact of the returning radiation (the radiation emitted by the disk and returning to the disk because of the strong light bending near the black hole) is likely the less understood source of uncertainty, since there are only partial studies in the literature; see \cite{Riaz:2020zqb} and reference therein.

XRS-based tests of GR in general, and the KRZ metric-based exploration in particular, are in early stages of development, with a lot of scope for the future. With the technique in general, significant progress is possible with MHD simulations of the BH neighborhood in non-GR backgrounds~\cite{Mizuno:2018lxz,Fromm:2021flr}, implementation of numerically evaluated BH solutions from modified theories of gravity, etc. With the KRZ metric in particular, it is possible to explore things like chaos~\cite{Destounis:2020kss,Destounis:2021mqv} as well as some BHs from modified theories of gravity that can be mapped to the KRZ metric~\cite{Konoplya:2020hyk}. 

{\bf Acknowledgments --}
This work was supported by the Innovation Program of the Shanghai Municipal Education Commission, Grant No.~2019-01-07-00-07-E00035, the National Natural Science Foundation of China (NSFC), Grant No.~11973019, and Fudan University, Grant No.~JIH1512604. D.A. is supported through the Teach@T{\"u}bingen Fellowship. S.N. acknowledges support from the Alexander von Humboldt Foundation. The authors acknowledge support by the High Performance and Cloud Computing Group at the Zentrum f\"{u}r Datenverarbeitung of the University of T\"{u}bingen, the state of Baden-W\"{u}rttemberg through bwHPC and the German Research Foundation (DFG) through Grant No.~INST 37/935-1 FUGG. 

\bibliographystyle{apsrev4-1}
\bibliography{references}

\begin{thebibliography}{89}%
\makeatletter
\providecommand \@ifxundefined [1]{%
 \@ifx{#1\undefined}
}%
\providecommand \@ifnum [1]{%
 \ifnum #1\expandafter \@firstoftwo
 \else \expandafter \@secondoftwo
 \fi
}%
\providecommand \@ifx [1]{%
 \ifx #1\expandafter \@firstoftwo
 \else \expandafter \@secondoftwo
 \fi
}%
\providecommand \natexlab [1]{#1}%
\providecommand \enquote  [1]{``#1''}%
\providecommand \bibnamefont  [1]{#1}%
\providecommand \bibfnamefont [1]{#1}%
\providecommand \citenamefont [1]{#1}%
\providecommand \href@noop [0]{\@secondoftwo}%
\providecommand \href [0]{\begingroup \@sanitize@url \@href}%
\providecommand \@href[1]{\@@startlink{#1}\@@href}%
\providecommand \@@href[1]{\endgroup#1\@@endlink}%
\providecommand \@sanitize@url [0]{\catcode `\\12\catcode `\$12\catcode
  `\&12\catcode `\#12\catcode `\^12\catcode `\_12\catcode `\%12\relax}%
\providecommand \@@startlink[1]{}%
\providecommand \@@endlink[0]{}%
\providecommand \url  [0]{\begingroup\@sanitize@url \@url }%
\providecommand \@url [1]{\endgroup\@href {#1}{\urlprefix }}%
\providecommand \urlprefix  [0]{URL }%
\providecommand \Eprint [0]{\href }%
\providecommand \doibase [0]{http://dx.doi.org/}%
\providecommand \selectlanguage [0]{\@gobble}%
\providecommand \bibinfo  [0]{\@secondoftwo}%
\providecommand \bibfield  [0]{\@secondoftwo}%
\providecommand \translation [1]{[#1]}%
\providecommand \BibitemOpen [0]{}%
\providecommand \bibitemStop [0]{}%
\providecommand \bibitemNoStop [0]{.\EOS\space}%
\providecommand \EOS [0]{\spacefactor3000\relax}%
\providecommand \BibitemShut  [1]{\csname bibitem#1\endcsname}%
\let\auto@bib@innerbib\@empty
\bibitem [{\citenamefont {Psaltis}(2008)}]{Psaltis:2008bb}%
  \BibitemOpen
  \bibfield  {author} {\bibinfo {author} {\bibfnamefont {D.}~\bibnamefont
  {Psaltis}},\ }\href {\doibase 10.12942/lrr-2008-9} {\bibfield  {journal}
  {\bibinfo  {journal} {Living Rev. Rel.}\ }\textbf {\bibinfo {volume} {11}},\
  \bibinfo {pages} {9} (\bibinfo {year} {2008})},\ \Eprint
  {http://arxiv.org/abs/0806.1531} {arXiv:0806.1531 [astro-ph]} \BibitemShut
  {NoStop}%
\bibitem [{\citenamefont {Baker}\ \emph {et~al.}(2015)\citenamefont {Baker},
  \citenamefont {Psaltis},\ and\ \citenamefont {Skordis}}]{Baker:2014zba}%
  \BibitemOpen
  \bibfield  {author} {\bibinfo {author} {\bibfnamefont {T.}~\bibnamefont
  {Baker}}, \bibinfo {author} {\bibfnamefont {D.}~\bibnamefont {Psaltis}}, \
  and\ \bibinfo {author} {\bibfnamefont {C.}~\bibnamefont {Skordis}},\ }\href
  {\doibase 10.1088/0004-637X/802/1/63} {\bibfield  {journal} {\bibinfo
  {journal} {Astrophys. J.}\ }\textbf {\bibinfo {volume} {802}},\ \bibinfo
  {pages} {63} (\bibinfo {year} {2015})},\ \Eprint
  {http://arxiv.org/abs/1412.3455} {arXiv:1412.3455 [astro-ph.CO]} \BibitemShut
  {NoStop}%
\bibitem [{\citenamefont {Yunes}\ \emph {et~al.}(2016)\citenamefont {Yunes},
  \citenamefont {Yagi},\ and\ \citenamefont {Pretorius}}]{Yunes:2016jcc}%
  \BibitemOpen
  \bibfield  {author} {\bibinfo {author} {\bibfnamefont {N.}~\bibnamefont
  {Yunes}}, \bibinfo {author} {\bibfnamefont {K.}~\bibnamefont {Yagi}}, \ and\
  \bibinfo {author} {\bibfnamefont {F.}~\bibnamefont {Pretorius}},\ }\href
  {\doibase 10.1103/PhysRevD.94.084002} {\bibfield  {journal} {\bibinfo
  {journal} {Phys. Rev. D}\ }\textbf {\bibinfo {volume} {94}},\ \bibinfo
  {pages} {084002} (\bibinfo {year} {2016})},\ \Eprint
  {http://arxiv.org/abs/1603.08955} {arXiv:1603.08955 [gr-qc]} \BibitemShut
  {NoStop}%
\bibitem [{\citenamefont {Cardenas-Avendano}\ \emph
  {et~al.}(2020{\natexlab{a}})\citenamefont {Cardenas-Avendano}, \citenamefont
  {Nampalliwar},\ and\ \citenamefont {Yunes}}]{Cardenas-Avendano:2019zxd}%
  \BibitemOpen
  \bibfield  {author} {\bibinfo {author} {\bibfnamefont {A.}~\bibnamefont
  {Cardenas-Avendano}}, \bibinfo {author} {\bibfnamefont {S.}~\bibnamefont
  {Nampalliwar}}, \ and\ \bibinfo {author} {\bibfnamefont {N.}~\bibnamefont
  {Yunes}},\ }\href {\doibase 10.1088/1361-6382/ab8f64} {\bibfield  {journal}
  {\bibinfo  {journal} {Class. Quant. Grav.}\ }\textbf {\bibinfo {volume}
  {37}},\ \bibinfo {pages} {135008} (\bibinfo {year} {2020}{\natexlab{a}})},\
  \Eprint {http://arxiv.org/abs/1912.08062} {arXiv:1912.08062 [gr-qc]}
  \BibitemShut {NoStop}%
\bibitem [{\citenamefont {Chrusciel}\ \emph {et~al.}(2012)\citenamefont
  {Chrusciel}, \citenamefont {Lopes~Costa},\ and\ \citenamefont
  {Heusler}}]{Chrusciel:2012jk}%
  \BibitemOpen
  \bibfield  {author} {\bibinfo {author} {\bibfnamefont {P.~T.}\ \bibnamefont
  {Chrusciel}}, \bibinfo {author} {\bibfnamefont {J.}~\bibnamefont
  {Lopes~Costa}}, \ and\ \bibinfo {author} {\bibfnamefont {M.}~\bibnamefont
  {Heusler}},\ }\href {\doibase 10.12942/lrr-2012-7} {\bibfield  {journal}
  {\bibinfo  {journal} {Living Rev. Rel.}\ }\textbf {\bibinfo {volume} {15}},\
  \bibinfo {pages} {7} (\bibinfo {year} {2012})},\ \Eprint
  {http://arxiv.org/abs/1205.6112} {arXiv:1205.6112 [gr-qc]} \BibitemShut
  {NoStop}%
\bibitem [{\citenamefont {Bambi}\ \emph {et~al.}(2009)\citenamefont {Bambi},
  \citenamefont {Dolgov},\ and\ \citenamefont {Petrov}}]{Bambi:2008hp}%
  \BibitemOpen
  \bibfield  {author} {\bibinfo {author} {\bibfnamefont {C.}~\bibnamefont
  {Bambi}}, \bibinfo {author} {\bibfnamefont {A.~D.}\ \bibnamefont {Dolgov}}, \
  and\ \bibinfo {author} {\bibfnamefont {A.~A.}\ \bibnamefont {Petrov}},\
  }\href {\doibase 10.1088/1475-7516/2009/09/013} {\bibfield  {journal}
  {\bibinfo  {journal} {JCAP}\ }\textbf {\bibinfo {volume} {09}},\ \bibinfo
  {pages} {013} (\bibinfo {year} {2009})},\ \Eprint
  {http://arxiv.org/abs/0806.3440} {arXiv:0806.3440 [astro-ph]} \BibitemShut
  {NoStop}%
\bibitem [{\citenamefont {Psaltis}\ \emph {et~al.}(2008)\citenamefont
  {Psaltis}, \citenamefont {Perrodin}, \citenamefont {Dienes},\ and\
  \citenamefont {Mocioiu}}]{Psaltis:2007cw}%
  \BibitemOpen
  \bibfield  {author} {\bibinfo {author} {\bibfnamefont {D.}~\bibnamefont
  {Psaltis}}, \bibinfo {author} {\bibfnamefont {D.}~\bibnamefont {Perrodin}},
  \bibinfo {author} {\bibfnamefont {K.~R.}\ \bibnamefont {Dienes}}, \ and\
  \bibinfo {author} {\bibfnamefont {I.}~\bibnamefont {Mocioiu}},\ }\href
  {\doibase 10.1103/PhysRevLett.100.091101} {\bibfield  {journal} {\bibinfo
  {journal} {Phys. Rev. Lett.}\ }\textbf {\bibinfo {volume} {100}},\ \bibinfo
  {pages} {091101} (\bibinfo {year} {2008})},\ \Eprint
  {http://arxiv.org/abs/0710.4564} {arXiv:0710.4564 [astro-ph]} \BibitemShut
  {NoStop}%
\bibitem [{\citenamefont {Bambi}\ \emph {et~al.}(2014)\citenamefont {Bambi},
  \citenamefont {Malafarina},\ and\ \citenamefont {Tsukamoto}}]{Bambi:2014koa}%
  \BibitemOpen
  \bibfield  {author} {\bibinfo {author} {\bibfnamefont {C.}~\bibnamefont
  {Bambi}}, \bibinfo {author} {\bibfnamefont {D.}~\bibnamefont {Malafarina}}, \
  and\ \bibinfo {author} {\bibfnamefont {N.}~\bibnamefont {Tsukamoto}},\ }\href
  {\doibase 10.1103/PhysRevD.89.127302} {\bibfield  {journal} {\bibinfo
  {journal} {Phys. Rev. D}\ }\textbf {\bibinfo {volume} {89}},\ \bibinfo
  {pages} {127302} (\bibinfo {year} {2014})},\ \Eprint
  {http://arxiv.org/abs/1406.2181} {arXiv:1406.2181 [gr-qc]} \BibitemShut
  {NoStop}%
\bibitem [{\citenamefont {Suvorov}(2021)}]{Suvorov:2020bvk}%
  \BibitemOpen
  \bibfield  {author} {\bibinfo {author} {\bibfnamefont {A.~G.}\ \bibnamefont
  {Suvorov}},\ }\href {\doibase 10.1007/s10714-020-02779-8} {\bibfield
  {journal} {\bibinfo  {journal} {Gen. Rel. Grav.}\ }\textbf {\bibinfo {volume}
  {53}},\ \bibinfo {pages} {6} (\bibinfo {year} {2021})},\ \Eprint
  {http://arxiv.org/abs/2008.02510} {arXiv:2008.02510 [gr-qc]} \BibitemShut
  {NoStop}%
\bibitem [{\citenamefont {Bambi}(2017)}]{Bambi2015}%
  \BibitemOpen
  \bibfield  {author} {\bibinfo {author} {\bibfnamefont {C.}~\bibnamefont
  {Bambi}},\ }\href {\doibase 10.1103/RevModPhys.89.025001} {\bibfield
  {journal} {\bibinfo  {journal} {Rev. Mod. Phys.}\ }\textbf {\bibinfo {volume}
  {89}},\ \bibinfo {pages} {025001} (\bibinfo {year} {2017})},\ \Eprint
  {http://arxiv.org/abs/1509.03884} {arXiv:1509.03884 [gr-qc]} \BibitemShut
  {NoStop}%
\bibitem [{\citenamefont {Abbott}\ \emph {et~al.}(2019)\citenamefont {Abbott}
  \emph {et~al.}}]{LIGOScientific:2019fpa}%
  \BibitemOpen
  \bibfield  {author} {\bibinfo {author} {\bibfnamefont {B.}~\bibnamefont
  {Abbott}} \emph {et~al.} (\bibinfo {collaboration} {LIGO Scientific,
  Virgo}),\ }\href {\doibase 10.1103/PhysRevD.100.104036} {\bibfield  {journal}
  {\bibinfo  {journal} {Phys. Rev. D}\ }\textbf {\bibinfo {volume} {100}},\
  \bibinfo {pages} {104036} (\bibinfo {year} {2019})},\ \Eprint
  {http://arxiv.org/abs/1903.04467} {arXiv:1903.04467 [gr-qc]} \BibitemShut
  {NoStop}%
\bibitem [{\citenamefont {Abbott}\ \emph {et~al.}(2020)\citenamefont {Abbott}
  \emph {et~al.}}]{Abbott:2020jks}%
  \BibitemOpen
  \bibfield  {author} {\bibinfo {author} {\bibfnamefont {R.}~\bibnamefont
  {Abbott}} \emph {et~al.} (\bibinfo {collaboration} {LIGO Scientific,
  Virgo}),\ }\href@noop {} {\  (\bibinfo {year} {2020})},\ \Eprint
  {http://arxiv.org/abs/2010.14529} {arXiv:2010.14529 [gr-qc]} \BibitemShut
  {NoStop}%
\bibitem [{\citenamefont {Cao}\ \emph {et~al.}(2018)\citenamefont {Cao},
  \citenamefont {Nampalliwar}, \citenamefont {Bambi}, \citenamefont {Dauser},\
  and\ \citenamefont {Garcia}}]{Cao:2017kdq}%
  \BibitemOpen
  \bibfield  {author} {\bibinfo {author} {\bibfnamefont {Z.}~\bibnamefont
  {Cao}}, \bibinfo {author} {\bibfnamefont {S.}~\bibnamefont {Nampalliwar}},
  \bibinfo {author} {\bibfnamefont {C.}~\bibnamefont {Bambi}}, \bibinfo
  {author} {\bibfnamefont {T.}~\bibnamefont {Dauser}}, \ and\ \bibinfo {author}
  {\bibfnamefont {J.~A.}\ \bibnamefont {Garcia}},\ }\href {\doibase
  10.1103/PhysRevLett.120.051101} {\bibfield  {journal} {\bibinfo  {journal}
  {Phys. Rev. Lett.}\ }\textbf {\bibinfo {volume} {120}},\ \bibinfo {pages}
  {051101} (\bibinfo {year} {2018})},\ \Eprint
  {http://arxiv.org/abs/1709.00219} {arXiv:1709.00219 [gr-qc]} \BibitemShut
  {NoStop}%
\bibitem [{\citenamefont {Tripathi}\ \emph
  {et~al.}(2021{\natexlab{a}})\citenamefont {Tripathi}, \citenamefont
  {Abdikamalov}, \citenamefont {Ayzenberg}, \citenamefont {Bambi},
  \citenamefont {Grinberg},\ and\ \citenamefont {Zhou}}]{Tripathi:2020dni}%
  \BibitemOpen
  \bibfield  {author} {\bibinfo {author} {\bibfnamefont {A.}~\bibnamefont
  {Tripathi}}, \bibinfo {author} {\bibfnamefont {A.~B.}\ \bibnamefont
  {Abdikamalov}}, \bibinfo {author} {\bibfnamefont {D.}~\bibnamefont
  {Ayzenberg}}, \bibinfo {author} {\bibfnamefont {C.}~\bibnamefont {Bambi}},
  \bibinfo {author} {\bibfnamefont {V.}~\bibnamefont {Grinberg}}, \ and\
  \bibinfo {author} {\bibfnamefont {M.}~\bibnamefont {Zhou}},\ }\href {\doibase
  10.3847/1538-4357/abccbd} {\bibfield  {journal} {\bibinfo  {journal}
  {Astrophys. J.}\ }\textbf {\bibinfo {volume} {907}},\ \bibinfo {pages} {31}
  (\bibinfo {year} {2021}{\natexlab{a}})},\ \Eprint
  {http://arxiv.org/abs/2010.13474} {arXiv:2010.13474 [astro-ph.HE]}
  \BibitemShut {NoStop}%
\bibitem [{\citenamefont {Tripathi}\ \emph
  {et~al.}(2021{\natexlab{b}})\citenamefont {Tripathi}, \citenamefont {Zhang},
  \citenamefont {Abdikamalov}, \citenamefont {Ayzenberg}, \citenamefont
  {Bambi}, \citenamefont {Jiang}, \citenamefont {Liu},\ and\ \citenamefont
  {Zhou}}]{Tripathi:2020yts}%
  \BibitemOpen
  \bibfield  {author} {\bibinfo {author} {\bibfnamefont {A.}~\bibnamefont
  {Tripathi}}, \bibinfo {author} {\bibfnamefont {Y.}~\bibnamefont {Zhang}},
  \bibinfo {author} {\bibfnamefont {A.~B.}\ \bibnamefont {Abdikamalov}},
  \bibinfo {author} {\bibfnamefont {D.}~\bibnamefont {Ayzenberg}}, \bibinfo
  {author} {\bibfnamefont {C.}~\bibnamefont {Bambi}}, \bibinfo {author}
  {\bibfnamefont {J.}~\bibnamefont {Jiang}}, \bibinfo {author} {\bibfnamefont
  {H.}~\bibnamefont {Liu}}, \ and\ \bibinfo {author} {\bibfnamefont
  {M.}~\bibnamefont {Zhou}},\ }\href {\doibase 10.3847/1538-4357/abf6cd}
  {\bibfield  {journal} {\bibinfo  {journal} {Astrophys. J.}\ }\textbf
  {\bibinfo {volume} {913}},\ \bibinfo {pages} {79} (\bibinfo {year}
  {2021}{\natexlab{b}})},\ \Eprint {http://arxiv.org/abs/2012.10669}
  {arXiv:2012.10669 [astro-ph.HE]} \BibitemShut {NoStop}%
\bibitem [{\citenamefont {Bambi}(2021)}]{Bambi:2021hxv}%
  \BibitemOpen
  \bibfield  {author} {\bibinfo {author} {\bibfnamefont {C.}~\bibnamefont
  {Bambi}}\ }(\bibinfo {year} {2021})\ \Eprint
  {http://arxiv.org/abs/2103.11365} {arXiv:2103.11365 [gr-qc]} \BibitemShut
  {NoStop}%
\bibitem [{\citenamefont {Psaltis}\ \emph
  {et~al.}(2020{\natexlab{a}})\citenamefont {Psaltis} \emph
  {et~al.}}]{PhysRevLett.125.141104}%
  \BibitemOpen
  \bibfield  {author} {\bibinfo {author} {\bibfnamefont {D.}~\bibnamefont
  {Psaltis}} \emph {et~al.} (\bibinfo {collaboration} {EHT Collaboration}),\
  }\href {\doibase 10.1103/PhysRevLett.125.141104} {\bibfield  {journal}
  {\bibinfo  {journal} {Phys. Rev. Lett.}\ }\textbf {\bibinfo {volume} {125}},\
  \bibinfo {pages} {141104} (\bibinfo {year} {2020}{\natexlab{a}})}\BibitemShut
  {NoStop}%
\bibitem [{\citenamefont {Psaltis}\ \emph
  {et~al.}(2020{\natexlab{b}})\citenamefont {Psaltis}, \citenamefont {Talbot},
  \citenamefont {Payne},\ and\ \citenamefont {Mandel}}]{Psaltis:2020ctj}%
  \BibitemOpen
  \bibfield  {author} {\bibinfo {author} {\bibfnamefont {D.}~\bibnamefont
  {Psaltis}}, \bibinfo {author} {\bibfnamefont {C.}~\bibnamefont {Talbot}},
  \bibinfo {author} {\bibfnamefont {E.}~\bibnamefont {Payne}}, \ and\ \bibinfo
  {author} {\bibfnamefont {I.}~\bibnamefont {Mandel}},\ }\href@noop {} {\
  (\bibinfo {year} {2020}{\natexlab{b}})},\ \Eprint
  {http://arxiv.org/abs/2012.02117} {arXiv:2012.02117 [gr-qc]} \BibitemShut
  {NoStop}%
\bibitem [{\citenamefont {V\"olkel}\ \emph {et~al.}(2020)\citenamefont
  {V\"olkel}, \citenamefont {Barausse}, \citenamefont {Franchini},\ and\
  \citenamefont {Broderick}}]{Volkel:2020xlc}%
  \BibitemOpen
  \bibfield  {author} {\bibinfo {author} {\bibfnamefont {S.~H.}\ \bibnamefont
  {V\"olkel}}, \bibinfo {author} {\bibfnamefont {E.}~\bibnamefont {Barausse}},
  \bibinfo {author} {\bibfnamefont {N.}~\bibnamefont {Franchini}}, \ and\
  \bibinfo {author} {\bibfnamefont {A.~E.}\ \bibnamefont {Broderick}},\
  }\href@noop {} {\  (\bibinfo {year} {2020})},\ \Eprint
  {http://arxiv.org/abs/2011.06812} {arXiv:2011.06812 [gr-qc]} \BibitemShut
  {NoStop}%
\bibitem [{\citenamefont {Abuter}\ \emph {et~al.}(2018)\citenamefont {Abuter}
  \emph {et~al.}}]{Abuter:2018drb}%
  \BibitemOpen
  \bibfield  {author} {\bibinfo {author} {\bibfnamefont {R.}~\bibnamefont
  {Abuter}} \emph {et~al.} (\bibinfo {collaboration} {GRAVITY}),\ }\href
  {\doibase 10.1051/0004-6361/201833718} {\bibfield  {journal} {\bibinfo
  {journal} {Astron. Astrophys.}\ }\textbf {\bibinfo {volume} {615}},\ \bibinfo
  {pages} {L15} (\bibinfo {year} {2018})},\ \Eprint
  {http://arxiv.org/abs/1807.09409} {arXiv:1807.09409 [astro-ph.GA]}
  \BibitemShut {NoStop}%
\bibitem [{\citenamefont {Tripathi}\ \emph {et~al.}(2019)\citenamefont
  {Tripathi}, \citenamefont {Nampalliwar}, \citenamefont {Abdikamalov},
  \citenamefont {Ayzenberg}, \citenamefont {Bambi}, \citenamefont {Dauser},
  \citenamefont {Garcia},\ and\ \citenamefont {Marinucci}}]{Tripathi2019}%
  \BibitemOpen
  \bibfield  {author} {\bibinfo {author} {\bibfnamefont {A.}~\bibnamefont
  {Tripathi}}, \bibinfo {author} {\bibfnamefont {S.}~\bibnamefont
  {Nampalliwar}}, \bibinfo {author} {\bibfnamefont {A.~B.}\ \bibnamefont
  {Abdikamalov}}, \bibinfo {author} {\bibfnamefont {D.}~\bibnamefont
  {Ayzenberg}}, \bibinfo {author} {\bibfnamefont {C.}~\bibnamefont {Bambi}},
  \bibinfo {author} {\bibfnamefont {T.}~\bibnamefont {Dauser}}, \bibinfo
  {author} {\bibfnamefont {J.~A.}\ \bibnamefont {Garcia}}, \ and\ \bibinfo
  {author} {\bibfnamefont {A.}~\bibnamefont {Marinucci}},\ }\href {\doibase
  10.3847/1538-4357/ab0e7e} {\bibfield  {journal} {\bibinfo  {journal}
  {Astrophys. J.}\ }\textbf {\bibinfo {volume} {875}},\ \bibinfo {pages} {56}
  (\bibinfo {year} {2019})},\ \Eprint {http://arxiv.org/abs/1811.08148}
  {arXiv:1811.08148 [gr-qc]} \BibitemShut {NoStop}%
\bibitem [{\citenamefont {Liu}\ \emph {et~al.}(2020)\citenamefont {Liu},
  \citenamefont {Wang}, \citenamefont {Abdikamalov}, \citenamefont
  {Ayzenberg},\ and\ \citenamefont {Bambi}}]{Liu:2020fpv}%
  \BibitemOpen
  \bibfield  {author} {\bibinfo {author} {\bibfnamefont {H.}~\bibnamefont
  {Liu}}, \bibinfo {author} {\bibfnamefont {H.}~\bibnamefont {Wang}}, \bibinfo
  {author} {\bibfnamefont {A.~B.}\ \bibnamefont {Abdikamalov}}, \bibinfo
  {author} {\bibfnamefont {D.}~\bibnamefont {Ayzenberg}}, \ and\ \bibinfo
  {author} {\bibfnamefont {C.}~\bibnamefont {Bambi}},\ }\href {\doibase
  10.3847/1538-4357/ab917a} {\bibfield  {journal} {\bibinfo  {journal}
  {Astrophys. J.}\ }\textbf {\bibinfo {volume} {896}},\ \bibinfo {pages} {160}
  (\bibinfo {year} {2020})},\ \Eprint {http://arxiv.org/abs/2004.11542}
  {arXiv:2004.11542 [gr-qc]} \BibitemShut {NoStop}%
\bibitem [{\citenamefont {Konoplya}\ \emph {et~al.}(2016)\citenamefont
  {Konoplya}, \citenamefont {Rezzolla},\ and\ \citenamefont
  {Zhidenko}}]{Konoplya2016}%
  \BibitemOpen
  \bibfield  {author} {\bibinfo {author} {\bibfnamefont {R.}~\bibnamefont
  {Konoplya}}, \bibinfo {author} {\bibfnamefont {L.}~\bibnamefont {Rezzolla}},
  \ and\ \bibinfo {author} {\bibfnamefont {A.}~\bibnamefont {Zhidenko}},\
  }\href {\doibase 10.1103/PhysRevD.93.064015} {\bibfield  {journal} {\bibinfo
  {journal} {Phys. Rev.}\ }\textbf {\bibinfo {volume} {D93}},\ \bibinfo {pages}
  {064015} (\bibinfo {year} {2016})},\ \Eprint
  {http://arxiv.org/abs/1602.02378} {arXiv:1602.02378 [gr-qc]} \BibitemShut
  {NoStop}%
\bibitem [{\citenamefont {Nampalliwar}\ \emph {et~al.}(2020)\citenamefont
  {Nampalliwar}, \citenamefont {Xin}, \citenamefont {Srivastava}, \citenamefont
  {Abdikamalov}, \citenamefont {Ayzenberg}, \citenamefont {Bambi},
  \citenamefont {Dauser}, \citenamefont {Garcia},\ and\ \citenamefont
  {Tripathi}}]{Nampalliwar:2019iti}%
  \BibitemOpen
  \bibfield  {author} {\bibinfo {author} {\bibfnamefont {S.}~\bibnamefont
  {Nampalliwar}}, \bibinfo {author} {\bibfnamefont {S.}~\bibnamefont {Xin}},
  \bibinfo {author} {\bibfnamefont {S.}~\bibnamefont {Srivastava}}, \bibinfo
  {author} {\bibfnamefont {A.~B.}\ \bibnamefont {Abdikamalov}}, \bibinfo
  {author} {\bibfnamefont {D.}~\bibnamefont {Ayzenberg}}, \bibinfo {author}
  {\bibfnamefont {C.}~\bibnamefont {Bambi}}, \bibinfo {author} {\bibfnamefont
  {T.}~\bibnamefont {Dauser}}, \bibinfo {author} {\bibfnamefont {J.~A.}\
  \bibnamefont {Garcia}}, \ and\ \bibinfo {author} {\bibfnamefont
  {A.}~\bibnamefont {Tripathi}},\ }\href {\doibase 10.1103/PhysRevD.102.124071}
  {\bibfield  {journal} {\bibinfo  {journal} {Phys. Rev. D}\ }\textbf {\bibinfo
  {volume} {102}},\ \bibinfo {pages} {124071} (\bibinfo {year} {2020})},\
  \Eprint {http://arxiv.org/abs/1903.12119} {arXiv:1903.12119 [gr-qc]}
  \BibitemShut {NoStop}%
\bibitem [{\citenamefont {Bertotti}\ \emph {et~al.}(2003)\citenamefont
  {Bertotti}, \citenamefont {Iess},\ and\ \citenamefont
  {Tortora}}]{Bertotti:2003rm}%
  \BibitemOpen
  \bibfield  {author} {\bibinfo {author} {\bibfnamefont {B.}~\bibnamefont
  {Bertotti}}, \bibinfo {author} {\bibfnamefont {L.}~\bibnamefont {Iess}}, \
  and\ \bibinfo {author} {\bibfnamefont {P.}~\bibnamefont {Tortora}},\ }\href
  {\doibase 10.1038/nature01997} {\bibfield  {journal} {\bibinfo  {journal}
  {Nature}\ }\textbf {\bibinfo {volume} {425}},\ \bibinfo {pages} {374}
  (\bibinfo {year} {2003})}\BibitemShut {NoStop}%
\bibitem [{\citenamefont {Will}(2014)}]{Will2014}%
  \BibitemOpen
  \bibfield  {author} {\bibinfo {author} {\bibfnamefont {C.~M.}\ \bibnamefont
  {Will}},\ }\href {\doibase 10.12942/lrr-2014-4} {\bibfield  {journal}
  {\bibinfo  {journal} {Living Rev. Rel.}\ }\textbf {\bibinfo {volume} {17}},\
  \bibinfo {pages} {4} (\bibinfo {year} {2014})},\ \Eprint
  {http://arxiv.org/abs/1403.7377} {arXiv:1403.7377 [gr-qc]} \BibitemShut
  {NoStop}%
\bibitem [{\citenamefont {Miller}\ \emph {et~al.}(2019)\citenamefont {Miller}
  \emph {et~al.}}]{Miller:2019cac}%
  \BibitemOpen
  \bibfield  {author} {\bibinfo {author} {\bibfnamefont {M.~C.}\ \bibnamefont
  {Miller}} \emph {et~al.},\ }\href {\doibase 10.3847/2041-8213/ab50c5}
  {\bibfield  {journal} {\bibinfo  {journal} {Astrophys. J. Lett.}\ }\textbf
  {\bibinfo {volume} {887}},\ \bibinfo {pages} {L24} (\bibinfo {year}
  {2019})},\ \Eprint {http://arxiv.org/abs/1912.05705} {arXiv:1912.05705
  [astro-ph.HE]} \BibitemShut {NoStop}%
\bibitem [{\citenamefont {Silva}\ \emph {et~al.}(2021)\citenamefont {Silva},
  \citenamefont {Holgado}, \citenamefont {C\'ardenas-Avenda\~no},\ and\
  \citenamefont {Yunes}}]{Silva:2020acr}%
  \BibitemOpen
  \bibfield  {author} {\bibinfo {author} {\bibfnamefont {H.~O.}\ \bibnamefont
  {Silva}}, \bibinfo {author} {\bibfnamefont {A.~M.}\ \bibnamefont {Holgado}},
  \bibinfo {author} {\bibfnamefont {A.}~\bibnamefont {C\'ardenas-Avenda\~no}},
  \ and\ \bibinfo {author} {\bibfnamefont {N.}~\bibnamefont {Yunes}},\ }\href
  {\doibase 10.1103/PhysRevLett.126.181101} {\bibfield  {journal} {\bibinfo
  {journal} {Phys. Rev. Lett.}\ }\textbf {\bibinfo {volume} {126}},\ \bibinfo
  {pages} {181101} (\bibinfo {year} {2021})},\ \Eprint
  {http://arxiv.org/abs/2004.01253} {arXiv:2004.01253 [gr-qc]} \BibitemShut
  {NoStop}%
\bibitem [{\citenamefont {Abbott}\ \emph {et~al.}(2016)\citenamefont {Abbott}
  \emph {et~al.}}]{TheLIGOScientific:2016src}%
  \BibitemOpen
  \bibfield  {author} {\bibinfo {author} {\bibfnamefont {B.~P.}\ \bibnamefont
  {Abbott}} \emph {et~al.} (\bibinfo {collaboration} {LIGO Scientific,
  Virgo}),\ }\href {\doibase 10.1103/PhysRevLett.116.221101} {\bibfield
  {journal} {\bibinfo  {journal} {Phys. Rev. Lett.}\ }\textbf {\bibinfo
  {volume} {116}},\ \bibinfo {pages} {221101} (\bibinfo {year} {2016})},\
  \bibinfo {note} {[Erratum: Phys.Rev.Lett. 121, 129902 (2018)]},\ \Eprint
  {http://arxiv.org/abs/1602.03841} {arXiv:1602.03841 [gr-qc]} \BibitemShut
  {NoStop}%
\bibitem [{\citenamefont {Garc\'{i}a}\ \emph {et~al.}(2014)\citenamefont
  {Garc\'{i}a} \emph {et~al.}}]{Garcia2013}%
  \BibitemOpen
  \bibfield  {author} {\bibinfo {author} {\bibfnamefont {J.}~\bibnamefont
  {Garc\'{i}a}} \emph {et~al.},\ }\href {\doibase 10.1088/0004-637X/782/2/76}
  {\bibfield  {journal} {\bibinfo  {journal} {Astrophys. J.}\ }\textbf
  {\bibinfo {volume} {782}},\ \bibinfo {pages} {76} (\bibinfo {year} {2014})},\
  \Eprint {http://arxiv.org/abs/1312.3231} {arXiv:1312.3231 [astro-ph.HE]}
  \BibitemShut {NoStop}%
\bibitem [{\citenamefont {Dauser}\ \emph {et~al.}(2014)\citenamefont {Dauser}
  \emph {et~al.}}]{Dauser2014}%
  \BibitemOpen
  \bibfield  {author} {\bibinfo {author} {\bibfnamefont {T.}~\bibnamefont
  {Dauser}} \emph {et~al.},\ }\href {\doibase 10.1093/mnrasl/slu125} {\bibfield
   {journal} {\bibinfo  {journal} {Mon. Not. Roy. Astron. Soc.}\ }\textbf
  {\bibinfo {volume} {444}},\ \bibinfo {pages} {100} (\bibinfo {year}
  {2014})},\ \Eprint {http://arxiv.org/abs/1408.2347} {arXiv:1408.2347
  [astro-ph.HE]} \BibitemShut {NoStop}%
\bibitem [{\citenamefont {Bambi}\ \emph {et~al.}(2017)\citenamefont {Bambi}
  \emph {et~al.}}]{relxillnk}%
  \BibitemOpen
  \bibfield  {author} {\bibinfo {author} {\bibfnamefont {C.}~\bibnamefont
  {Bambi}} \emph {et~al.},\ }\href {\doibase 10.3847/1538-4357/aa74c0}
  {\bibfield  {journal} {\bibinfo  {journal} {Astrophys. J.}\ }\textbf
  {\bibinfo {volume} {842}},\ \bibinfo {pages} {76} (\bibinfo {year} {2017})},\
  \Eprint {http://arxiv.org/abs/1607.00596} {arXiv:1607.00596 [gr-qc]}
  \BibitemShut {NoStop}%
\bibitem [{\citenamefont {Abdikamalov}\ \emph {et~al.}(2019)\citenamefont
  {Abdikamalov}, \citenamefont {Ayzenberg}, \citenamefont {Bambi},
  \citenamefont {Dauser}, \citenamefont {Garcia},\ and\ \citenamefont
  {Nampalliwar}}]{Abdikamalov:2019yrr}%
  \BibitemOpen
  \bibfield  {author} {\bibinfo {author} {\bibfnamefont {A.~B.}\ \bibnamefont
  {Abdikamalov}}, \bibinfo {author} {\bibfnamefont {D.}~\bibnamefont
  {Ayzenberg}}, \bibinfo {author} {\bibfnamefont {C.}~\bibnamefont {Bambi}},
  \bibinfo {author} {\bibfnamefont {T.}~\bibnamefont {Dauser}}, \bibinfo
  {author} {\bibfnamefont {J.~A.}\ \bibnamefont {Garcia}}, \ and\ \bibinfo
  {author} {\bibfnamefont {S.}~\bibnamefont {Nampalliwar}},\ }\href {\doibase
  10.3847/1538-4357/ab1f89} {\bibfield  {journal} {\bibinfo  {journal}
  {Astrophys. J.}\ }\textbf {\bibinfo {volume} {878}},\ \bibinfo {pages} {91}
  (\bibinfo {year} {2019})},\ \Eprint {http://arxiv.org/abs/1902.09665}
  {arXiv:1902.09665 [gr-qc]} \BibitemShut {NoStop}%
\bibitem [{\citenamefont {Abdikamalov}\ \emph {et~al.}(2020)\citenamefont
  {Abdikamalov}, \citenamefont {Ayzenberg}, \citenamefont {Bambi},
  \citenamefont {Dauser}, \citenamefont {Garcia}, \citenamefont {Nampalliwar},
  \citenamefont {Tripathi},\ and\ \citenamefont {Zhou}}]{Abdikamalov:2020oci}%
  \BibitemOpen
  \bibfield  {author} {\bibinfo {author} {\bibfnamefont {A.~B.}\ \bibnamefont
  {Abdikamalov}}, \bibinfo {author} {\bibfnamefont {D.}~\bibnamefont
  {Ayzenberg}}, \bibinfo {author} {\bibfnamefont {C.}~\bibnamefont {Bambi}},
  \bibinfo {author} {\bibfnamefont {T.}~\bibnamefont {Dauser}}, \bibinfo
  {author} {\bibfnamefont {J.~A.}\ \bibnamefont {Garcia}}, \bibinfo {author}
  {\bibfnamefont {S.}~\bibnamefont {Nampalliwar}}, \bibinfo {author}
  {\bibfnamefont {A.}~\bibnamefont {Tripathi}}, \ and\ \bibinfo {author}
  {\bibfnamefont {M.}~\bibnamefont {Zhou}},\ }\href {\doibase
  10.3847/1538-4357/aba625} {\bibfield  {journal} {\bibinfo  {journal}
  {Astrophys. J.}\ }\textbf {\bibinfo {volume} {899}},\ \bibinfo {pages} {80}
  (\bibinfo {year} {2020})},\ \Eprint {http://arxiv.org/abs/2003.09663}
  {arXiv:2003.09663 [astro-ph.HE]} \BibitemShut {NoStop}%
\bibitem [{\citenamefont {Garc{\'\i}a}\ and\ \citenamefont
  {Kallman}(2010)}]{Garcia2010}%
  \BibitemOpen
  \bibfield  {author} {\bibinfo {author} {\bibfnamefont {J.}~\bibnamefont
  {Garc{\'\i}a}}\ and\ \bibinfo {author} {\bibfnamefont {T.~R.}\ \bibnamefont
  {Kallman}},\ }\href {\doibase 10.1088/0004-637X/718/2/695} {\bibfield
  {journal} {\bibinfo  {journal} {Astrophys. J.}\ }\textbf {\bibinfo {volume}
  {718}},\ \bibinfo {pages} {695} (\bibinfo {year} {2010})},\ \Eprint
  {http://arxiv.org/abs/1006.0485} {arXiv:1006.0485 [astro-ph.HE]} \BibitemShut
  {NoStop}%
\bibitem [{\citenamefont {Dauser}\ \emph {et~al.}(2010)\citenamefont {Dauser},
  \citenamefont {Wilms}, \citenamefont {Reynolds},\ and\ \citenamefont
  {Brenneman}}]{Dauser:2010ne}%
  \BibitemOpen
  \bibfield  {author} {\bibinfo {author} {\bibfnamefont {T.}~\bibnamefont
  {Dauser}}, \bibinfo {author} {\bibfnamefont {J.}~\bibnamefont {Wilms}},
  \bibinfo {author} {\bibfnamefont {C.}~\bibnamefont {Reynolds}}, \ and\
  \bibinfo {author} {\bibfnamefont {L.}~\bibnamefont {Brenneman}},\ }\href
  {\doibase 10.1111/j.1365-2966.2010.17393.x} {\bibfield  {journal} {\bibinfo
  {journal} {Mon. Not. Roy. Astron. Soc.}\ }\textbf {\bibinfo {volume} {409}},\
  \bibinfo {pages} {1534} (\bibinfo {year} {2010})},\ \Eprint
  {http://arxiv.org/abs/1007.4937} {arXiv:1007.4937 [astro-ph.HE]} \BibitemShut
  {NoStop}%
\bibitem [{\citenamefont {Dauser}\ \emph {et~al.}(2013)\citenamefont {Dauser},
  \citenamefont {Garcia}, \citenamefont {Wilms}, \citenamefont {Bock},
  \citenamefont {Brenneman}, \citenamefont {Falanga}, \citenamefont
  {Fukumura},\ and\ \citenamefont {Reynolds}}]{Dauser:2013xv}%
  \BibitemOpen
  \bibfield  {author} {\bibinfo {author} {\bibfnamefont {T.}~\bibnamefont
  {Dauser}}, \bibinfo {author} {\bibfnamefont {J.}~\bibnamefont {Garcia}},
  \bibinfo {author} {\bibfnamefont {J.}~\bibnamefont {Wilms}}, \bibinfo
  {author} {\bibfnamefont {M.}~\bibnamefont {Bock}}, \bibinfo {author}
  {\bibfnamefont {L.~W.}\ \bibnamefont {Brenneman}}, \bibinfo {author}
  {\bibfnamefont {M.}~\bibnamefont {Falanga}}, \bibinfo {author} {\bibfnamefont
  {K.}~\bibnamefont {Fukumura}}, \ and\ \bibinfo {author} {\bibfnamefont
  {C.~S.}\ \bibnamefont {Reynolds}},\ }\href {\doibase 10.1093/mnras/sts710}
  {\bibfield  {journal} {\bibinfo  {journal} {Mon. Not. Roy. Astron. Soc.}\
  }\textbf {\bibinfo {volume} {430}},\ \bibinfo {pages} {1694} (\bibinfo {year}
  {2013})},\ \Eprint {http://arxiv.org/abs/1301.4922} {arXiv:1301.4922
  [astro-ph.HE]} \BibitemShut {NoStop}%
\bibitem [{\citenamefont {Bambi}\ \emph {et~al.}(2020)\citenamefont {Bambi}
  \emph {et~al.}}]{Bambi:2020jpe}%
  \BibitemOpen
  \bibfield  {author} {\bibinfo {author} {\bibfnamefont {C.}~\bibnamefont
  {Bambi}} \emph {et~al.},\ }\href@noop {} {\  (\bibinfo {year} {2020})},\
  \Eprint {http://arxiv.org/abs/2011.04792} {arXiv:2011.04792 [astro-ph.HE]}
  \BibitemShut {NoStop}%
\bibitem [{\citenamefont {Novikov}\ and\ \citenamefont
  {Thorne}(1973)}]{Novikov1973}%
  \BibitemOpen
  \bibfield  {author} {\bibinfo {author} {\bibfnamefont {I.~D.}\ \bibnamefont
  {Novikov}}\ and\ \bibinfo {author} {\bibfnamefont {K.~S.}\ \bibnamefont
  {Thorne}},\ }in\ \href@noop {} {\emph {\bibinfo {booktitle} {{Proceedings,
  Ecole d'Et\'{e} de Physique Th\'{e}orique: Les Astres Occlus: Les Houches,
  France, August, 1972}}}}\ (\bibinfo {year} {1973})\ pp.\ \bibinfo {pages}
  {343--550}\BibitemShut {NoStop}%
\bibitem [{\citenamefont {Cunningham}(1975)}]{Cunningham1975}%
  \BibitemOpen
  \bibfield  {author} {\bibinfo {author} {\bibfnamefont {C.~T.}\ \bibnamefont
  {Cunningham}},\ }\href {\doibase 10.1086/154033} {\bibfield  {journal}
  {\bibinfo  {journal} {Astrophys. J.}\ }\textbf {\bibinfo {volume} {202}},\
  \bibinfo {pages} {788} (\bibinfo {year} {1975})}\BibitemShut {NoStop}%
\bibitem [{\citenamefont {Ni}\ \emph {et~al.}(2016)\citenamefont {Ni},
  \citenamefont {Jiang},\ and\ \citenamefont {Bambi}}]{Ni2016}%
  \BibitemOpen
  \bibfield  {author} {\bibinfo {author} {\bibfnamefont {Y.}~\bibnamefont
  {Ni}}, \bibinfo {author} {\bibfnamefont {J.}~\bibnamefont {Jiang}}, \ and\
  \bibinfo {author} {\bibfnamefont {C.}~\bibnamefont {Bambi}},\ }\href
  {\doibase 10.1088/1475-7516/2016/09/014} {\bibfield  {journal} {\bibinfo
  {journal} {JCAP}\ }\textbf {\bibinfo {volume} {1609}},\ \bibinfo {pages}
  {014} (\bibinfo {year} {2016})},\ \Eprint {http://arxiv.org/abs/1607.04893}
  {arXiv:1607.04893 [gr-qc]} \BibitemShut {NoStop}%
\bibitem [{\citenamefont {Tanaka}\ \emph {et~al.}(1995)\citenamefont {Tanaka}
  \emph {et~al.}}]{Tanaka1995}%
  \BibitemOpen
  \bibfield  {author} {\bibinfo {author} {\bibfnamefont {Y.}~\bibnamefont
  {Tanaka}} \emph {et~al.},\ }\href {\doibase 10.1038/375659a0} {\bibfield
  {journal} {\bibinfo  {journal} {Nature}\ }\textbf {\bibinfo {volume} {375}},\
  \bibinfo {pages} {659} (\bibinfo {year} {1995})}\BibitemShut {NoStop}%
\bibitem [{\citenamefont {Iwasawa}\ \emph {et~al.}(1996)\citenamefont {Iwasawa}
  \emph {et~al.}}]{Iwasawa1996}%
  \BibitemOpen
  \bibfield  {author} {\bibinfo {author} {\bibfnamefont {K.}~\bibnamefont
  {Iwasawa}} \emph {et~al.},\ }\href {\doibase 10.1093/mnras/282.3.1038}
  {\bibfield  {journal} {\bibinfo  {journal} {Mon. Not. Roy. Astron. Soc.}\
  }\textbf {\bibinfo {volume} {282}},\ \bibinfo {pages} {1038} (\bibinfo {year}
  {1996})},\ \Eprint {http://arxiv.org/abs/astro-ph/9606103}
  {arXiv:astro-ph/9606103} \BibitemShut {NoStop}%
\bibitem [{\citenamefont {Iwasawa}\ \emph {et~al.}(1999)\citenamefont
  {Iwasawa}, \citenamefont {Fabian}, \citenamefont {Young}, \citenamefont
  {Inoue},\ and\ \citenamefont {Matsumoto}}]{Iwasawa1999}%
  \BibitemOpen
  \bibfield  {author} {\bibinfo {author} {\bibfnamefont {K.}~\bibnamefont
  {Iwasawa}}, \bibinfo {author} {\bibfnamefont {A.~C.}\ \bibnamefont {Fabian}},
  \bibinfo {author} {\bibfnamefont {A.~J.}\ \bibnamefont {Young}}, \bibinfo
  {author} {\bibfnamefont {H.}~\bibnamefont {Inoue}}, \ and\ \bibinfo {author}
  {\bibfnamefont {C.}~\bibnamefont {Matsumoto}},\ }\href {\doibase
  10.1046/j.1365-8711.1999.02671.x} {\bibfield  {journal} {\bibinfo  {journal}
  {Mon. Not. Roy. Astron. Soc.}\ }\textbf {\bibinfo {volume} {306}},\ \bibinfo
  {pages} {19} (\bibinfo {year} {1999})},\ \Eprint
  {http://arxiv.org/abs/astro-ph/9904078} {arXiv:astro-ph/9904078} \BibitemShut
  {NoStop}%
\bibitem [{\citenamefont {Guainazzi}\ \emph {et~al.}(1999)\citenamefont
  {Guainazzi} \emph {et~al.}}]{Guainazzi1998}%
  \BibitemOpen
  \bibfield  {author} {\bibinfo {author} {\bibfnamefont {M.}~\bibnamefont
  {Guainazzi}} \emph {et~al.},\ }\href@noop {} {\bibfield  {journal} {\bibinfo
  {journal} {Astron. Astrophys.}\ }\textbf {\bibinfo {volume} {341}},\ \bibinfo
  {pages} {L27} (\bibinfo {year} {1999})},\ \Eprint
  {http://arxiv.org/abs/astro-ph/9811246} {arXiv:astro-ph/9811246} \BibitemShut
  {NoStop}%
\bibitem [{\citenamefont {Lee}\ \emph {et~al.}(2000)\citenamefont {Lee},
  \citenamefont {Fabian}, \citenamefont {Reynolds}, \citenamefont {Brandt},\
  and\ \citenamefont {Iwasawa}}]{Lee2000}%
  \BibitemOpen
  \bibfield  {author} {\bibinfo {author} {\bibfnamefont {J.~C.}\ \bibnamefont
  {Lee}}, \bibinfo {author} {\bibfnamefont {A.~C.}\ \bibnamefont {Fabian}},
  \bibinfo {author} {\bibfnamefont {C.~S.}\ \bibnamefont {Reynolds}}, \bibinfo
  {author} {\bibfnamefont {W.~N.}\ \bibnamefont {Brandt}}, \ and\ \bibinfo
  {author} {\bibfnamefont {K.}~\bibnamefont {Iwasawa}},\ }\href {\doibase
  10.1046/j.1365-8711.2000.03835.x} {\bibfield  {journal} {\bibinfo  {journal}
  {Mon. Not. Roy. Astron. Soc.}\ }\textbf {\bibinfo {volume} {318}},\ \bibinfo
  {pages} {857} (\bibinfo {year} {2000})},\ \Eprint
  {http://arxiv.org/abs/astro-ph/9909239} {arXiv:astro-ph/9909239} \BibitemShut
  {NoStop}%
\bibitem [{\citenamefont {Vaughan}\ and\ \citenamefont
  {Edelson}(2001)}]{Vaughan2001}%
  \BibitemOpen
  \bibfield  {author} {\bibinfo {author} {\bibfnamefont {S.}~\bibnamefont
  {Vaughan}}\ and\ \bibinfo {author} {\bibfnamefont {R.}~\bibnamefont
  {Edelson}},\ }\href {\doibase 10.1086/319028} {\bibfield  {journal} {\bibinfo
   {journal} {Astrophys. J.}\ }\textbf {\bibinfo {volume} {548}},\ \bibinfo
  {pages} {694} (\bibinfo {year} {2001})},\ \Eprint
  {http://arxiv.org/abs/astro-ph/0010274} {arXiv:astro-ph/0010274} \BibitemShut
  {NoStop}%
\bibitem [{\citenamefont {Wilms}\ \emph {et~al.}(2001)\citenamefont {Wilms},
  \citenamefont {Reynolds}, \citenamefont {Begelman}, \citenamefont {Reeves},
  \citenamefont {Molendi}, \citenamefont {Stuabert},\ and\ \citenamefont
  {Kendziorra}}]{Wilms2001}%
  \BibitemOpen
  \bibfield  {author} {\bibinfo {author} {\bibfnamefont {J.}~\bibnamefont
  {Wilms}}, \bibinfo {author} {\bibfnamefont {C.~S.}\ \bibnamefont {Reynolds}},
  \bibinfo {author} {\bibfnamefont {M.~C.}\ \bibnamefont {Begelman}}, \bibinfo
  {author} {\bibfnamefont {J.}~\bibnamefont {Reeves}}, \bibinfo {author}
  {\bibfnamefont {S.}~\bibnamefont {Molendi}}, \bibinfo {author} {\bibfnamefont
  {R.}~\bibnamefont {Stuabert}}, \ and\ \bibinfo {author} {\bibfnamefont
  {E.}~\bibnamefont {Kendziorra}},\ }\href {\doibase
  10.1046/j.1365-8711.2001.05066.x} {\bibfield  {journal} {\bibinfo  {journal}
  {Mon. Not. Roy. Astron. Soc.}\ }\textbf {\bibinfo {volume} {328}},\ \bibinfo
  {pages} {L27} (\bibinfo {year} {2001})},\ \Eprint
  {http://arxiv.org/abs/astro-ph/0110520} {arXiv:astro-ph/0110520} \BibitemShut
  {NoStop}%
\bibitem [{\citenamefont {Fabian}\ \emph {et~al.}(2002)\citenamefont {Fabian},
  \citenamefont {Vaughan}, \citenamefont {Nandra}, \citenamefont {Iwasawa},
  \citenamefont {Ballantyne}, \citenamefont {Lee}, \citenamefont {De~Rosa},
  \citenamefont {Turner},\ and\ \citenamefont {Young}}]{Fabian2002}%
  \BibitemOpen
  \bibfield  {author} {\bibinfo {author} {\bibfnamefont {A.}~\bibnamefont
  {Fabian}}, \bibinfo {author} {\bibfnamefont {S.}~\bibnamefont {Vaughan}},
  \bibinfo {author} {\bibfnamefont {K.}~\bibnamefont {Nandra}}, \bibinfo
  {author} {\bibfnamefont {K.}~\bibnamefont {Iwasawa}}, \bibinfo {author}
  {\bibfnamefont {D.}~\bibnamefont {Ballantyne}}, \bibinfo {author}
  {\bibfnamefont {J.}~\bibnamefont {Lee}}, \bibinfo {author} {\bibfnamefont
  {A.}~\bibnamefont {De~Rosa}}, \bibinfo {author} {\bibfnamefont
  {A.}~\bibnamefont {Turner}}, \ and\ \bibinfo {author} {\bibfnamefont
  {A.}~\bibnamefont {Young}},\ }\href {\doibase
  10.1046/j.1365-8711.2002.05740.x} {\bibfield  {journal} {\bibinfo  {journal}
  {Mon. Not. Roy. Astron. Soc.}\ }\textbf {\bibinfo {volume} {335}},\ \bibinfo
  {pages} {L1} (\bibinfo {year} {2002})},\ \Eprint
  {http://arxiv.org/abs/astro-ph/0206095} {arXiv:astro-ph/0206095} \BibitemShut
  {NoStop}%
\bibitem [{\citenamefont {Fabian}\ and\ \citenamefont
  {Vaughan}(2003)}]{Fabian2003}%
  \BibitemOpen
  \bibfield  {author} {\bibinfo {author} {\bibfnamefont {A.~C.}\ \bibnamefont
  {Fabian}}\ and\ \bibinfo {author} {\bibfnamefont {S.}~\bibnamefont
  {Vaughan}},\ }\href {\doibase 10.1046/j.1365-8711.2003.06465.x} {\bibfield
  {journal} {\bibinfo  {journal} {Mon. Not. Roy. Astron. Soc.}\ }\textbf
  {\bibinfo {volume} {340}},\ \bibinfo {pages} {L28} (\bibinfo {year}
  {2003})},\ \Eprint {http://arxiv.org/abs/astro-ph/0301588}
  {arXiv:astro-ph/0301588} \BibitemShut {NoStop}%
\bibitem [{\citenamefont {Vaughan}\ and\ \citenamefont
  {Fabian}(2004)}]{Vaughan2004}%
  \BibitemOpen
  \bibfield  {author} {\bibinfo {author} {\bibfnamefont {S.}~\bibnamefont
  {Vaughan}}\ and\ \bibinfo {author} {\bibfnamefont {A.~C.}\ \bibnamefont
  {Fabian}},\ }\href {\doibase 10.1111/j.1365-2966.2004.07456.x} {\bibfield
  {journal} {\bibinfo  {journal} {Mon. Not. Roy. Astron. Soc.}\ }\textbf
  {\bibinfo {volume} {348}},\ \bibinfo {pages} {1415} (\bibinfo {year}
  {2004})},\ \Eprint {http://arxiv.org/abs/astro-ph/0311473}
  {arXiv:astro-ph/0311473} \BibitemShut {NoStop}%
\bibitem [{\citenamefont {Brenneman}\ and\ \citenamefont
  {Reynolds}(2006)}]{Brenneman2006}%
  \BibitemOpen
  \bibfield  {author} {\bibinfo {author} {\bibfnamefont {L.~W.}\ \bibnamefont
  {Brenneman}}\ and\ \bibinfo {author} {\bibfnamefont {C.~S.}\ \bibnamefont
  {Reynolds}},\ }\href {\doibase 10.1086/508146} {\bibfield  {journal}
  {\bibinfo  {journal} {Astrophys. J.}\ }\textbf {\bibinfo {volume} {652}},\
  \bibinfo {pages} {1028} (\bibinfo {year} {2006})},\ \Eprint
  {http://arxiv.org/abs/astro-ph/0608502} {arXiv:astro-ph/0608502} \BibitemShut
  {NoStop}%
\bibitem [{\citenamefont {Miniutti}\ \emph {et~al.}(2007)\citenamefont
  {Miniutti} \emph {et~al.}}]{Miniutti2007}%
  \BibitemOpen
  \bibfield  {author} {\bibinfo {author} {\bibfnamefont {G.}~\bibnamefont
  {Miniutti}} \emph {et~al.},\ }\href {\doibase 10.1093/pasj/59.sp1.S315}
  {\bibfield  {journal} {\bibinfo  {journal} {Publ. Astron. Soc. Jap.}\
  }\textbf {\bibinfo {volume} {59}},\ \bibinfo {pages} {315} (\bibinfo {year}
  {2007})},\ \Eprint {http://arxiv.org/abs/astro-ph/0609521}
  {arXiv:astro-ph/0609521} \BibitemShut {NoStop}%
\bibitem [{\citenamefont {Noda}\ \emph {et~al.}(2011)\citenamefont {Noda},
  \citenamefont {Makishima}, \citenamefont {Yamada}, \citenamefont {Torii},
  \citenamefont {Sakurai},\ and\ \citenamefont {Nakazawa}}]{Noda2011}%
  \BibitemOpen
  \bibfield  {author} {\bibinfo {author} {\bibfnamefont {H.}~\bibnamefont
  {Noda}}, \bibinfo {author} {\bibfnamefont {K.}~\bibnamefont {Makishima}},
  \bibinfo {author} {\bibfnamefont {S.}~\bibnamefont {Yamada}}, \bibinfo
  {author} {\bibfnamefont {S.}~\bibnamefont {Torii}}, \bibinfo {author}
  {\bibfnamefont {S.}~\bibnamefont {Sakurai}}, \ and\ \bibinfo {author}
  {\bibfnamefont {K.}~\bibnamefont {Nakazawa}},\ }\href {\doibase
  10.1093/pasj/63.sp3.S925} {\bibfield  {journal} {\bibinfo  {journal} {Publ.
  Astron. Soc. Jap.}\ }\textbf {\bibinfo {volume} {63}},\ \bibinfo {pages}
  {925} (\bibinfo {year} {2011})},\ \Eprint {http://arxiv.org/abs/1109.0457}
  {arXiv:1109.0457 [astro-ph.CO]} \BibitemShut {NoStop}%
\bibitem [{\citenamefont {Marinucci}\ \emph {et~al.}(2014)\citenamefont
  {Marinucci} \emph {et~al.}}]{Marinucci2014}%
  \BibitemOpen
  \bibfield  {author} {\bibinfo {author} {\bibfnamefont {A.}~\bibnamefont
  {Marinucci}} \emph {et~al.},\ }\href {\doibase 10.1088/0004-637X/787/1/83}
  {\bibfield  {journal} {\bibinfo  {journal} {Astrophys. J.}\ }\textbf
  {\bibinfo {volume} {787}},\ \bibinfo {pages} {83} (\bibinfo {year} {2014})},\
  \Eprint {http://arxiv.org/abs/1404.3561} {arXiv:1404.3561 [astro-ph.HE]}
  \BibitemShut {NoStop}%
\bibitem [{\citenamefont {Johannsen}(2013)}]{Johannsen2015}%
  \BibitemOpen
  \bibfield  {author} {\bibinfo {author} {\bibfnamefont {T.}~\bibnamefont
  {Johannsen}},\ }\href {\doibase 10.1103/PhysRevD.88.044002} {\bibfield
  {journal} {\bibinfo  {journal} {Phys. Rev.}\ }\textbf {\bibinfo {volume}
  {D88}},\ \bibinfo {pages} {044002} (\bibinfo {year} {2013})},\ \Eprint
  {http://arxiv.org/abs/1501.02809} {arXiv:1501.02809 [gr-qc]} \BibitemShut
  {NoStop}%
\bibitem [{\citenamefont {Otani}\ \emph {et~al.}(1996)\citenamefont {Otani}
  \emph {et~al.}}]{Otani1996}%
  \BibitemOpen
  \bibfield  {author} {\bibinfo {author} {\bibfnamefont {C.}~\bibnamefont
  {Otani}} \emph {et~al.},\ }\href {\doibase 10.1093/pasj/48.2.211} {\bibfield
  {journal} {\bibinfo  {journal} {Publ. Astron. Soc. Jap.}\ }\textbf {\bibinfo
  {volume} {48}},\ \bibinfo {pages} {211} (\bibinfo {year} {1996})},\ \Eprint
  {http://arxiv.org/abs/astro-ph/9511063} {arXiv:astro-ph/9511063} \BibitemShut
  {NoStop}%
\bibitem [{\citenamefont {Branduardi-Raymont}\ \emph
  {et~al.}(2001)\citenamefont {Branduardi-Raymont}, \citenamefont {Sako},
  \citenamefont {Kahn}, \citenamefont {Brinkman}, \citenamefont {Kaastra},\
  and\ \citenamefont {Page}}]{Brand2001}%
  \BibitemOpen
  \bibfield  {author} {\bibinfo {author} {\bibfnamefont {G.}~\bibnamefont
  {Branduardi-Raymont}}, \bibinfo {author} {\bibfnamefont {M.}~\bibnamefont
  {Sako}}, \bibinfo {author} {\bibfnamefont {S.~M.}\ \bibnamefont {Kahn}},
  \bibinfo {author} {\bibfnamefont {A.~C.}\ \bibnamefont {Brinkman}}, \bibinfo
  {author} {\bibfnamefont {J.~S.}\ \bibnamefont {Kaastra}}, \ and\ \bibinfo
  {author} {\bibfnamefont {M.~J.}\ \bibnamefont {Page}},\ }\href {\doibase
  10.1051/0004-6361:20000209} {\bibfield  {journal} {\bibinfo  {journal}
  {Astron. Astrophys.}\ }\textbf {\bibinfo {volume} {365}},\ \bibinfo {pages}
  {L140} (\bibinfo {year} {2001})},\ \Eprint
  {http://arxiv.org/abs/astro-ph/0011167} {arXiv:astro-ph/0011167} \BibitemShut
  {NoStop}%
\bibitem [{\citenamefont {Lee}\ \emph {et~al.}(2001)\citenamefont {Lee},
  \citenamefont {Ogle}, \citenamefont {Canizares}, \citenamefont {Marshall},
  \citenamefont {Schulz}, \citenamefont {Morales}, \citenamefont {Fabian},\
  and\ \citenamefont {Iwasawa}}]{Lee2001}%
  \BibitemOpen
  \bibfield  {author} {\bibinfo {author} {\bibfnamefont {J.~C.}\ \bibnamefont
  {Lee}}, \bibinfo {author} {\bibfnamefont {P.~M.}\ \bibnamefont {Ogle}},
  \bibinfo {author} {\bibfnamefont {C.~R.}\ \bibnamefont {Canizares}}, \bibinfo
  {author} {\bibfnamefont {H.~L.}\ \bibnamefont {Marshall}}, \bibinfo {author}
  {\bibfnamefont {N.~S.}\ \bibnamefont {Schulz}}, \bibinfo {author}
  {\bibfnamefont {R.}~\bibnamefont {Morales}}, \bibinfo {author} {\bibfnamefont
  {A.~C.}\ \bibnamefont {Fabian}}, \ and\ \bibinfo {author} {\bibfnamefont
  {K.}~\bibnamefont {Iwasawa}},\ }\href {\doibase 10.1086/320912} {\bibfield
  {journal} {\bibinfo  {journal} {Astrophys. J. Lett.}\ }\textbf {\bibinfo
  {volume} {554}},\ \bibinfo {pages} {L13} (\bibinfo {year} {2001})},\ \Eprint
  {http://arxiv.org/abs/astro-ph/0101065} {arXiv:astro-ph/0101065} \BibitemShut
  {NoStop}%
\bibitem [{\citenamefont {Young}\ \emph {et~al.}(2005)\citenamefont {Young},
  \citenamefont {Lee}, \citenamefont {Fabian}, \citenamefont {Reynolds},
  \citenamefont {Gibson},\ and\ \citenamefont {Canizares}}]{Young2005}%
  \BibitemOpen
  \bibfield  {author} {\bibinfo {author} {\bibfnamefont {A.~J.}\ \bibnamefont
  {Young}}, \bibinfo {author} {\bibfnamefont {J.~C.}\ \bibnamefont {Lee}},
  \bibinfo {author} {\bibfnamefont {A.~C.}\ \bibnamefont {Fabian}}, \bibinfo
  {author} {\bibfnamefont {C.~S.}\ \bibnamefont {Reynolds}}, \bibinfo {author}
  {\bibfnamefont {R.~R.}\ \bibnamefont {Gibson}}, \ and\ \bibinfo {author}
  {\bibfnamefont {C.~R.}\ \bibnamefont {Canizares}},\ }\href {\doibase
  10.1086/432607} {\bibfield  {journal} {\bibinfo  {journal} {Astrophys. J.}\
  }\textbf {\bibinfo {volume} {631}},\ \bibinfo {pages} {733} (\bibinfo {year}
  {2005})},\ \Eprint {http://arxiv.org/abs/astro-ph/0506082}
  {arXiv:astro-ph/0506082} \BibitemShut {NoStop}%
\bibitem [{\citenamefont {Turner}\ \emph {et~al.}(2003)\citenamefont {Turner},
  \citenamefont {Fabian}, \citenamefont {Vaughan},\ and\ \citenamefont
  {Lee}}]{Turner2003}%
  \BibitemOpen
  \bibfield  {author} {\bibinfo {author} {\bibfnamefont {A.~K.}\ \bibnamefont
  {Turner}}, \bibinfo {author} {\bibfnamefont {A.~C.}\ \bibnamefont {Fabian}},
  \bibinfo {author} {\bibfnamefont {S.}~\bibnamefont {Vaughan}}, \ and\
  \bibinfo {author} {\bibfnamefont {J.~C.}\ \bibnamefont {Lee}},\ }\href
  {\doibase 10.1111/j.1365-2966.2003.07127.x} {\bibfield  {journal} {\bibinfo
  {journal} {Mon. Not. Roy. Astron. Soc.}\ }\textbf {\bibinfo {volume} {346}},\
  \bibinfo {pages} {833} (\bibinfo {year} {2003})},\ \Eprint
  {http://arxiv.org/abs/astro-ph/0303418} {arXiv:astro-ph/0303418} \BibitemShut
  {NoStop}%
\bibitem [{\citenamefont {Turner}\ \emph {et~al.}(2004)\citenamefont {Turner},
  \citenamefont {Fabian}, \citenamefont {Lee},\ and\ \citenamefont
  {Vaughan}}]{Turner2004}%
  \BibitemOpen
  \bibfield  {author} {\bibinfo {author} {\bibfnamefont {A.~K.}\ \bibnamefont
  {Turner}}, \bibinfo {author} {\bibfnamefont {A.~C.}\ \bibnamefont {Fabian}},
  \bibinfo {author} {\bibfnamefont {J.~C.}\ \bibnamefont {Lee}}, \ and\
  \bibinfo {author} {\bibfnamefont {S.}~\bibnamefont {Vaughan}},\ }\href
  {\doibase 10.1111/j.1365-2966.2004.08075.x} {\bibfield  {journal} {\bibinfo
  {journal} {Mon. Not. Roy. Astron. Soc.}\ }\textbf {\bibinfo {volume} {353}},\
  \bibinfo {pages} {319} (\bibinfo {year} {2004})},\ \Eprint
  {http://arxiv.org/abs/astro-ph/0405570} {arXiv:astro-ph/0405570} \BibitemShut
  {NoStop}%
\bibitem [{\citenamefont {Jansen}\ \emph {et~al.}(2001)\citenamefont {Jansen}
  \emph {et~al.}}]{Jansen2001}%
  \BibitemOpen
  \bibfield  {author} {\bibinfo {author} {\bibfnamefont {F.}~\bibnamefont
  {Jansen}} \emph {et~al.},\ }\href {\doibase 10.1051/0004-6361:20000036}
  {\bibfield  {journal} {\bibinfo  {journal} {Astron. Astrophys.}\ }\textbf
  {\bibinfo {volume} {365}},\ \bibinfo {pages} {L1} (\bibinfo {year}
  {2001})}\BibitemShut {NoStop}%
\bibitem [{\citenamefont {Struder}\ \emph {et~al.}(2001)\citenamefont {Struder}
  \emph {et~al.}}]{struder2001}%
  \BibitemOpen
  \bibfield  {author} {\bibinfo {author} {\bibfnamefont {L.}~\bibnamefont
  {Struder}} \emph {et~al.},\ }\href {\doibase 10.1051/0004-6361:20000066}
  {\bibfield  {journal} {\bibinfo  {journal} {Astron. Astrophys.}\ }\textbf
  {\bibinfo {volume} {365}},\ \bibinfo {pages} {L18} (\bibinfo {year}
  {2001})}\BibitemShut {NoStop}%
\bibitem [{\citenamefont {Turner}\ \emph {et~al.}(2001)\citenamefont {Turner}
  \emph {et~al.}}]{Turner2001}%
  \BibitemOpen
  \bibfield  {author} {\bibinfo {author} {\bibfnamefont {M.~J.~L.}\
  \bibnamefont {Turner}} \emph {et~al.},\ }\href {\doibase
  10.1051/0004-6361:20000087} {\bibfield  {journal} {\bibinfo  {journal}
  {Astron. Astrophys.}\ }\textbf {\bibinfo {volume} {365}},\ \bibinfo {pages}
  {L27} (\bibinfo {year} {2001})},\ \Eprint
  {http://arxiv.org/abs/astro-ph/0011498} {arXiv:astro-ph/0011498} \BibitemShut
  {NoStop}%
\bibitem [{\citenamefont {Harrison}\ \emph {et~al.}(2013)\citenamefont
  {Harrison} \emph {et~al.}}]{Harrison2013}%
  \BibitemOpen
  \bibfield  {author} {\bibinfo {author} {\bibfnamefont {F.~A.}\ \bibnamefont
  {Harrison}} \emph {et~al.},\ }\href {\doibase 10.1088/0004-637X/770/2/103}
  {\bibfield  {journal} {\bibinfo  {journal} {Astrophys. J.}\ }\textbf
  {\bibinfo {volume} {770}},\ \bibinfo {pages} {103} (\bibinfo {year}
  {2013})},\ \Eprint {http://arxiv.org/abs/1301.7307} {arXiv:1301.7307
  [astro-ph.IM]} \BibitemShut {NoStop}%
\bibitem [{\citenamefont {{Arnaud}}(1996)}]{Arnaud1996}%
  \BibitemOpen
  \bibfield  {author} {\bibinfo {author} {\bibfnamefont {K.~A.}\ \bibnamefont
  {{Arnaud}}},\ }in\ \href@noop {} {\emph {\bibinfo {booktitle} {Astronomical
  Data Analysis Software and Systems V}}},\ \bibinfo {series} {Astronomical
  Society of the Pacific Conference Series}, Vol.\ \bibinfo {volume} {101},\
  \bibinfo {editor} {edited by\ \bibinfo {editor} {\bibfnamefont {G.~H.}\
  \bibnamefont {{Jacoby}}}\ and\ \bibinfo {editor} {\bibfnamefont
  {J.}~\bibnamefont {{Barnes}}}}\ (\bibinfo {year} {1996})\ p.~\bibinfo {pages}
  {17}\BibitemShut {NoStop}%
\bibitem [{\citenamefont {Wilms}\ \emph {et~al.}(2000)\citenamefont {Wilms},
  \citenamefont {Allen},\ and\ \citenamefont {McCray}}]{Wilms2000}%
  \BibitemOpen
  \bibfield  {author} {\bibinfo {author} {\bibfnamefont {J.}~\bibnamefont
  {Wilms}}, \bibinfo {author} {\bibfnamefont {A.}~\bibnamefont {Allen}}, \ and\
  \bibinfo {author} {\bibfnamefont {R.}~\bibnamefont {McCray}},\ }\href
  {\doibase 10.1086/317016} {\bibfield  {journal} {\bibinfo  {journal}
  {Astrophys. J.}\ }\textbf {\bibinfo {volume} {542}},\ \bibinfo {pages} {914}
  (\bibinfo {year} {2000})},\ \Eprint {http://arxiv.org/abs/astro-ph/0008425}
  {arXiv:astro-ph/0008425} \BibitemShut {NoStop}%
\bibitem [{\citenamefont {Verner}\ \emph {et~al.}(1996)\citenamefont {Verner},
  \citenamefont {Ferland}, \citenamefont {Korista},\ and\ \citenamefont
  {Yakovlev}}]{Verner1996}%
  \BibitemOpen
  \bibfield  {author} {\bibinfo {author} {\bibfnamefont {D.~A.}\ \bibnamefont
  {Verner}}, \bibinfo {author} {\bibfnamefont {G.~J.}\ \bibnamefont {Ferland}},
  \bibinfo {author} {\bibfnamefont {K.~T.}\ \bibnamefont {Korista}}, \ and\
  \bibinfo {author} {\bibfnamefont {D.~G.}\ \bibnamefont {Yakovlev}},\ }\href
  {\doibase 10.1086/177435} {\bibfield  {journal} {\bibinfo  {journal}
  {Astrophys. J.}\ }\textbf {\bibinfo {volume} {465}},\ \bibinfo {pages} {487}
  (\bibinfo {year} {1996})},\ \Eprint {http://arxiv.org/abs/astro-ph/9601009}
  {arXiv:astro-ph/9601009} \BibitemShut {NoStop}%
\bibitem [{\citenamefont {Sako}\ \emph {et~al.}(2003)\citenamefont {Sako} \emph
  {et~al.}}]{Sako2003}%
  \BibitemOpen
  \bibfield  {author} {\bibinfo {author} {\bibfnamefont {M.}~\bibnamefont
  {Sako}} \emph {et~al.},\ }\href {\doibase 10.1086/377575} {\bibfield
  {journal} {\bibinfo  {journal} {Astrophys. J.}\ }\textbf {\bibinfo {volume}
  {596}},\ \bibinfo {pages} {114} (\bibinfo {year} {2003})},\ \Eprint
  {http://arxiv.org/abs/astro-ph/0112436} {arXiv:astro-ph/0112436} \BibitemShut
  {NoStop}%
\bibitem [{\citenamefont {Gierlinski}\ and\ \citenamefont
  {Done}(2004)}]{Gierlinski2004}%
  \BibitemOpen
  \bibfield  {author} {\bibinfo {author} {\bibfnamefont {M.}~\bibnamefont
  {Gierlinski}}\ and\ \bibinfo {author} {\bibfnamefont {C.}~\bibnamefont
  {Done}},\ }\href {\doibase 10.1111/j.1365-2966.2004.07687.x} {\bibfield
  {journal} {\bibinfo  {journal} {Mon. Not. Roy. Astron. Soc.}\ }\textbf
  {\bibinfo {volume} {349}},\ \bibinfo {pages} {L7} (\bibinfo {year} {2004})},\
  \Eprint {http://arxiv.org/abs/astro-ph/0312271} {arXiv:astro-ph/0312271}
  \BibitemShut {NoStop}%
\bibitem [{\citenamefont {Crummy}\ \emph {et~al.}(2006)\citenamefont {Crummy},
  \citenamefont {Fabian}, \citenamefont {Gallo},\ and\ \citenamefont
  {Ross}}]{Crummy2006}%
  \BibitemOpen
  \bibfield  {author} {\bibinfo {author} {\bibfnamefont {J.}~\bibnamefont
  {Crummy}}, \bibinfo {author} {\bibfnamefont {A.~C.}\ \bibnamefont {Fabian}},
  \bibinfo {author} {\bibfnamefont {L.}~\bibnamefont {Gallo}}, \ and\ \bibinfo
  {author} {\bibfnamefont {R.}~\bibnamefont {Ross}},\ }\href {\doibase
  10.1111/j.1365-2966.2005.09844.x} {\bibfield  {journal} {\bibinfo  {journal}
  {Mon. Not. Roy. Astron. Soc.}\ }\textbf {\bibinfo {volume} {365}},\ \bibinfo
  {pages} {1067} (\bibinfo {year} {2006})},\ \Eprint
  {http://arxiv.org/abs/astro-ph/0511457} {arXiv:astro-ph/0511457} \BibitemShut
  {NoStop}%
\bibitem [{\citenamefont {Miniutti}\ \emph {et~al.}(2009)\citenamefont
  {Miniutti}, \citenamefont {Ponti}, \citenamefont {Greene}, \citenamefont
  {Ho}, \citenamefont {Fabian},\ and\ \citenamefont {Iwasawa}}]{Miniutti2009}%
  \BibitemOpen
  \bibfield  {author} {\bibinfo {author} {\bibfnamefont {G.}~\bibnamefont
  {Miniutti}}, \bibinfo {author} {\bibfnamefont {G.}~\bibnamefont {Ponti}},
  \bibinfo {author} {\bibfnamefont {J.~E.}\ \bibnamefont {Greene}}, \bibinfo
  {author} {\bibfnamefont {L.~C.}\ \bibnamefont {Ho}}, \bibinfo {author}
  {\bibfnamefont {A.~C.}\ \bibnamefont {Fabian}}, \ and\ \bibinfo {author}
  {\bibfnamefont {K.}~\bibnamefont {Iwasawa}},\ }\href {\doibase
  10.1111/j.1365-2966.2008.14334.x} {\bibfield  {journal} {\bibinfo  {journal}
  {Mon. Not. Roy. Astron. Soc.}\ }\textbf {\bibinfo {volume} {394}},\ \bibinfo
  {pages} {443} (\bibinfo {year} {2009})},\ \Eprint
  {http://arxiv.org/abs/0812.1652} {arXiv:0812.1652 [astro-ph]} \BibitemShut
  {NoStop}%
\bibitem [{\citenamefont {Walton}\ \emph {et~al.}(2014)\citenamefont {Walton}
  \emph {et~al.}}]{Walton2014}%
  \BibitemOpen
  \bibfield  {author} {\bibinfo {author} {\bibfnamefont {D.~J.}\ \bibnamefont
  {Walton}} \emph {et~al.},\ }\href {\doibase 10.1088/0004-637X/788/1/76}
  {\bibfield  {journal} {\bibinfo  {journal} {Astrophys. J.}\ }\textbf
  {\bibinfo {volume} {788}},\ \bibinfo {pages} {76} (\bibinfo {year} {2014})},\
  \Eprint {http://arxiv.org/abs/1404.5620} {arXiv:1404.5620 [astro-ph.HE]}
  \BibitemShut {NoStop}%
\bibitem [{\citenamefont {Dickey}\ and\ \citenamefont
  {Lockman}(1990)}]{Dickey1990}%
  \BibitemOpen
  \bibfield  {author} {\bibinfo {author} {\bibfnamefont {J.~M.}\ \bibnamefont
  {Dickey}}\ and\ \bibinfo {author} {\bibfnamefont {F.~J.}\ \bibnamefont
  {Lockman}},\ }\href {\doibase 10.1146/annurev.aa.28.090190.001243} {\bibfield
   {journal} {\bibinfo  {journal} {Ann. Rev. Astron. Astrophys.}\ }\textbf
  {\bibinfo {volume} {28}},\ \bibinfo {pages} {215} (\bibinfo {year}
  {1990})}\BibitemShut {NoStop}%
\bibitem [{\citenamefont {Leighly}\ \emph {et~al.}(1997)\citenamefont
  {Leighly}, \citenamefont {Mushotzky}, \citenamefont {Nandra},\ and\
  \citenamefont {Forster}}]{Leighly1997}%
  \BibitemOpen
  \bibfield  {author} {\bibinfo {author} {\bibfnamefont {K.~M.}\ \bibnamefont
  {Leighly}}, \bibinfo {author} {\bibfnamefont {R.~F.}\ \bibnamefont
  {Mushotzky}}, \bibinfo {author} {\bibfnamefont {K.}~\bibnamefont {Nandra}}, \
  and\ \bibinfo {author} {\bibfnamefont {K.}~\bibnamefont {Forster}},\ }\href
  {\doibase 10.1086/310950} {\bibfield  {journal} {\bibinfo  {journal}
  {Astrophys. J. Lett.}\ }\textbf {\bibinfo {volume} {489}},\ \bibinfo {pages}
  {L25} (\bibinfo {year} {1997})},\ \Eprint
  {http://arxiv.org/abs/astro-ph/9708112} {arXiv:astro-ph/9708112} \BibitemShut
  {NoStop}%
\bibitem [{\citenamefont {Nandra}\ \emph {et~al.}(2013)\citenamefont {Nandra}
  \emph {et~al.}}]{Nandra:2013jka}%
  \BibitemOpen
  \bibfield  {author} {\bibinfo {author} {\bibfnamefont {K.}~\bibnamefont
  {Nandra}} \emph {et~al.},\ }\href@noop {} {\  (\bibinfo {year} {2013})},\
  \Eprint {http://arxiv.org/abs/1306.2307} {arXiv:1306.2307 [astro-ph.HE]}
  \BibitemShut {NoStop}%
\bibitem [{\citenamefont {Zhang}\ \emph {et~al.}(2016)\citenamefont {Zhang}
  \emph {et~al.}}]{Zhang:2016ach}%
  \BibitemOpen
  \bibfield  {author} {\bibinfo {author} {\bibfnamefont {S.~N.}\ \bibnamefont
  {Zhang}} \emph {et~al.} (\bibinfo {collaboration} {eXTP}),\ }\href {\doibase
  10.1117/12.2232034} {\bibfield  {journal} {\bibinfo  {journal} {Proc. SPIE
  Int. Soc. Opt. Eng.}\ }\textbf {\bibinfo {volume} {9905}},\ \bibinfo {pages}
  {99051Q} (\bibinfo {year} {2016})},\ \Eprint
  {http://arxiv.org/abs/1607.08823} {arXiv:1607.08823 [astro-ph.IM]}
  \BibitemShut {NoStop}%
\bibitem [{\citenamefont {Kinch}\ \emph {et~al.}(2016)\citenamefont {Kinch},
  \citenamefont {Schnittman}, \citenamefont {Kallman},\ and\ \citenamefont
  {Krolik}}]{Kinch:2016ipi}%
  \BibitemOpen
  \bibfield  {author} {\bibinfo {author} {\bibfnamefont {B.~E.}\ \bibnamefont
  {Kinch}}, \bibinfo {author} {\bibfnamefont {J.~D.}\ \bibnamefont
  {Schnittman}}, \bibinfo {author} {\bibfnamefont {T.~R.}\ \bibnamefont
  {Kallman}}, \ and\ \bibinfo {author} {\bibfnamefont {J.~H.}\ \bibnamefont
  {Krolik}},\ }\href {\doibase 10.3847/0004-637X/826/1/52} {\bibfield
  {journal} {\bibinfo  {journal} {Astrophys. J.}\ }\textbf {\bibinfo {volume}
  {826}},\ \bibinfo {pages} {52} (\bibinfo {year} {2016})},\ \Eprint
  {http://arxiv.org/abs/1604.01126} {arXiv:1604.01126 [astro-ph.HE]}
  \BibitemShut {NoStop}%
\bibitem [{\citenamefont {Kinch}\ \emph {et~al.}(2019)\citenamefont {Kinch},
  \citenamefont {Schnittman}, \citenamefont {Kallman},\ and\ \citenamefont
  {Krolik}}]{Kinch:2018ceh}%
  \BibitemOpen
  \bibfield  {author} {\bibinfo {author} {\bibfnamefont {B.~E.}\ \bibnamefont
  {Kinch}}, \bibinfo {author} {\bibfnamefont {J.~D.}\ \bibnamefont
  {Schnittman}}, \bibinfo {author} {\bibfnamefont {T.~R.}\ \bibnamefont
  {Kallman}}, \ and\ \bibinfo {author} {\bibfnamefont {J.~H.}\ \bibnamefont
  {Krolik}},\ }\href {\doibase 10.3847/1538-4357/ab05d5} {\bibfield  {journal}
  {\bibinfo  {journal} {Astrophys. J.}\ }\textbf {\bibinfo {volume} {873}},\
  \bibinfo {pages} {71} (\bibinfo {year} {2019})},\ \Eprint
  {http://arxiv.org/abs/1810.13099} {arXiv:1810.13099 [astro-ph.HE]}
  \BibitemShut {NoStop}%
\bibitem [{\citenamefont {Tripathi}\ \emph
  {et~al.}(2021{\natexlab{c}})\citenamefont {Tripathi}, \citenamefont
  {Abdikamalov}, \citenamefont {Ayzenberg}, \citenamefont {Bambi},\ and\
  \citenamefont {Liu}}]{Tripathi:2021wap}%
  \BibitemOpen
  \bibfield  {author} {\bibinfo {author} {\bibfnamefont {A.}~\bibnamefont
  {Tripathi}}, \bibinfo {author} {\bibfnamefont {A.~B.}\ \bibnamefont
  {Abdikamalov}}, \bibinfo {author} {\bibfnamefont {D.}~\bibnamefont
  {Ayzenberg}}, \bibinfo {author} {\bibfnamefont {C.}~\bibnamefont {Bambi}}, \
  and\ \bibinfo {author} {\bibfnamefont {H.}~\bibnamefont {Liu}},\ }\href
  {\doibase 10.3847/1538-4357/abf6c5} {\bibfield  {journal} {\bibinfo
  {journal} {Astrophys. J.}\ }\textbf {\bibinfo {volume} {913}},\ \bibinfo
  {pages} {129} (\bibinfo {year} {2021}{\natexlab{c}})},\ \Eprint
  {http://arxiv.org/abs/2102.04695} {arXiv:2102.04695 [astro-ph.HE]}
  \BibitemShut {NoStop}%
\bibitem [{\citenamefont {Cardenas-Avendano}\ \emph
  {et~al.}(2020{\natexlab{b}})\citenamefont {Cardenas-Avendano}, \citenamefont
  {Zhou},\ and\ \citenamefont {Bambi}}]{Cardenas-Avendano:2020xtw}%
  \BibitemOpen
  \bibfield  {author} {\bibinfo {author} {\bibfnamefont {A.}~\bibnamefont
  {Cardenas-Avendano}}, \bibinfo {author} {\bibfnamefont {M.}~\bibnamefont
  {Zhou}}, \ and\ \bibinfo {author} {\bibfnamefont {C.}~\bibnamefont {Bambi}},\
  }\href {\doibase 10.1103/PhysRevD.101.123014} {\bibfield  {journal} {\bibinfo
   {journal} {Phys. Rev. D}\ }\textbf {\bibinfo {volume} {101}},\ \bibinfo
  {pages} {123014} (\bibinfo {year} {2020}{\natexlab{b}})},\ \Eprint
  {http://arxiv.org/abs/2005.06719} {arXiv:2005.06719 [astro-ph.HE]}
  \BibitemShut {NoStop}%
\bibitem [{\citenamefont {Abdikamalov}\ \emph {et~al.}(2021)\citenamefont
  {Abdikamalov}, \citenamefont {Ayzenberg}, \citenamefont {Bambi},
  \citenamefont {Liu},\ and\ \citenamefont {Zhang}}]{Abdikamalov:2021rty}%
  \BibitemOpen
  \bibfield  {author} {\bibinfo {author} {\bibfnamefont {A.~B.}\ \bibnamefont
  {Abdikamalov}}, \bibinfo {author} {\bibfnamefont {D.}~\bibnamefont
  {Ayzenberg}}, \bibinfo {author} {\bibfnamefont {C.}~\bibnamefont {Bambi}},
  \bibinfo {author} {\bibfnamefont {H.}~\bibnamefont {Liu}}, \ and\ \bibinfo
  {author} {\bibfnamefont {Y.}~\bibnamefont {Zhang}},\ }\href {\doibase
  10.1103/PhysRevD.103.103023} {\bibfield  {journal} {\bibinfo  {journal}
  {Phys. Rev. D}\ }\textbf {\bibinfo {volume} {103}},\ \bibinfo {pages}
  {103023} (\bibinfo {year} {2021})},\ \Eprint
  {http://arxiv.org/abs/2101.10100} {arXiv:2101.10100 [astro-ph.HE]}
  \BibitemShut {NoStop}%
\bibitem [{\citenamefont {Riaz}\ \emph {et~al.}(2021)\citenamefont {Riaz},
  \citenamefont {Szanecki}, \citenamefont {Nied\'zwiecki}, \citenamefont
  {Ayzenberg},\ and\ \citenamefont {Bambi}}]{Riaz:2020zqb}%
  \BibitemOpen
  \bibfield  {author} {\bibinfo {author} {\bibfnamefont {S.}~\bibnamefont
  {Riaz}}, \bibinfo {author} {\bibfnamefont {M.~L.}\ \bibnamefont {Szanecki}},
  \bibinfo {author} {\bibfnamefont {A.}~\bibnamefont {Nied\'zwiecki}}, \bibinfo
  {author} {\bibfnamefont {D.}~\bibnamefont {Ayzenberg}}, \ and\ \bibinfo
  {author} {\bibfnamefont {C.}~\bibnamefont {Bambi}},\ }\href {\doibase
  10.3847/1538-4357/abe2a3} {\bibfield  {journal} {\bibinfo  {journal}
  {Astrophys. J.}\ }\textbf {\bibinfo {volume} {910}},\ \bibinfo {pages} {49}
  (\bibinfo {year} {2021})},\ \Eprint {http://arxiv.org/abs/2006.15838}
  {arXiv:2006.15838 [astro-ph.HE]} \BibitemShut {NoStop}%
\bibitem [{\citenamefont {Mizuno}\ \emph {et~al.}(2018)\citenamefont {Mizuno},
  \citenamefont {Younsi}, \citenamefont {Fromm}, \citenamefont {Porth},
  \citenamefont {De~Laurentis}, \citenamefont {Olivares}, \citenamefont
  {Falcke}, \citenamefont {Kramer},\ and\ \citenamefont
  {Rezzolla}}]{Mizuno:2018lxz}%
  \BibitemOpen
  \bibfield  {author} {\bibinfo {author} {\bibfnamefont {Y.}~\bibnamefont
  {Mizuno}}, \bibinfo {author} {\bibfnamefont {Z.}~\bibnamefont {Younsi}},
  \bibinfo {author} {\bibfnamefont {C.~M.}\ \bibnamefont {Fromm}}, \bibinfo
  {author} {\bibfnamefont {O.}~\bibnamefont {Porth}}, \bibinfo {author}
  {\bibfnamefont {M.}~\bibnamefont {De~Laurentis}}, \bibinfo {author}
  {\bibfnamefont {H.}~\bibnamefont {Olivares}}, \bibinfo {author}
  {\bibfnamefont {H.}~\bibnamefont {Falcke}}, \bibinfo {author} {\bibfnamefont
  {M.}~\bibnamefont {Kramer}}, \ and\ \bibinfo {author} {\bibfnamefont
  {L.}~\bibnamefont {Rezzolla}},\ }\href {\doibase 10.1038/s41550-018-0449-5}
  {\bibfield  {journal} {\bibinfo  {journal} {Nature Astron.}\ }\textbf
  {\bibinfo {volume} {2}},\ \bibinfo {pages} {585} (\bibinfo {year} {2018})},\
  \Eprint {http://arxiv.org/abs/1804.05812} {arXiv:1804.05812 [astro-ph.GA]}
  \BibitemShut {NoStop}%
\bibitem [{\citenamefont {Fromm}\ \emph {et~al.}(2021)\citenamefont {Fromm},
  \citenamefont {Mizuno}, \citenamefont {Younsi}, \citenamefont {Olivares},
  \citenamefont {Porth}, \citenamefont {De~Laurentis}, \citenamefont {Falcke},
  \citenamefont {Kramer},\ and\ \citenamefont {Rezzolla}}]{Fromm:2021flr}%
  \BibitemOpen
  \bibfield  {author} {\bibinfo {author} {\bibfnamefont {C.~M.}\ \bibnamefont
  {Fromm}}, \bibinfo {author} {\bibfnamefont {Y.}~\bibnamefont {Mizuno}},
  \bibinfo {author} {\bibfnamefont {Z.}~\bibnamefont {Younsi}}, \bibinfo
  {author} {\bibfnamefont {H.}~\bibnamefont {Olivares}}, \bibinfo {author}
  {\bibfnamefont {O.}~\bibnamefont {Porth}}, \bibinfo {author} {\bibfnamefont
  {M.}~\bibnamefont {De~Laurentis}}, \bibinfo {author} {\bibfnamefont
  {H.}~\bibnamefont {Falcke}}, \bibinfo {author} {\bibfnamefont
  {M.}~\bibnamefont {Kramer}}, \ and\ \bibinfo {author} {\bibfnamefont
  {L.}~\bibnamefont {Rezzolla}},\ }\href@noop {} {\  (\bibinfo {year}
  {2021})},\ \Eprint {http://arxiv.org/abs/2101.08618} {arXiv:2101.08618
  [astro-ph.HE]} \BibitemShut {NoStop}%
\bibitem [{\citenamefont {Destounis}\ \emph {et~al.}(2020)\citenamefont
  {Destounis}, \citenamefont {Suvorov},\ and\ \citenamefont
  {Kokkotas}}]{Destounis:2020kss}%
  \BibitemOpen
  \bibfield  {author} {\bibinfo {author} {\bibfnamefont {K.}~\bibnamefont
  {Destounis}}, \bibinfo {author} {\bibfnamefont {A.~G.}\ \bibnamefont
  {Suvorov}}, \ and\ \bibinfo {author} {\bibfnamefont {K.~D.}\ \bibnamefont
  {Kokkotas}},\ }\href {\doibase 10.1103/PhysRevD.102.064041} {\bibfield
  {journal} {\bibinfo  {journal} {Phys. Rev. D}\ }\textbf {\bibinfo {volume}
  {102}},\ \bibinfo {pages} {064041} (\bibinfo {year} {2020})},\ \Eprint
  {http://arxiv.org/abs/2009.00028} {arXiv:2009.00028 [gr-qc]} \BibitemShut
  {NoStop}%
\bibitem [{\citenamefont {Destounis}\ \emph {et~al.}(2021)\citenamefont
  {Destounis}, \citenamefont {Suvorov},\ and\ \citenamefont
  {Kokkotas}}]{Destounis:2021mqv}%
  \BibitemOpen
  \bibfield  {author} {\bibinfo {author} {\bibfnamefont {K.}~\bibnamefont
  {Destounis}}, \bibinfo {author} {\bibfnamefont {A.~G.}\ \bibnamefont
  {Suvorov}}, \ and\ \bibinfo {author} {\bibfnamefont {K.~D.}\ \bibnamefont
  {Kokkotas}},\ }\href@noop {} {\  (\bibinfo {year} {2021})},\ \Eprint
  {http://arxiv.org/abs/2103.05643} {arXiv:2103.05643 [gr-qc]} \BibitemShut
  {NoStop}%
\bibitem [{\citenamefont {Konoplya}\ and\ \citenamefont
  {Zhidenko}(2020)}]{Konoplya:2020hyk}%
  \BibitemOpen
  \bibfield  {author} {\bibinfo {author} {\bibfnamefont {R.~A.}\ \bibnamefont
  {Konoplya}}\ and\ \bibinfo {author} {\bibfnamefont {A.}~\bibnamefont
  {Zhidenko}},\ }\href {\doibase 10.1103/PhysRevD.101.124004} {\bibfield
  {journal} {\bibinfo  {journal} {Phys. Rev. D}\ }\textbf {\bibinfo {volume}
  {101}},\ \bibinfo {pages} {124004} (\bibinfo {year} {2020})},\ \Eprint
  {http://arxiv.org/abs/2001.06100} {arXiv:2001.06100 [gr-qc]} \BibitemShut
  {NoStop}%
\end{thebibliography}%
\end{document}